\documentclass[conference,onecolumn]{IEEEtran}

\IEEEoverridecommandlockouts                              
\usepackage{enumerate}

\usepackage{graphicx, epsfig}\usepackage{epstopdf}

\newtheorem{Remark}{Remark}

\usepackage{algorithmic}
\usepackage{algorithm}

\usepackage{amsmath}
\usepackage{wasysym}

\usepackage{stfloats} 

\usepackage{url}

\usepackage{amsfonts}
\usepackage{amssymb}

\usepackage{multirow}

\makeatletter

\newcommand{\Rmnum}[1]{\expandafter\@slowromancap\romannumeral #1@}
\makeatother

\usepackage{authblk}
\title{\LARGE \bf Close-Form Design of Antenna-Constrained  Multi-Cell Multi-User Downlink Interference Alignment}
\author{Haichuan Zhou, Tharm Ratnarajah$\dag$\\
 The Institute of Electronics, Communications and Information Technology (ECIT),\\
Queen's University Belfast, Queen's Road, Belfast, UK\\
$\dag$The University of Edinburgh, Edinburgh, UK\\
Email: hzhou01@qub.ac.uk }

\begin{document}

\maketitle \thispagestyle{empty} \pagestyle{empty}

\begin {abstract}
This paper investigates the downlink channels in multi-cell
multi-user interfering networks. The goal is to propose close-form
designs to obtain degrees of freedom (DoF) in high SNR region for
the network composed of base stations (BS) as transmitters and
mobile stations (MS) as receivers. Consider the realistic system,
both BS and MS have finite antennas, so that the design of
interference alignment is highly constrained by the feasibility
conditions. The focus of design is to explore potential
opportunities of alignment in the subspace both from the BS transmit
side and from the MS receive side. The new IA schemes for cellular
downlink channels are in the form of causal dynamic processes in
contrary to conventional static IA schemes. For different
implementations, system conditions are compared from all aspects,
which include antenna usage, CSI overhead and computational
complexity. This research scope covers a wide range of typical
multi-cell multi-user network models. The first one is a $K$-cell
fully connected cellular network; the second one is a Wyner cyclic
cellular network with two adjacent interfering links; the third one
is a Wyner cyclic cellular network with single adjacent interfering
link considering cell-edge and cell-interior users respectively.
\end {abstract}

\IEEEpeerreviewmaketitle

\section{Introduction}






Cellular networks have been intensively studied for years from all
kinds of theoretical models to various complicated realistic
settings. The single cell model has been investigated in the
prevalent scenario of multi-user MIMO (MU-MIMO) networks
\cite{cellular-network-MIMO-archi-MU-efficiency},
\cite{MU-MIMO-achievrates-downlink-training-CSI},
\cite{adapt-random-BF-IS-selection-cellular}. In MU-MIMO, the base
stations (BS) simultaneously transmit to multiple receivers, i.e.
mobile stations (MS), which form a typical MIMO broadcast downlink
channel. The self-interferences among multiple streams cause
performance degradation for each MS. While with perfect channel
state information both at the transmitters (CSIT) and at the
receivers (CSIR), capacity could be achieved by using Gaussian
codes, linear beamforming as well as dirty-paper-coding (DPC). Since
DPC is hard to implement, opportunistic beamforming combined with
user selection could also asymptotically approach the gain. The
maximum number of degrees of freedom (DoF) in MU-MIMO is
$\min(N_t,KN_r)$, where $N_t$ is the number of BS transmit antennas,
$K$ is the number of MS users, and $N_r$ is the number of MS receive
antennas. Recently, the model of multi-cell network also receives a
rich body of research
\cite{multi-cell-cooperative-newlook-interference},
\cite{MU-limited-inter-cell-approximate-IA-time-frequency-space},
\cite{multi-cell-downlink-capacity-coordinated-processing}. Its
downlink scenario is coined as the interfering broadcast channel
(IFBC or IBC), while the uplink channel is called the interfering
multiple access (IMAC) channel. From a practical perspective, the
multi-cell and multi-user downlink transmission schemes have been
actively discussed for fourth-generation (4G) cellular systems such
as network MIMO and coordinated multi-point communication (CoMP) in
the interference-limited environment
\cite{future-wireless-broadband-challenge-possibility}. In
particular, systems such as LTE-Advanced as well as IEEE802.16m
require a great increase in cell-edge spectral efficiency over
previous 3G systems \cite{physical-layer-aspects-EUTRA},
\cite{IEEE80216mEMD}. Interferences between cells and users are the
key bottleneck for the whole network to boost system capacity, so
that a common goal of these schemes is to effectively mitigate
interferences. Thus interference management techniques are
intensively investigated to improving cell-edge throughput
\cite{multi-cell-cooperative-newlook-interference}. It was shown
that near interference-free throughput performance can be achieved
in the cellular network \cite{IA-cellular}. In our following work,
only downlink channels are considered, and all the parts focus on
the model of IFBC channels.




\subsection{Multi-Cell Downlink Interference Alignment Methods}

Interference management based on multi-cell cooperation can
dramatically improve the system performance. The recently emerging
technique of interference alignment (IA) is a promising interference
management scheme to mitigate interferences in term of DoF in high
SNR region. However, since all the scenarios and schemes only
consider time-invariant/constant channels regarding practical
requirements, there are finite spatial signal dimensions created by
multiple antennas on each node without any frequency or time
extensions. Then it is hard to achieve interference alignment with
only finite dimensions either by using numerical algorithms or
analytical designs.

Regarding the methods of iterative algorithms, there are not many
results from initial studies. \cite{IA-MIMO-cellular} explicitly
explores the feasibility condition of linear interference alignment
in constant MIMO cellular networks.
\cite{IA-algorithm-quasi-static-cellular,iterative-IA-cellular}
extend the iterative interference alignment algorithm in
interference channels to IFBC channels, but the alignment is
{implicit} and requires a large amount of iterations.
\cite{MU-limited-inter-cell-approximate-IA-time-frequency-space}
looks into a special clustered shifting approach.

Regarding the methods of analytical designs, there are in general
three significant approaches. The first approach is in
\cite{downlink-IA,IA-cellular} where two cascaded precoders are set
for each transmitter (BS) node. The preceding precoder is designed
to represent inter-cell interference (ICI) subspace with respect to
other users, while the subsequent precoder is designed to represent
inter-user interference (IUI) subspace. Because they are cascaded,
the IUI subspace is contained in the ICI subspace, i.e. aligned.
This scheme is only for a two-cell network. The second approach is
in
\cite{mixed-rank-compound-X-IA-MU,Downlink-MU-IA-TwoCell-Waterloo,downlink-MU-IA-compound-X}
where a predetermined reference subspace is set for every user
receiver to align the interference coming from the other BSs. Thus
all the effective subspaces generated from one BS interfering with
receivers in the other cell are aligned and orthogonal to the
desired signal subspace, which is a duality design process compared
with \cite{downlink-IA,IA-cellular}. No back-and-forth signalling is
needed between the BS side and MS side. The third approach is in
\cite{design-IA-two-cell-IBC} where an intricate and meticulous DoF
design is realized in a two-cell and two-user per cell MIMO IFBC
channel. By a novel jointly design of transmit and receive
beamforming vectors in a closed-form expression without any
iterative computation, it outperforms previous method in
\cite{downlink-IA,IA-cellular} and
\cite{DoF-sumRate-two-IFBC,sumRate-twocell-BC-spatial-MG} in terms
of antenna usage and DoF. The process includes two main steps. The
receive beamforming vectors first align effective inter-cell
interferences, and then the transmit beamforming vectors remove both
inter-cell interferences and inter-user interferences.

\subsection{Multi-Cell Downlink Interference Alignment Issues}

It is absolutely challenging to design general interference
alignment schemes for multi-cell downlink channels. Broadly
speaking, there are three key issues worthy attention:

First, the above three existing approaches of \cite{downlink-IA},
\cite{mixed-rank-compound-X-IA-MU} and \cite{design-IA-two-cell-IBC}
mainly consider the configuration of two cells with two users in
each cell as well as single data stream for each user. It is highly
non-trivial to extend the case to a general scenario in which more
than two cells, more than two users in each cell, and multiple data
streams are involved in the network. So the new scenario is distinct
from either the original IA problem in the $K$-pair interference
channel \cite{IA-DOF-Kuser-Interference} or the two-user cellular
channel \cite{design-IA-two-cell-IBC,downlink-IA}. In brief, the
multi-cell network has more complicated {interfering} structures to
proceed alignment. For each user in each cell, it receives
interferences from: multiple streams of its own; interfering streams
from other intra-cell users; multiple interfering links from BSs in
other cells. Accordingly various levels of alignment are necessary
for different sources of interferences.

Second, multiple antenna (MIMO) configuration is becoming a critical
issue in term of DoF obtainable by IA. Conventional research works
usually adopt finite transmit antennas and single receive antenna,
e.g. MU-MIMO, so that it only needs to design the transmit
beamforming vectors to remove IUI. However, multiple antennas are
nowadays introduced at the mobile stations in the 3GPP LTE, IEEE
802.16e systems \cite{physical-layer-aspects-EUTRA},
\cite{IEEE80216mEMD}. For example, the configuration of 4 transmit
antennas and 4 receive antennas is one option of antennas in the 4G
standards. Hence a lot of recent works take multiple receive
antennas as well as transmitter antennas into account in the
coordinated scheme for MIMO IFBC channels. \cite{downlink-IA} and
\cite{DoF-K-MIMO-IFC-constant-downlink} assume sufficient number of
transmitter antennas so that explicit IA conditions are guaranteed
to achieve;
\cite{mixed-rank-compound-X-IA-MU,Downlink-MU-IA-TwoCell-Waterloo,downlink-MU-IA-compound-X}
particularly rely on multiple receive antennas to construct the
predefined interference subspace; \cite{design-IA-two-cell-IBC}
delicately confines a critical range of the number of transmit
antennas $N_t$ and the number of receive antennas $N_r$, satisfying
$3N_r/2<N_t<2N_r$. The range of transmit and receive antennas
reflects a very subtle balance that both transmit side and receive
side are capable of contributing to the construction of IA via a
precise and dedicated cooperation, although it needs large
overhead of CSI exchange between BSs and MSs in the cells. 



Third, as just mentioned, the transmitters require perfect channel
state information (CSI) specified as CSIT, and receivers may also
require channel CSI as CSIR. The huge amount of coordinations
between nodes are supported by all the CSI so that IA schemes are
implemented in realistic cellular networks. To acquire CSI, a
relatively modest amount of backhaul communication is needed
\cite{multi-cell-cooperative-newlook-interference}; training and
feedback are also carried out within the cell and across cells
\cite{IA-training-feedback-constraints}. However, as CSI
requirements increase, the corresponding system overhead may cause
bottleneck that limits the growth of dimensionality and spectral
efficiency.

\subsection{Aims and Contributions of This Work}

Regarding the mentioned scenarios, methods and issues, the aims and
contributions of this work could be summarized into the following
aspects:

1) An explicit close-form structure is built for interference
alignment in general multi-cell (more than two cells) networks. The
IA structure is constructed in a standardized dynamic procedure,
where different parts in the interfering structure interplay with
each other step by step and alignments are made in diverse levels.

2) Three types of models of multi-cell networks are analyzed
respectively. One is an ordinary full connected symmetric network,
which has $K$ cells, $M$ users per cell, and $d$ streams per user.
The other is a cyclic symmetric Wyner-type network
\cite{Wyner-Model}, in which the MSs in each cell suffer
interferences only from BSs in two adjacent cells. The last is a
cyclic asymmetric Wyner-type network, also known as the cascaded
Z-interference channel
\cite{sum-capacity-K-user-cascade-Gaussian-Z-IFC}. In this network,
MSs in each cell suffer interferences only from the BS in one
adjacent cell, and moreover the distinction between cell-interior
MSs and cell-edge MSs is considered.





3) Both transmit side and receive side are involved and balanced to
satisfy
zero-forcing conditions for IA. 
Regarding finite antennas, CSI overhead and computational
complexity, three options are provided for IA design: only on the
transmit side; only on the receive side; tradeoff of two sides.


4) Cascaded precoders or receive filters could simplify cooperations
between nodes, however they consume a number of antennas for every
node. So this work looks into the balance between the complicacy of
cascaded coders and antenna numbers as well as CSI overhead and
computational complexity. Meanwhile the performance loss due to
residue interference of
imperfect alignment is also evaluated. 





5) Due to concerns of the implementation of practical systems, a
robust measure of coder selection method is also provided.
Conventional schemes adopt fixed random precoders in
\cite{downlink-IA} on the transmit side and identical predetermined
filters in \cite{mixed-rank-compound-X-IA-MU} on the receive side.
Therefore, the system performance could be improved by selecting
appropriate coders as in
\cite{IA-oppo-selection-3-user-IFC,oppo-IA-receiver-selection-K-1x3-SIMO,oppo-IA-user-selection-MU-IFC}.
For example, on the transmit side, the fixed random precoder matrix
is set to be finitely optional to be aligned. The solution is robust
and the antenna usage could be saved to some extent. 

6) Instead of static structure designs in conventional IA schemes,
dynamic procedures are introduced due to the complicacy of the
network. High DoF are reachable from the direct interplay between
interfering cells. Alignments are arranged in a natural way between
the transmit side and the receive side.

The following parts are organized as: Section \Rmnum{2} introduces
and describes all the scenarios and models; Section \Rmnum{3}
designs and analyzes the full connected network; Section \Rmnum{4}
discusses the Wyner-type network with two adjacent interfering
links; Section \Rmnum{5} discusses the Wyner-type network with one
adjacent interfering link; Section \Rmnum{6} proposes advanced
designs for the Wyner-type networks; Section \Rmnum{7} gives a
summary and conclusion.

\section{Problem Statement and Model Description}


Before the discussion of the three models of cellular networks in
this work, common settings are described for all these scenarios.
The network contains $K$ cells. Each cell has one base stations
(BS), and $M$ users, i.e. mobile stations (MS). The BSs are denoted
in a set as $\mathcal{K}=\{1,2,\ldots,K\}$ and the MSs in an
individual cell are denoted in a set as
$\mathcal{M}=\{1,2,\ldots,M\}$. Each BS is equipped with $N_t$
antennas and each MS is equipped with $N_r$ antennas. Denote
$\mathbf{H}_{k'}^{k:m}\in\mathbb{C}^{N_r\times N_t}$ as the channel
from $k'$-th BS to to $m$-th MS in the $k$-th cell. $d$ is the
number of datastreams from arbitrary $k$-th BS to $m$-th MS in its
own cell. So that define $\mathbf{V}_{k:m}\in\mathbb{C}^{N_t\times
d}$ as the precoder at $k$-th BS for the $m$-th MS in its own cell,
and define $\mathbf{x}_{k:m}\in\mathbb{C}^{d\times 1}$ as the
corresponding input streams at BS. Similarly, define
$\mathbf{U}_{k:m}\in\mathbb{C}^{N_r\times d}$,
$\mathbf{y}_{k:m}\in\mathbb{C}^{N_r\times 1}$ and $\mathbf{\hat
y}_{k:m}=\mathbf{U}_{k:m}^{\dag}\mathbf{y}_{k:m}\in\mathbb{C}^{d\times
1}$ as the receive filter, received signal and output streams at
$m$-th MS in $k$-th cell.

\textit{Notice: since all discussions are regarding DoF design in
high SNR region, the noise terms are ignored in following
expressions.}

\subsection{Model 1: Full Connected Network}

\begin{figure}[htpb]
  \begin{center}
    \includegraphics[width=4.1in]{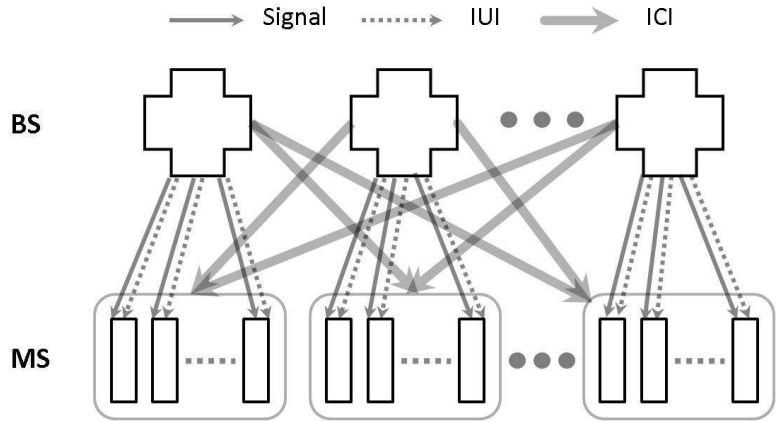}\\
  \end{center}
  \caption{Model 1: Full Connected Network}\label{Cellular-IFBC-IA}
\end{figure}

Model 1 is shown as in Fig. \ref{Cellular-IFBC-IA}, which is a
symmetric full connected network with $K$ cells interfering to each
other. For simplicity of expressions, a special model of three cells
with three users in each cell is used for illustration. However, all
the schemes and results shown could be directly applied to $K$-cell
networks with losing generality. Compared with the conventional
scenario of two cells in the network as in
\cite{design-IA-two-cell-IBC,downlink-IA,mixed-rank-compound-X-IA-MU},
the three-cell scenario is intrinsically different and complicated,
because MSs in each cell observe more than two interfering cells
which have a potential to cooperate.

The received signal at $m$-th MS of $k$-th cell is shown in equation
(\ref{cellular-model-1-eq1}):

\begin{equation}
\begin{aligned}\label{cellular-model-1-eq1}
\mathbf{y}_{k:m}=\mathbf{H}_{k}^{k:m}\mathbf{V}_{k:m}\mathbf{x}_{k:m}&+\sum_{n\neq
m}^{n\in\mathcal{M}}\mathbf{H}_{k}^{k:m}\mathbf{V}_{k:n}\mathbf{x}_{k:n}\\
&+\sum_{j\neq
k}^{j\in\mathcal{K}}\sum_{n\in\mathcal{M}}\mathbf{H}_{j}^{k:m}\mathbf{V}_{j:n}\mathbf{x}_{j:n}
\end{aligned}
\end{equation}

While the received signal $\mathbf{y}_{k:m}$ in
(\ref{cellular-model-1-eq1}) is composed of three parts: desired
signal, intra-cell interference (IUI) from users in the same cell,
and inter-cell interference (ICI) from all other cells.

\subsection{Model 2: Cyclic Network with Two-Side Adjacent Links}



\begin{figure}[htpb]
  \begin{center}
    \includegraphics[width=4.5in]{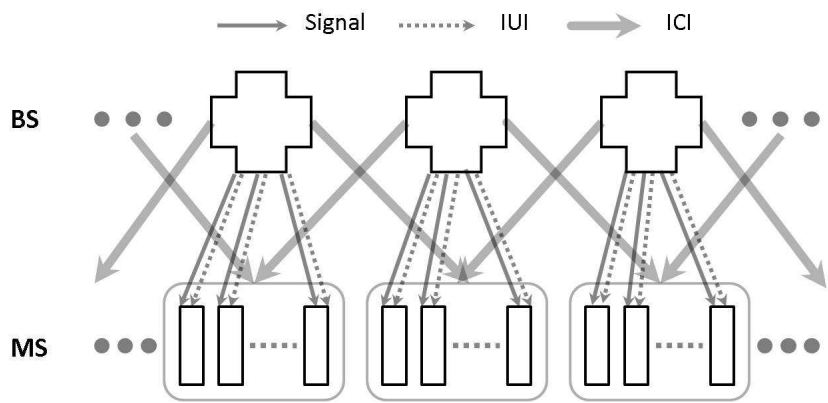}\\
  \end{center}
  \caption{Model 2: Cyclic Network with Two-Side Adjacent Links}\label{Cellular-IFBC-IA-2}
\end{figure}

Model 2 is shown as in Fig. \ref{Cellular-IFBC-IA-2}, which is a
cyclic network with users in each cell only interfered by two
adjacent cells. It is a simple and tractable model proposed by Wyner
\cite{Wyner-Model}. The cells are arranged in an infinite linear
array or equivalently on a circle
\cite{limited-capacity-backhaul-inter-users-links-cooperative-multicell}
, \cite{symmetric-feedback-capacity-K-user-cyclic-IFC}.

The received signal at $m$-th MS of $k$-th cell is shown in equation
(\ref{cellular-model-2-eq1}):

\begin{equation}
\begin{aligned}\label{cellular-model-2-eq1}
&\mathbf{y}_{k:m}=\mathbf{H}_{k}^{k:m}\mathbf{V}_{k:m}\mathbf{x}_{k:m}+\sum_{n\neq
m}^{n\in\mathcal{M}}\mathbf{H}_{k}^{k:m}\mathbf{V}_{k:n}\mathbf{x}_{k:n}\\
&+\sum_{n\in\mathcal{M}}\mathbf{H}_{k-1}^{k:m}\mathbf{V}_{k-1:n}\mathbf{x}_{k-1:n}+\sum_{n\in\mathcal{M}}\mathbf{H}_{k+1}^{k:m}\mathbf{V}_{k+1:n}\mathbf{x}_{k+1:n}
\end{aligned}
\end{equation}

While the received signal $\mathbf{y}_{k:m}$ in
(\ref{cellular-model-2-eq1}) is composed of four parts: desired
signal, intra-cell interferences (IUI) from users in the same cell,
inter-cell interferences (ICI) from one adjacent cell on one side,
and ICI from the other adjacent cell on the other side.

%

\subsection{Model 3: Cyclic Network with One-Side Link only at Edge}

\begin{figure}[htpb]
  \begin{center}
    \includegraphics[width=4.5in]{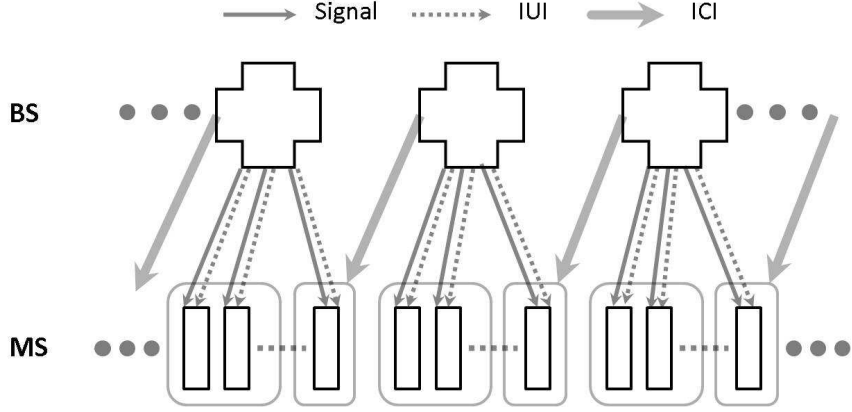}\\
  \end{center}
  \caption{Model 3: Cyclic Network with One-Side Link only at Edge}\label{Cellular-IFBC-IA-3}
\end{figure}

Model 3 is shown in Fig. \ref{Cellular-IFBC-IA-3}, which is a cyclic
network where each cell has both cell-interior users and cell-edge
users. Only the cell-edge users are interfered by only the adjacent
cell on one side. It is motivated by the modified Wyner model as in
\cite{capacity-K-user-cyclic-Gaussian-IFC,symmetric-feedback-capacity-K-user-cyclic-IFC}
and the cascaded Z-interference channel as in
\cite{sum-capacity-K-user-cascade-Gaussian-Z-IFC}. The different
roles of cell-interior users and cell-edge users are also worth
attention in practical systems
\cite{multi-cell-downlink-capacity-coordinated-processing}.

Denote
$\mathcal{{M}^{\star}},\mathcal{{M}^{\circ}}\subset\mathcal{M}$ as
the sets of cell-interior users and cell-edge users respectively.
$|\mathcal{{M}^{\star}}|={{M}^{\star}}$,
$|\mathcal{{M}^{\circ}}|={{M}^{\circ}}$ and
${{M}^{\star}}+{{M}^{\circ}}=M$. Each cell-interior MS has
$N_r^{\star}$ antennas, while each cell-edge MS has $N_r^{\circ}$
antennas.

For cell-interior users, the received signal at ${m}^{\star}$-th
(${m}^{\star}\in\mathcal{{M}^{\star}}$) MS of $k$-th cell is shown
in equation (\ref{cellular-model-3-eq1a}):

\begin{equation}
\begin{aligned}\label{cellular-model-3-eq1a}
&\mathbf{y}_{k:{m}^{\star}}=\mathbf{H}_{k}^{k:{m}^{\star}}\mathbf{V}_{k:{m}^{\star}}\mathbf{x}_{k:{m}^{\star}}+\sum_{{n}^{\star}\neq
{m}^{\star}}^{{n}^{\star}\in\mathcal{{M}^{\star}}}\mathbf{H}_{k}^{k:{m}^{\star}}\mathbf{V}_{k:{n}^{\star}}\mathbf{x}_{k:{n}^{\star}}\\
&\hspace{35mm}+\sum_{{n}^{\circ}\in\mathcal{{M}^{\circ}}}\mathbf{H}_{k}^{k:{m}^{\star}}\mathbf{V}_{k:{n}^{\circ}}\mathbf{x}_{k:{n}^{\circ}}\\
\end{aligned}
\end{equation}

Then the received signal $\mathbf{y}_{k:m^{\star}}$ in
(\ref{cellular-model-3-eq1a}) is composed of three parts: desired
signal; IUI intended for other cell-interior users in the same cell;
IUI intended for all cell-edge users in the same cell.

For cell-edge users, the received signal at ${m}^{\circ}$-th
(${m}^{\circ}\in\mathcal{{M}^{\circ}}$) MS of $k$-th cell is shown
in equation (\ref{cellular-model-3-eq1b}):

\begin{equation}
\begin{aligned}\label{cellular-model-3-eq1b}
&\mathbf{y}_{k:{m}^{\circ}}=\mathbf{H}_{k}^{k:{m}^{\circ}}\mathbf{V}_{k:{m}^{\circ}}\mathbf{x}_{k:{m}^{\circ}}\\
&+\sum_{{n}^{\star}\in\mathcal{{M}^{\star}}}\mathbf{H}_{k}^{k:{m}^{\circ}}\mathbf{V}_{k:{n}^{\star}}\mathbf{x}_{k:{n}^{\star}}+\sum_{{n}^{\circ}\neq{m}^{\circ}}^{{n}^{\circ}\in\mathcal{{M}^{\circ}}}\mathbf{H}_{k}^{k:{m}^{\circ}}\mathbf{V}_{k:{n}^{\circ}}\mathbf{x}_{k:{n}^{\circ}}\\
&+\sum_{{n}^{\star}\in\mathcal{{M}^{\star}}}\mathbf{H}_{k+1}^{k:{m}^{\circ}}\mathbf{V}_{k+1:{n}^{\star}}\mathbf{x}_{k+1:{n}^{\star}}+\sum_{{n}^{\circ}\in\mathcal{{M}^{\circ}}}\mathbf{H}_{k+1}^{k:{m}^{\circ}}\mathbf{V}_{k+1:{n}^{\circ}}\mathbf{x}_{k+1:{n}^{\circ}}\\
\end{aligned}
\end{equation}

Then the received signal $\mathbf{y}_{k:m^{\circ}}$ in
(\ref{cellular-model-3-eq1b}) is composed of five parts: desired
signal; IUI intended for all cell-interior users in the same cell;
IUI intended for other cell-edge users in the same cell; ICI from
the adjacent cell intended for all its own cell-interior users; ICI
from the adjacent cell intended for all its own cell-edge users.

\subsection{Characteristics of Dealing with Multi-Cell Downlink IA}

Since the models of multi-cell downlink channels are complicated to
directly reach the status of interference alignment, the designs
should be to generalized to be along three most critical threads.
The three threads include interference decomposition, alignment
planning, and process control, which are described as following.

\subsubsection{Interference Decomposition}

In the conventional $K$-pair peer-peer interference channel, all the
interferences are seen at the same level symmetrically from equal
sources to equal destinations. While in the downlink channels of
cellular networks, all the interferences are distinguished to
different levels according to their different sources and
destinations. They are mainly categorized into three levels: from
one BS to one MS in its own cell, the link has multiple streams
which cause interferences between each other; from one BS to
multiple MSs of its own cell, streams to one MS cause interferences
to streams to another MS, which is defined as inter-user
interference (IUI), i.e. with cross links overlapping the direct
links \cite{IA-algorithm-quasi-static-cellular}; from one BS to all
MSs in another cell, all the streams (including desired signals,
inter-stream interference and IUI in the cell) cause interferences
to the other cell.

\subsubsection{Alignment Planning}

Since all the interference is decomposed into different levels and
different parts, accordingly, it is necessary to make reasonable
alignment, not simply overlapping all the interference. In
principle, the more interferences are possibly aligned together, the
less dimensionality of subspace they occupy, and the more DoF the
network could obtain. In detail, it gives rise to complicated
options to plan the alignment. Primarily, alignment are made at the
same levels. Align inter-stream interference with inter-stream
interference, IUI with IUI, ICI with ICI from one same cell.
Secondarily, consider alignment between different levels of
interferences. Usually, IUI could be aligned with ICI as shown in
\cite{downlink-IA,mixed-rank-compound-X-IA-MU,design-IA-two-cell-IBC};
Third, notice this work is looking into the new scenario and model
of multi-cell networks, so the new alignment between ICI and ICI
from different cells are considered deliberately. Nevertheless,
there also exist other potential options than the ones just
mentioned.

\subsubsection{Process Control}

Even if the alignment planning is perfect made, the coding for all
nodes are not possible to be determined and set in one time. All
different levels and different sources of interferences are aligned
step by step. A new issue worth attention is that alignment could be
made either at the BS side or at the MS side, i.e. the transmitted
signals are aligned or the received signals are aligned. Another
issue is the acquisition of CSI to satisfy the causality conditions
between the steps of alignment. Profound cooperations are needed
between BS and BS, MS and MS, intra-cell BS and MS, inter-cell BS
and MS, etc. as in
\cite{limited-capacity-backhaul-inter-users-links-cooperative-multicell}.
In particular, a \textit{chicken-and-egg} problem arises in this
whole process, because all the steps are related or coupled back and
forth and here and there, between desired signals and interferences,
BSs and MSs, intra-cell and inter-cell.



\section{Basic Design and Analysis of Model 1}

%


This section looks into model 1 as shown in Fig.
\ref{Cellular-IFBC-IA} and equation (\ref{cellular-model-1-eq1}). In
general, there are three angles to implement the design. Recent
works \cite{downlink-IA}, \cite{mixed-rank-compound-X-IA-MU}, and
\cite{design-IA-two-cell-IBC} implement the alignment through
reference subspace on the BS side, subspace on the MS side, and
meticulous interplay on both BS and MS sides, respectively.
According these three angles, the following work proposes five
approaches to implement the alignment.

\textit{Clarification}: Be advised that, in the following
expressions of the design process, when a same BS or MS in a set is
referred, the index may change between successive steps according to
its subject or object position in the context of the current step.
In each step, the index refers to all equal nodes. Specifically, $k$
is used to denote the subject, and $j$ is used to denote the object.
Besides, in the corresponding tables, the index may also change
according to the context of discussion.

\textit{Notations}: with a little abuse, denote $\mathbf{A}^{-1}$ as
both inverse and pseudo inverse of the matrix $\mathbf A$, as long
as the dimensions are satisfied to proceed the operation. Also
denote $[\mathbf{A}_1\ \mathbf{A}_2\ \cdots\ \mathbf{A}_N]$ as the
horizontal concatenation of matrices from $\mathbf{A}_1$ to
$\mathbf{A}_N$.

\subsection{Approach on Transmitter Side with Cascaded Coder}

This approach is similar to the design of \cite{downlink-IA}. Each
BS uses a cascaded precoder to zero-forcing all inter-cell
interferences (ICI) from other cells, and then deal with intra-cell
interferences (IUI) in its own cell. The whole process takes four
steps as following:

Step 1: Choose an auxiliary intermediate precoder
$\mathbf{\Phi}_{k}\in\mathbb{C}^{N_t\times (M\cdot d)}$ at the
$k$-th BS (transmitter), and a second precoder $\mathbf{\tilde
V}_{k:m}\in\mathbb{C}^{(M\cdot d)\times d}$ preceding the first
chosen precoder. Then the actual precoder $\mathbf{V}_{k:m}$ is
composed of:

\begin{equation}
\begin{aligned}\label{cellular-model-1-eq1-transmit-V}
\mathbf{V}_{k:m}=\mathbf{\Phi}_{k}\mathbf{\tilde{V}}_{k:m} \ \ \
m\in\mathcal{M}, k\in\mathcal{K}
\end{aligned}
\end{equation}

(Notice the index $k$ and $m$ refer to all BSs and MSs, so that they
could change accordingly.)

Step 2: When the first precoder $\mathbf{\Phi}_{j}$ is randomly
picked, to enable zero-forcing the ICI from arbitrary $k$-th cell,
all $M$ users in the $k$-th cell should determine $\mathbf{U}_{k:m}$
to satisfy the following conditions (as mentioned, the index has
changed for consistency of expressions):

\begin{equation}
\begin{aligned}\label{cellular-model-1-eq1-transmit-ZF-u}
\mathbf{U}_{k:m}^{\dag}[\underbrace{\cdots\
\mathbf{{H}}_j^{k:m}\mathbf{\Phi}_{j}\
\cdots}_{j\in\mathcal{K}\backslash\{k\}}]=\mathbf{0} \ \ \
m\in\mathcal{M}
\end{aligned}
\end{equation}

To be feasible to find $\mathbf{U}_{k:m}$, the dimensions should
satisfy $N_r\geq (K-1)Md+d$.

Step 3: Now observe each $\mathbf{\Phi}_{k}$ again, notice there're
$(K-1)M$ zero-forcing conditions (notice the index as well):

\begin{equation}
\begin{aligned}\label{cellular-model-1-eq1-transmit-ZF-phi}
\left[
\begin{aligned}
&\hspace{7mm}\vdots\\
&\mathbf{U}_{j:m}^{\dag}\mathbf{{H}}_k^{j:m}\\
&\hspace{7mm}\vdots
\end{aligned}
\right]\mathbf{\Phi}_{k}=\mathbf{0} \ \ \ m\in\mathcal{M},
j\in\mathcal{K}\backslash\{k\}
\end{aligned}
\end{equation}

\begin{Remark}
Check the equation (\ref{cellular-model-1-eq1-transmit-ZF-phi}).
Normally, it is feasible to find $\mathbf{\Phi}_{k}$ only when the
dimensions satisfy $N_t\geq(K-1)Md+Md$. However, since
(\ref{cellular-model-1-eq1-transmit-ZF-u}) already guarantees this
condition, it is only required that $N_t\geq Md$ in this specific
design process. So this process naturally contains an
\textit{implicit} and \textit{inherent} alignment.
\end{Remark}

Step 4: Then each $k$-th BS knows its equivalent downlink channel
$\mathbf{U}_{k:m}\mathbf{{H}}_k^{k:m}\mathbf{\Phi}_{k}$ from each
intra-cell MS. It forms zero-forcing transmit streams which could
eliminate IUI between users, while it does not need to know the
actual interfering streams. The second precoder matrix is obtained
by an inverse operation (notice the index as well):

\begin{equation}\label{cellular-model-1-eq1-transmit-Vtilde}
\begin{aligned}
&\left[\mathbf{\tilde{V}}_{k:1}\ \mathbf{\tilde{V}}_{k:2}\ \cdots\
\mathbf{\tilde{V}}_{k:M}\right]= \left[
\begin{aligned}
&\mathbf{U}_{k:1}^{\dag}\mathbf{{H}}_k^{k:1}\mathbf{\Phi}_{k}\\&\mathbf{U}_{k:2}^{\dag}\mathbf{{H}}_k^{k:2}\mathbf{\Phi}_{k}\\&\hspace{10mm}\vdots\\&\mathbf{U}_{k:M}^{\dag}\mathbf{{H}}_k^{k:M}\mathbf{\Phi}_{k}
\end{aligned}
\right]^{-1}
\end{aligned}
\end{equation}


In summary, the four steps determine all intermediate precoders,
then receive filters, and finally second precoders, in sequence.
Most important, alignment is implicitly contained in step 3 from the
view of BS, which reflects the complicacy of IA in cellular
networks.

\subsection{Approach on Receiver Side with Cascaded Coder: I}

This approach is proposed as  a dual solution of the above approach
on transmitter side. Set the cascaded coders on the receiver side.
The whole process takes the following four steps:

Step 1: Choose an auxiliary or intermediate receive filter matrix
$\mathbf{\Psi}_{k:m}\in\mathbb{C}^{N_r\times Md}$ at the $m$-th MS
in the $k$-th cell, and a second receive filter $\mathbf{\tilde
U}_{k:m}\in\mathbb{C}^{Md\times d}$ which precedes the first filter.
Then the actual receive filter $\mathbf{U}_{k:m}$ is composed of:

\begin{equation}
\begin{aligned}\label{cellular-model-1-eq1-receiver-U}
\mathbf{U}_{k:m}=\mathbf{\Psi}_{k:m}\mathbf{\tilde{U}}_{k:m} \ \ \
m\in\mathcal{M}, k\in\mathcal{K}
\end{aligned}
\end{equation}

Step 2: When every first receive filter $\mathbf{\Psi}_{j:n}$ is
randomly picked, to enable zero-forcing ICI, the $k$-th BS should
determine $\mathbf{V}_{k:m}, m\in\mathcal{M}$ which satisfies the
condition:

\begin{equation}
\begin{aligned}\label{cellular-model-1-eq1-receiver-ZF-V}
\left[
\begin{aligned}
&\hspace{7mm}\vdots\\
&\mathbf{\Psi}_{j:n}^{\dag}\mathbf{{H}}_k^{j:n}\\
&\hspace{7mm}\vdots
\end{aligned}
\right][\mathbf{V}_{k:1}\ \mathbf{V}_{k:2}\ \cdots\
\mathbf{V}_{k:M}]=\mathbf{0} \ \ \ n\in\mathcal{M},
j\in\mathcal{K}\backslash\{k\}
\end{aligned}
\end{equation}

To be feasible to determine $[\mathbf{V}_{k:1}\ \mathbf{V}_{k:2}\
\cdots\ \mathbf{V}_{k:M}]$, the dimensions should satisfy $N_t\geq
(K-1)M^2d+Md$.

Step 3: Now observe each $\mathbf{\Psi}_{k:m}$ again, notice
there're $(K-1)M$ number of zero-forcing conditions:

\begin{equation}
\begin{aligned}\label{cellular-model-1-eq1-receiver-ZF-psi}
\mathbf{\Psi}_{k:m}^{\dag}[\underbrace{\cdots\
\mathbf{{H}}_j^{k:m}\mathbf{V}_{j:n}\ \cdots}_{n\in\mathcal{M},
j\in\mathcal{K}\backslash\{k\}}]=\mathbf{0} \ \ \ m\in\mathcal{M}
\end{aligned}
\end{equation}

\begin{Remark}
Check the condition of (\ref{cellular-model-1-eq1-receiver-ZF-psi}).
Notice in general it is only feasible to find $\mathbf{\Psi}_{k:m}$
when the dimensions satisfy $N_r\geq(K-1)Md+Md$. However, since
(\ref{cellular-model-1-eq1-receiver-ZF-V}) already guarantees the
condition, it is only required that $N_r\geq Md$ in this specific
design process. So this process naturally contains an
\textit{implicit} and \textit{inherent} alignment.
\end{Remark}

Step 4: Then each intra-cell MS in the $k$-th cell knows its
equivalent channel
$\mathbf{\Psi}_{k:m}^{\dag}\mathbf{{H}}_k^{k:m}\mathbf{V}_{k:m}$,
while it does not know the actual interfering beams. It forms the
zero-forcing receive filter to eliminate IUI. The second receive
filter $\mathbf{\tilde{U}}_{k:m}$ is determined by an operation to
find the null space:

\begin{equation}\label{cellular-model-1-eq1-receiver-Utilde}
\begin{aligned}
&\mathbf{\tilde{U}}_{k:m}^{\dag}\left[\ \underbrace{\cdots\
\mathbf{\Psi}_{k:m}^{\dag}\mathbf{{H}}_k^{k:m}\mathbf{V}_{k:n}\
\cdots}_{n\in\mathcal{M}\backslash\{m\}}\ \right]=\mathbf{0}
\end{aligned}
\end{equation}

In summary, the four steps determine all intermediate receive
filters, then precoders, and finally second receive filters, in
sequence. Most important, alignment is implicitly contained in step
3 from the view of MS, which reflects the complicacy of IA in
cellular networks.

%

\subsection{Approach on Receiver Side with  Cascaded Coder: II}

This approach is similar to the design of
\cite{mixed-rank-compound-X-IA-MU}. Each MS uses a cascaded receive
filter to align all inter-cell interferences (ICI) from other cells,
and then deal with ICI and intra-cell interferences (IUI) in its own
cell at the same time. The whole process takes four steps as
following:

Step 1: For each $m$-th MS in $k$-th cell, the receive filter
$\mathbf{U}_{k:m}$ is defined as two cascaded filter matrices:

\begin{equation}
\begin{aligned}\label{cellular-model-1-eq1-receiver-ref-1}
&\mathbf{U}_{k:m}=\mathbf{G}_{k:m}^{\dag}{\Lambda}_{\mathcal{K}:m}\\
&\mathbf{G}_{k:m}=[\underbrace{\mathbf{I}_{N_t}\cdots
\mathbf{I}_{N_t}}_{K-1}][\underbrace{\cdots\ \mathbf{H}_{j}^{k:m}
\ \cdots}_{j\in\mathcal{K}\backslash\{k\}}]^{-1}\\
&\text{So that}\ \ \mathbf{G}_{k:m}\in\mathbb{C}^{N_t\times
N_r},\mathbf{G}_{k:m}\mathbf{H}_j^{k:m}=\mathbf{I}_{N_t}\ \ \forall
j\in\mathcal{K}\backslash\{k\}
\end{aligned}
\end{equation}

$\mathbf{G}_{k:m}$ is set that all ICI from any other $j$-th cell
are aligned together. While ${\Lambda}_{\mathcal{K}:m}$ is a common
filter set as the same for all cells in the set $\mathcal{K}$.

To be feasible to determine $\mathbf{G}_{k:m}$ by the inverse
operation, the dimensions should satisfy $N_r\geq (K-1)N_t$.

Step 2: Since the common receive filter
$\Lambda_{\mathcal{K}:m}\in\mathbb{C}^{N_t\times d}$ is set to be
same for all cells of $\mathcal{K}$, the output stream as in the
system equation (\ref{cellular-model-1-eq1}) is presented as
following:

\begin{equation}
\begin{aligned}\label{cellular-model-1-eq1-receiver-ref-2}
&\Lambda_{\mathcal{K}:m}^{\dag}\mathbf{G}_{k:m}\mathbf{y}_{k:m}=\\
&\Lambda_{\mathcal{K}:m}^{\dag}\mathbf{G}_{k:m}\mathbf{H}_{k}^{k:m}\mathbf{V}_{k:m}\mathbf{x}_{k:m}
+\sum_{n\neq
m}^{n\in\mathcal{M}}\Lambda_{\mathcal{K}:m}^{\dag}\mathbf{G}_{k:m}\mathbf{H}_{k}^{k:m}\mathbf{V}_{k:n}\mathbf{x}_{k:n}\\
&\hspace{38mm}+\sum_{j\neq
k}^{j\in\mathcal{K}}\sum_{n\in\mathcal{M}}\Lambda_{\mathcal{K}:m}^{\dag}\mathbf{V}_{j:n}\mathbf{x}_{j:n}
\end{aligned}
\end{equation}

To eliminate ICI and IUI terms in the equation
(\ref{cellular-model-1-eq1-receiver-ref-2}) respectively, the
zero-forcing conditions should be set as:

\begin{equation}
\begin{aligned}\label{cellular-model-1-eq1-receiver-ref-2zf}
\Lambda_{\mathcal{K}:m}^{\dag}\mathbf{G}_{k:m}\mathbf{H}_{k}^{k:m}\mathbf{V}_{k:n}&=\mathbf{0}\ \ n\in\mathcal{M}\backslash\{m\}\\
\Lambda_{\mathcal{K}:m}^{\dag}\mathbf{V}_{j:n}&=\mathbf{0}\ \
j\in\mathcal{K}\backslash\{k\},n\in\mathcal{M}
\end{aligned}
\end{equation}

Step 3: Then observe from $k$-th BS to choose the precoder $[\cdots\
\mathbf{V}_{k:m}\ \cdots]$, the zero-forcing conditions in equation
(\ref{cellular-model-1-eq1-receiver-ref-2zf}) are equivalently
presented with new indices in the following expression:

\begin{equation}
\begin{aligned}\label{cellular-model-1-eq1-receiver-ref-3}
\Lambda_{\mathcal{K}:n}^{\dag}\mathbf{G}_{k:n}\mathbf{H}_{k}^{k:n}\mathbf{V}_{k:m}&=\mathbf{0}\ \ n\in\mathcal{M}\backslash\{m\}\\
\Lambda_{\mathcal{K}:n}^{\dag}\mathbf{V}_{k:m}&=\mathbf{0}\ \
n\in\mathcal{M}
\end{aligned}
\end{equation}

Step 4: So for each $k$-th BS, the precoder could be determined by
only one inverse operation:

\begin{equation}
\begin{aligned}\label{cellular-model-1-eq1-receiver-ref-4}
\mathbf{B}_k&=\left[\begin{aligned}\left.
\begin{aligned}
&\ \ \ \ \ \ \ \ \ \ \vdots\\
&\Lambda_{\mathcal{K}:m}^{\dag}\mathbf{G}_{k:m}\mathbf{H}_{k}^{k:m}\\
&\ \ \ \ \ \ \ \ \ \ \vdots
\end{aligned}
\right\}{M}\\\left.
\begin{aligned}
&\vdots\\
&\Lambda_{\mathcal{K}:m}^{\dag}\\
&\vdots
\end{aligned}
\right\}{M}\end{aligned}
\right]\\
[\mathbf{V}_{k:1}\ \mathbf{V}_{k:2}\ \cdots\
\mathbf{V}_{k:M}]&=\text{The first}\ Md\ \text{columns of}\
\mathbf{B}_k^{-1}
\end{aligned}
\end{equation}

Check that $[\mathbf{V}_{k:1}\ \mathbf{V}_{k:2}\ \cdots\
\mathbf{V}_{k:M}]$ indeed satisfies the conditions in
(\ref{cellular-model-1-eq1-receiver-ref-3}).

\begin{Remark}
To be feasible to obtain $[\mathbf{V}_{k:1}\ \mathbf{V}_{k:2}\
\cdots\ \mathbf{V}_{k:M}]$ by an inverse operation, the dimensions
should satisfy $N_t\geq 2Md$. While in general cases, it is required
that $N_t\geq KMd$. So this process naturally contains an
\textit{implicit} and \textit{inherent} alignment.
\end{Remark}

In summary, the four steps determine all intermediate receive
filters, and then precoders as well as second receive filters, in
sequence. Most important, alignment is explicitly contained in step
1 from the view of BS.

\subsection{Approach from Receiver Side with Direct Interplay}

This approach is similar to the design of
\cite{design-IA-two-cell-IBC}. The alignment is jointly implemented
on both BS(transmitter) side and MS(receiver) side via intricate and
meticulous interplays and cooperations. The whole process takes two
steps as following:

Step 1: Regarding channels from the $k$-th BS to all MSs in $j$-th
cell, the equivalent receive filtered channel subspaces are set to
be equal to a random predetermined reference subspace:

\begin{equation}
\begin{aligned}\label{cellular-model-1-eq1-joint-1}
\mathbf{U}_{j:1}^{\dag}\mathbf{H}_{k}^{j:1}=\mathbf{U}_{j:2}^{\dag}\mathbf{H}_{k}^{j:2}=\cdots=\mathbf{U}_{j:M}^{\dag}\mathbf{H}_{k}^{j:M}=(\mathbf{\Omega}_{k}^j)^{\dag}\\
\hspace{0mm}j\in\mathcal{K}\backslash\{k\}
\end{aligned}
\end{equation}

\begin{Remark}
Notice the step of (\ref{cellular-model-1-eq1-joint-1}) is an
\textit{explicit} alignment of ICI from the $k$-th BS to all the MSs
in the other $j$-th cell.
\end{Remark}

Observe from the MSs in the $k$-th cell, (change the indices
accordingly) to satisfy the condition
(\ref{cellular-model-1-eq1-joint-1}), $\mathbf{U}_{k:m}$ and
$\mathbf{\Omega}_{j}^{k}$ could be found by solving a null space
problem:

\begin{equation}\label{cellular-model-1-eq1-joint-2}
\begin{aligned}
&\left[\underbrace{\cdots\ (\mathbf{\Omega}_{j}^k)^{\dag}\
\cdots}_{j\in\mathcal{K}\backslash\{k\}}\ \mathbf{U}_{k:1}^{\dag}\
\mathbf{U}_{k:2}^{\dag}\ \cdots\
\mathbf{U}_{k:M}^{\dag}\right]\bullet\\
&\ \ \left[
\begin{aligned}
&\left.\begin{aligned}
&\mathbf{I}_{N_t} \cdots \mathbf{I}_{N_t}        \\
&\ \ \ \ \ \ \ \ \ \ \ \ \mathbf{I}_{N_t} \cdots \mathbf{I}_{N_t}       \\
&\ \ \ \ \ \ \ \ \ \ \ \ \ \ \ \ \ \ \ \ \ \ \ \ \ \ \ \ \ \ddots              \\
&\ \ \ \ \ \ \ \ \ \ \ \ \ \ \ \ \ \ \ \ \ \ \ \ \ \ \ \ \ \ \ \ \
\ \ \ \ \ \ \ \mathbf{I}_{N_t} \cdots \mathbf{I}_{N_t}\end{aligned}\right\}K-1\\
&-\mathbf{H}_{1}^{k:1}\ \ \ \ \  \cdots\ \ \ \ \ \   -\mathbf{H}_{j}^{k:1}\ \ \cdots\   \ -\mathbf{H}_{K}^{k:1}\\
&\ \ -\mathbf{H}_{1}^{k:2}\ \ \ \ \    \cdots\ \ \ \   \ \ -\mathbf{H}_{j}^{k:2}\ \ \cdots\     \  -\mathbf{H}_{K}^{k:2}   \\
&\ \ \ \ \ \ddots\ \   \ \ \ \ \ \ \ \ddots\ \    \ \ \ \ \ \ \ddots\ \     \ \ \ \ \ddots  \ \ \ \  \ddots\\
&\ \ \ \ \ \ -\mathbf{H}_{1}^{k:M} \ \ \ \cdots \ \ \ \ \ \
-\mathbf{H}_{j}^{k:M} \ \cdots\ \ \ -\mathbf{H}_{K}^{k:M}
\end{aligned}
\right]=\mathbf{0}
\end{aligned}
\end{equation}

To be feasible to obtain $\mathbf{U}_{k:m}$ and
$\mathbf{\Omega}_{j}^{k}$ via a null space operation, the dimensions
should follow $(K-1)N_t+MN_r\geq (K-1)MN_t+d$.

Step 2: Thus for the $k$-th BS, the precoders are determined by the
following inverse operation to eliminate both IUI and ICI:

\begin{equation}
\begin{aligned}\label{cellular-model-1-eq1-joint-3}
[\ \cdots\ \mathbf{V}_{k:m}\ \cdots\ ]=\text{first }Md\text{ columns of }\\
\left[\begin{aligned}&\left.\begin{aligned}&\hspace{9mm}\vdots\\
&\mathbf{U}_{k:m}^{\dag}\mathbf{H}_{k}^{k:m}\\
&\hspace{9mm}\vdots\end{aligned}\right\}{m\in\mathcal{M}} \\
&\left.
\begin{aligned}
&\hspace{9mm}\vdots\\
&\hspace{8mm}{\mathbf{\Omega}_{k}^j}^{\dag}\\
&\hspace{9mm}\vdots\end{aligned}\right\}j\in\mathcal{K}\backslash\{k\}\end{aligned}
\right]^{-1}
\end{aligned}
\end{equation}

To be feasible to obtain $[\ \cdots\ \mathbf{V}_{k:m}\ \cdots\ ]$
via the inverse operation, the dimensions should satisfy $N_t\geq
Md+(K-1)d$.

Check the equation (\ref{cellular-model-1-eq1-joint-3}) and find
that $[\ \cdots\ \mathbf{V}_{k:m}\ \cdots\ ]$ indeed eliminate the
IUI between each other in the cell and ICI to other cells.

In summary, the two steps determine all receive filters and aligned
subspace and then precoders in sequence. Most important, alignment
is explicitly implemented in step 1 on the MS side, and implicitly
implemented by an inverse operation in step 2 on the BS side.
Compared with previous approaches, notice their alignment are only
on one side. So that this approach has the advantage of saving
antenna usage and obtaining higher DoF.  The disadvantage is the
corresponding increased CSI exchange.


\subsection{Approach from Transmitter Side with Direct Interplay}

This approach is proposed as a dual solution of the above approach
on receiver side. Here make the explicit alignment from the BS side.
The whole process takes the following two steps:

Step 1: Regarding channels from arbitrary $j$-th BS to all MSs in
the $k$-th cell, all the equivalent precoded channel subspaces are
set to be equal to a random predetermined reference subspace:

\begin{equation}
\begin{aligned}\label{cellular-model-1-eq1-jointII-1}
&\underbrace{\cdots=\mathbf{H}_{j}^{k:m}[\mathbf{V}_{j:1}\cdots\mathbf{V}_{j:M}]=\cdots}_{
j\in\mathcal{K}\backslash\{k\}}=\mathbf{\Theta}^{k:m}
\end{aligned}
\end{equation}

\begin{Remark}
Notice the step of (\ref{cellular-model-1-eq1-jointII-1}) is an
\textit{explicit} alignment of ICI from all the BSs in other cells
to each $m$-th MS in the $k$-th cell.
\end{Remark}

Observe from all the BSs, (change the indices accordingly) to
satisfy the condition (\ref{cellular-model-1-eq1-jointII-1}),
$[\mathbf{V}_{k:1}\cdots\mathbf{V}_{k:M}]$ and
$\mathbf{\Theta}^{k:m}$ could be found by solving the following null
space problem:

\begin{equation}
\begin{aligned}\label{cellular-model-1-eq1-jointII-2}
\left[\begin{aligned}
&\left.\begin{aligned}&\begin{aligned}&\mathbf{I}_{N_r}\\&\
\vdots\\&\mathbf{I}_{N_r}\end{aligned}
\hspace{18mm}{\left.\begin{aligned}&\ddots\\\hspace{0mm}&-\mathbf{H}_{j}^{1:m}\\&\hspace{8mm}\ddots\end{aligned}\right\}{\begin{aligned}&j\in\mathcal{K}\backslash\{1\}\end{aligned}}}\\
&\ \ \ \ddots\hspace{25mm}\vdots\end{aligned}\right\}m\in\mathcal{M}\\
&\left.\begin{aligned}&\ \ \ \ \ \
\begin{aligned}&\mathbf{I}_{N_r}\\&\
\vdots\\&\mathbf{I}_{N_r}\end{aligned}
\hspace{11mm}{\left.\begin{aligned}&\ddots\\\hspace{0mm}&-\mathbf{H}_{j}^{k:m}\\&\hspace{8mm}\ddots\end{aligned}\right\}{\begin{aligned}&j\in\mathcal{K}\backslash\{k\}\end{aligned}}}\\
&\ \ \ \ \ \ \ \ \ \ddots\hspace{20mm}\vdots\end{aligned}\right\}m\in\mathcal{M}\\
&\left.\begin{aligned}&\ \ \ \ \ \ \ \ \ \ \ \
\begin{aligned}&\mathbf{I}_{N_r}\\&\
\vdots\\&\mathbf{I}_{N_r}\end{aligned}
\hspace{4mm}{\left.\begin{aligned}&\ddots\\\hspace{0mm}&-\mathbf{H}_{j}^{K:m}\\&\hspace{8mm}\ddots\end{aligned}\right\}{\begin{aligned}&j\in\mathcal{K}\backslash\{K\}\end{aligned}}}\\
&\hspace{19mm}\ddots\hspace{14mm}\vdots\end{aligned}\right\}m\in\mathcal{M}\end{aligned}\right]
\\\bullet\left[\begin{aligned}&\left.\begin{aligned}&\left.\begin{aligned}&\hspace{8mm}\vdots\\&\hspace{7mm}\mathbf{\Theta}^{k:m}\\&\hspace{8mm}\vdots\end{aligned}\right\}k\in\mathcal{K}\\&\hspace{10mm}\vdots\end{aligned}\right\}m\in\mathcal{M}\\&\left.\begin{aligned}&\hspace{10mm}\vdots\\&\left[\mathbf{V}_{k:1}\cdots\mathbf{V}_{k:M}\right]\\&\hspace{10mm}\vdots\end{aligned}\right\}k\in\mathcal{K}\end{aligned}\right]=\mathbf{0}
\end{aligned}
\end{equation}

To be feasible to obtain $[\mathbf{V}_{k:1}\cdots\mathbf{V}_{k:M}]$
and $\mathbf{\Theta}^{k:m}$ via a null space operation, the
dimensions should follow $KMN_r+KN_t\geq K(K-1)MN_r+Md$.

Step 2: Thus for the $m$-th MS in $k$-th cell, its receive filter is
determined via a null space operation to eliminate both IUI and ICI:

\begin{equation}
\begin{aligned}\label{cellular-model-1-eq1-jointII-3}
\mathbf{U}_{k:m}^{\dag}\left[\
\mathbf{H}_{k}^{k:m}\big[\underbrace{\cdots\ \mathbf{V}_{k:n}\
\cdots}_{n\in\mathcal{M}\backslash\{m\}}\big] \ \
{\mathbf{\Theta}^{k:m}}\right]=\mathbf{0}
\end{aligned}
\end{equation}

To be feasible to obtain $\mathbf{U}_{k:m}$, the dimensions should
follow $N_r\geq (M-1)d+Md+d$. It is also easy to check that the
condition (\ref{cellular-model-1-eq1-jointII-3}) directly eliminates
both IUI and ICI.

In summary, the two steps determine all precoders and aligned
subspace and then receive filters in sequence. Most important,
alignment is explicitly implemented in step 1 on the BS side by a
null space operation. Notice that when the MS chooses a receive
filter $\mathbf{U}_{k:m}$ it needs not to concern about independency
with receive filters in other MSs thanks to its distributed nature,
while when the BS chooses the precoders $\mathbf{V}_{k:n}$
designated to multiple MSs it needs to concern their independency to
form the subspace.

\subsection{Comparison of System Conditions}


For all the proposed new approaches of multi-cell downlink
interference alignment, the system require different level of
conditions to support the implementations. There are three main
aspects to look into and compare. The first aspect is the usage of
antennas for both BS and MS nodes. The number of antennas limit the
achievable DoF for each node. The second aspect is the CSI for the
nodes to obtain channel knowledge. The CSI causes system overhead
which would degrade the performance to some extent, and proper
tradeoff could be made to treat the system as a whole. The third
aspect is the computational complexity for the nodes to calculate
necessary coders. The complexity impact on the system factors such
as delay, power, hardware etc., and it is expected to be the lower
the better.

\subsubsection{DoF and Antenna Range}

In general, the feasibility conditions of cellular networks are
still not solved so far, i.e. the minimum number of antennas
required to obtain intended DoF is not known. For example,
\cite{IA-MIMO-cellular} provides a special case in which each MS
receives a single beam so that the antennas should satisfy
$N_r+N_t\geq M(K+1)$ in a $K$-cell network with $M$ users in each
cell where $N_r$ and $N_t$ are the numbers of MS(receive) antennas
and BS(transmit) antennas respectively.
While in the example of \cite{design-IA-two-cell-IBC}, each BS and
MS have $N_t$ and $N_r$ antennas respectively, and satisfy the
condition $3N_r/2<N_t<2N_r$ to achieve a final $2N_r$ DoF in the
two-cell network. It shows that the number of antennas plays a key
role to implement alignment. Moreover, in an extreme case, if all
nodes are provided with unlimited number of antennas, each node
could trivially proceed interference cancellation via perfect
zero-forcing with sufficient antennas.

The aim of this work is to exploit interference alignment in the
network with limited provision of antennas at both BS and MS nodes,
and also to look into the balance between the BS side and MS side to
implement alignment to obtain DoF with a reasonable number antennas
in the system. For the above five approaches, all the antenna
configurations are listed in Table
\ref{cellular-model1-system-condition}, in which the notations of
'A', 'B', 'C', 'D', and 'E' denote the proposed five approaches in
sequence.

\begin{table}
\centering
\begin{tabular}{|c|c|c|}
  \hline
  Approach & Minimum BS Antenna. & Minimum MS Antenna  \\
  \hline
  A & $Md$ & $(K-1)Md+d$  \\
  \hline
  B & $(K-1)M^2d+Md$ & $Md$  \\
  \hline
  C & $2Md$ & $(K-1)2Md$  \\
  \hline
  D & $(M+K-1)d$ & $\scriptsize\begin{aligned}(K-1)\frac{(M-1)}{M}(M+K-1)d\\+d/M\end{aligned}\normalsize$  \\
  \hline
  E & $\scriptsize\begin{aligned}(KM-2M)2Md\\+(M/K)d\end{aligned}\normalsize$ & $2Md$  \\
  \hline
\end{tabular}\caption{Minimum Antenna Usage for BSs and MSs in All Approaches in Model 1}\label{cellular-model1-system-condition}
\end{table}

In summary, as shown in Table
\ref{cellular-model1-system-condition}, approach 'B' and approach
'E' need a large number of antennas at BS nodes, while approach 'A',
'C' and 'D' need a large number of antennas at MS nodes.

\subsubsection{CSI and Overhead}

Although the design of interference alignment provides a promising
DoF in theory, it still has to face the challenge of realistic
implementation. It is required that all BSs and all MSs should know
a certain level of channel state information (CSI), so that the
alignment could be made with operations of the channel matrices.
Usually, CSI could be acquired by five key steps: common forward
link training \cite{IA-training-feedback-constraints},
\cite{MU-MIMO-achievrates-downlink-training-CSI}; channel
quantization \cite{new-quantization-IA-limited-feedback}; channel
feedback \cite{IA-under-limited-feedback-MIMO-IFC}; beamformer
selection; dedicated training. Furthermore, there is a tradeoff
between the overhead to obtain CSI and the effective DoF.

The aim of this work is to provide a comparison of CSI requirements
for all nodes in different approaches. For the above five
approaches, all the CSI configurations are listed in Table
\ref{cellular-model1-system-condition-2}, in which the notations of
'A', 'B', 'C', 'D', and 'E' denote the proposed five approaches in
sequence. The CSI configurations for BSs and MSs are displayed
separately, and the corresponding descriptions include its source,
its content, and the quantity.

\begin{table}[htbp]\scriptsize
\centering
\begin{tabular}{|c|c|c|c|}
\hline
\multirow{2}{*}{Approach} & \multicolumn{3}{|c|}{required CSI at $k$-th BS} \\
\cline{2-4}
& Source & Content & Quantity\\
\hline
A & all intra-cell MSs & $\mathbf{U}_{k:m}^{\dag}\mathbf{H}_k^{k:m}$ & $M$ \\
\hline
B & all inter-cell MSs & $\mathbf{\Psi}_{j:n}^{\dag}\mathbf{H}_k^{j:n}$ & $(K-1)M$ \\
\hline
C & all intra-cell MSs & $\mathbf{G}_{k:m}\mathbf{H}_{k}^{k:m}$ & $M$ \\
\hline
D & $\begin{aligned}\text{all intra-cell MSs} \\\text{other inter-cell MSs}\end{aligned}$ & $\mathbf{U}_{k:m}^{\dag}\mathbf{H}_k^{k:m}$,$\mathbf{\Omega}_k^j$ & $M$,$K-1$ \\
\hline
E & all inter-cell MSs & $\mathbf{H}_k^{j:m}$ & $\begin{aligned}&\ \ (K-1)M\\&(\text{excl. backhaul})\end{aligned}$ \\
\hline \hline
\multirow{2}{*}{Approach} & \multicolumn{3}{|c|}{required CSI at $m$-th MS in $k$-th cell} \\
\cline{2-4}
& Source & Content & Quantity\\
\hline
A & all inter-cell BSs & $\mathbf{H}_j^{k:m}\mathbf{\Phi}_j$ & $(K-1)$ \\
\hline
B & intra-cell BS & $\mathbf{H}_k^{k:m}\mathbf{V}_{k:n}$ & $(M-1)$ \\
\hline
C & all inter-cell BSs & $\mathbf{H}_j^{k:m}$ & $(K-1)$ \\
\hline
D & all inter-cell BSs & $\mathbf{H}_j^{k:m}$ & $\begin{aligned}&\ \ \ \ \ (K-1)\\&(\text{excl. conferencing})\end{aligned}$ \\
\hline
E & intra-cell BS & $\mathbf{H}_k^{k:m}\mathbf{V}_{k:n}$,$\mathbf{\Theta}^{k:m}$ & $(M-1),1$ \\
\hline
\end{tabular}\caption{CSI for BSs and MSs in All
Approaches in Model
1}\label{cellular-model1-system-condition-2}\normalsize
\end{table}

In summary, as shown in Table
\ref{cellular-model1-system-condition-2}, approach 'B' and approach
'E' need a large amount of CSI. In addition, approach 'B', 'D' and
'E' need inter-cell CSI for BSs, and approach 'A', 'C', 'D' need
inter-cell CSI for MSs. In general, inter-cell CSI are much harder
to obtain than intra-cell CSI. So it is suggested that BSs and MSs
only utilize feedback within a cell, i.e. local CSI as in
\cite{downlink-IA}, with a few changes to the existing intra-cell
feedback mechanism for multi-user MIMO. When BSs need global CSI or
inter-cell CSI, direct feedback from inter-cell communication
sessions or backhaul cooperations are necessary. Due to the nature
of broadcast channels, BSs and MSs have different demands for CSI.
The MSs could share CSI in the cell. BSs and MSs could exchange CSI
via back-and-forth signalling. Collaborative BSs could be
coordinated with the multi-cell processing (MCP) strategy as in
\cite{limited-capacity-backhaul-inter-users-links-cooperative-multicell}.
MSs are also allowed to cooperate via conferencing
\cite{limited-capacity-backhaul-inter-users-links-cooperative-multicell}.


\subsubsection{Computational Complexity}

To implement the design of interference alignment, all nodes need to
calculate their coding matrices or filters with obtained channel
state information (CSI). To obtain different kinds of results with
different types of operations, all the nodes need to evaluate their
system cost by a universal metric, i.e. computational complexity.
The computational complexity is counted as the number of flops.
While a flop usually refers to a floating point operation as
addition, multiplication, or division of real numbers, so that a
complex addition and multiplication have two flops and six flops,
respectively \cite{scientific-computing-survey}
\cite{matrix-computation}.

The aim of this work is to provide a comparison of requirements of
computational complexities for all the nodes in different
approaches. For the above five approaches, all the complexity
configurations are listed in Table
\ref{cellular-model1-system-condition-3}. The notations 'A', 'B',
'C', 'D', 'E' denote the proposed five approaches in sequence. The
requirements for BS and MS are separately displayed. To calculate
the target, two operations are available: matrix inverse and null
space. The amount of calculation is also affected by the scales or
sizes of the matrices.

\begin{table}[htbp]
\centering
\begin{tabular}{|c|c|c|c|}
\hline
\multirow{2}{*}{Approach} & \multicolumn{3}{|c|}{expended Complexity at $k$-th BS } \\
\cline{2-4}
& Target & Operation & Scale \\
\hline
A & $\mathbf{\tilde V}_{k:m}$ & matrix inverse & $Md\times Md$ \\
\hline
B & $\mathbf{V}_{k:m}$ & null space & $N_t\times Md$ \\
\hline
C & $\mathbf{V}_{k:m}$ & matrix inverse & $2Md\times N_t$ \\
\hline
D & $\mathbf{V}_{k:m}$ & matrix inverse & $(M+K-1)d\times N_t$ \\
\hline
E & $\mathbf{V}_{k:m},\mathbf{\Theta}^{k:m}$ & null space & $(KMN_r+KN_t)\times Md$ \\
\hline \hline
\multirow{2}{*}{Approach} & \multicolumn{3}{|c|}{expended Complexity at $m$-th MS in $k$-th cell} \\
\cline{2-4}
& Target & Operation & Scale \\
\hline
A & $\mathbf{U}_{k:m}$ & null space & $N_r\times d$ \\
\hline
B & $\mathbf{\tilde U}_{k:m}$ & null space & $Md\times d$ \\
\hline
C & $\mathbf{G}_{k:m}$ & matrix inverse & $N_r\times (K-1)N_t$ \\
\hline
D & $\mathbf{U}_{k:m}$,$\mathbf{\Omega}_j^k$ & null space & $[(K-1)N_t+MN_r]\times d$ \\
\hline
E & $\mathbf{U}_{k:m}$ & null space & $N_r\times d$ \\
\hline
\end{tabular}\caption{Complexity for BSs and MSs in All
Approaches in Model 1}\label{cellular-model1-system-condition-3}
\end{table}

In summary, as shown in Table
\ref{cellular-model1-system-condition-3}, approach 'D' and 'E' need
a large amount of complexity at BSs for operations with large
scales, and approach 'C' and 'D' need a large amount of complexity
at MSs for operations with large scales.


\section{Basic Design and Analysis of Model 2}


This section looks into model 2 as shown in Fig.
\ref{Cellular-IFBC-IA-2} and equation (\ref{cellular-model-2-eq1}).
Each BS only affects MSs in two adjacent cells and vice versa each
MS only receives inter-cell interference from two BSs in adjacent
cells. In practical cellular networks, if the BS in one cell is not
nearby MSs in another cell, then the interference could be
negligible considering large path loss. So that the network is not
fully connected, and could be modeled as an infinite cyclic network,
in particular a Wyner model \cite{Wyner-Model}. The Wyner model
provides a simple and analytical tractable framework for a cellular
system to capture the essence of inter-cell interference and fading.
Previous extensive studies on the Wyner model mostly treat
interference as noise
\cite{limited-capacity-backhaul-inter-users-links-cooperative-multicell},
\cite{symmetric-capacity-cellular-cooperative-BS}. However in this
work, the focus is on the DoF structure of interference in this
cellular network. Similar to the analysis for model 1, in general,
there are three angles to implement the design, i.e. on the BS side,
on the MS side, and interplay on both BS and MS sides. Accordingly
the following work proposes five approaches to implement the
alignment.

\textit{Clarification}: Be advised that, the notation settings are
similar to the analysis for model 1, and some symbols are reused so
long as they do not cause any confusion. In the following
expressions of the design process, when a same BS or MS in a set is
referred, the index may change between successive steps according to
its subject or object position in the context of the current step.
In each step, the index refers to all equal nodes. Specifically, $k$
is used to denote the subject, and $j$ is used to denote the object.
Besides, in the corresponding tables, the index may also change
according to the context of discussion.

\subsection{Approach on Transmitter Side with Cascaded Coder}

This approach is similar to the design of \cite{downlink-IA} as used
in model 1. Each BS uses a cascaded precoder to zero-forcing ICI,
and then deal with its own IUI. The whole process takes four steps
as following:

Step 1: Choose an auxiliary intermediate precoder
$\mathbf{\Phi}_{k}\in\mathbb{C}^{N_t\times (M\cdot d)}$ at $k$-th BS
(transmitter), and a second precoder $\mathbf{\tilde
V}_{k:m}\in\mathbb{C}^{(M\cdot d)\times d}$ preceding the first
chosen precoder. Then the actual precoder $\mathbf{V}_{k:m}$ is
composed of:

\begin{equation}
\begin{aligned}\label{cellular-model-2-eq1-transmit-V}
\mathbf{V}_{k:m}=\mathbf{\Phi}_{k}\mathbf{\tilde{V}}_{k:m} \ \ \
m\in\mathcal{M}, k\in\mathcal{K}
\end{aligned}
\end{equation}

(Notice the index refers to all equal nodes.)

Step 2: When the first precoder $\mathbf{\Phi}_{j}$ is randomly
picked, to enable zero-forcing the ICI from the two adjacent cells,
all $M$ users in $k$-th cell should determine $\mathbf{U}_{k:m}$ to
satisfy the following conditions (index changed):

\begin{equation}
\begin{aligned}\label{cellular-model-2-eq1-transmit-ZF-u}
\mathbf{U}_{k:m}^{\dag}[\
\mathbf{{H}}_{k-1}^{k:m}\mathbf{\Phi}_{k-1}\
\mathbf{{H}}_{k+1}^{k:m}\mathbf{\Phi}_{k+1}\ ]=\mathbf{0} \ \ \
m\in\mathcal{M}
\end{aligned}
\end{equation}

To be feasible to find $\mathbf{U}_{k:m}$, the dimensions should
satisfy $N_r\geq 2Md+d$. Obviously, this requirement is greatly
loosened compared with the analysis for model 1, the full connected
network.

Step 3: Now observe each $\mathbf{\Phi}_{k}$ again, notice there're
$2M$ zero-forcing conditions to satisfy (notice index):

\begin{equation}
\begin{aligned}\label{cellular-model-2-eq1-transmit-ZF-phi}
\left[
\begin{aligned}
&\hspace{10mm}\vdots\\
&\mathbf{U}_{k-1:m}^{\dag}\mathbf{{H}}_k^{k-1:m}\\
&\hspace{10mm}\vdots \\
&\hspace{10mm}\vdots \\
&\mathbf{U}_{k+1:m}^{\dag}\mathbf{{H}}_k^{k+1:m}\\
&\hspace{10mm}\vdots\\
\end{aligned}
\right]\mathbf{\Phi}_{k}=\mathbf{0} \ \ \ m\in\mathcal{M}
\end{aligned}
\end{equation}

\begin{Remark}
Check the equation (\ref{cellular-model-2-eq1-transmit-ZF-phi}).
Normally it is feasible to find $\mathbf{\Phi}_{k}$ only when
$N_t\geq2Md+Md$. However, since
(\ref{cellular-model-2-eq1-transmit-ZF-u}) already guarantees this
condition, it is only required that $N_t\geq Md$ in this specific
design process. So this process naturally contains an
\textit{implicit} and \textit{inherent} alignment.
\end{Remark}

Step 4: Then each $k$-th BS knows its equivalent downlink channel
$\mathbf{U}_{k:m}\mathbf{{H}}_k^{k:m}\mathbf{\Phi}_{k}$ from each
intra-cell MS. It forms zero-forcing transmit streams eliminating
IUI, without knowing the actual interfering streams. The second
precoder matrix is obtained by an inverse operation (notice index):

\begin{equation}\label{cellular-model-2-eq1-transmit-Vtilde}
\begin{aligned}
&\left[\mathbf{\tilde{V}}_{k:1}\ \mathbf{\tilde{V}}_{k:2}\ \cdots\
\mathbf{\tilde{V}}_{k:M}\right]= \left[
\begin{aligned}
&\mathbf{U}_{k:1}^{\dag}\mathbf{{H}}_k^{k:1}\mathbf{\Phi}_{k}\\&\mathbf{U}_{k:2}^{\dag}\mathbf{{H}}_k^{k:2}\mathbf{\Phi}_{k}\\&\hspace{10mm}\vdots\\&\mathbf{U}_{k:M}^{\dag}\mathbf{{H}}_k^{k:M}\mathbf{\Phi}_{k}
\end{aligned}
\right]^{-1}
\end{aligned}
\end{equation}

In summary, the four steps determine intermediate precoders, receive
filters, and second precoders in sequence. Alignment is implicitly
contained in step 3. Compared with model 1, it is evident that less
antennas are used here to implement alignment because of the
connectivity only between neighboring cells.


\subsection{Approach on Receiver Side with Cascaded Coder: I}

This approach is proposed as a dual solution of the above approach
on transmitter side. Set the cascaded coders on the receiver side.
Then the whole process takes the following four steps:

Step 1: Choose an auxiliary or intermediate receive filter matrix
$\mathbf{\Psi}_{k:m}\in\mathbb{C}^{N_r\times Md}$ at the $m$-th MS
in the $k$-th cell, and a second receive filter $\mathbf{\tilde
U}_{k:m}\in\mathbb{C}^{Md\times d}$ which precedes the first filter.
Then the actual receive filter $\mathbf{U}_{k:m}$ is composed of:

\begin{equation}
\begin{aligned}\label{cellular-model-2-eq1-receiver-U}
\mathbf{U}_{k:m}=\mathbf{\Psi}_{k:m}\mathbf{\tilde{U}}_{k:m} \ \ \
m\in\mathcal{M}, k\in\mathcal{K}
\end{aligned}
\end{equation}

Step 2: When every first receive filter $\mathbf{\Psi}_{j:n}$ is
randomly picked, to enable zero-forcing ICI, the $k$-th BS should
determine $\mathbf{V}_{k:m}, m\in\mathcal{M}$ which satisfies the
condition:

\begin{equation}
\begin{aligned}\label{cellular-model-2-eq1-receiver-ZF-V}
\left[
\begin{aligned}
&\hspace{10mm}\vdots\\
&\mathbf{\Psi}_{k-1:n}^{\dag}\mathbf{{H}}_k^{k-1:n}\\
&\hspace{10mm}\vdots\\
&\hspace{10mm}\vdots\\
&\mathbf{\Psi}_{k+1:n}^{\dag}\mathbf{{H}}_k^{k+1:n}\\
&\hspace{10mm}\vdots
\end{aligned}
\right][\mathbf{V}_{k:1}\ \mathbf{V}_{k:2}\ \cdots\
\mathbf{V}_{k:M}]=\mathbf{0} \ \ \ n\in\mathcal{M}
\end{aligned}
\end{equation}

To be feasible to determine $[\mathbf{V}_{k:1}\ \mathbf{V}_{k:2}\
\cdots\ \mathbf{V}_{k:M}]$, the dimensions should satisfy $N_t\geq
2M^2d+Md$. Compared with model 1, the requirement for antennas is
greatly reduced.

Step 3: Now observe each $\mathbf{\Psi}_{k:m}$ again, notice
there're $2M$ zero-forcing conditions:

\begin{equation}
\begin{aligned}\label{cellular-model-2-eq1-receiver-ZF-psi}
&\mathbf{\Psi}_{k:m}^{\dag}[\underbrace{\cdots\
\mathbf{{H}}_{k-1}^{k:m}\mathbf{V}_{k-1:n}\ \cdots\ \cdots\
\mathbf{{H}}_{k+1}^{k:m}\mathbf{V}_{k+1:n}\
\cdots}_{n\in\mathcal{M}}]=\mathbf{0} \ \ \ \\
&\hspace{60mm}m\in\mathcal{M}
\end{aligned}
\end{equation}

\begin{Remark}
Check the condition of (\ref{cellular-model-2-eq1-receiver-ZF-psi}).
Notice in general it is only feasible to find $\mathbf{\Psi}_{k:m}$
when the dimensions satisfy $N_r\geq2Md+Md$. However, since
(\ref{cellular-model-2-eq1-receiver-ZF-V}) already guarantees the
condition, it is only required that $N_r\geq Md$ in this specific
design process. So this process naturally contains an
\textit{implicit} and \textit{inherent} alignment.
\end{Remark}

Step 4: Then each intra-cell MS in the $k$-th cell knows its
equivalent channel
$\mathbf{\Psi}_{k:m}^{\dag}\mathbf{{H}}_k^{k:m}\mathbf{V}_{k:m}$,
while it does not know the actual interfering beams. It forms the
zero-forcing receive filter to eliminate IUI. The second receive
filter $\mathbf{\tilde{U}}_{k:m}$ is determined by an operation to
find the null space:

\begin{equation}\label{cellular-model-2-eq1-receiver-Utilde}
\begin{aligned}
&\mathbf{\tilde{U}}_{k:m}^{\dag}\left[\ \underbrace{\cdots\
\mathbf{\Psi}_{k:m}^{\dag}\mathbf{{H}}_k^{k:m}\mathbf{V}_{k:n}\
\cdots}_{n\in\mathcal{M}\backslash\{m\}}\ \right]=\mathbf{0}
\end{aligned}
\end{equation}

In summary, the four steps determine intermediate receive filters,
precoders, second receive filters in sequence. Alignment is
implicitly contained in step 3. Compared with model 1, the usage of
antennas is evidently reduced because of the loose connectivity.

%

\subsection{Approach on Receiver Side with  Cascaded Coder: II}

This approach is similar to the design of
\cite{mixed-rank-compound-X-IA-MU}. Each MS uses a cascaded receive
filter to align ICI from other cells, and deal with ICI and IUI in
its own cell. The whole process takes four steps as following:

Step 1: For each $m$-th MS in $k$-th cell, the receive filter
$\mathbf{U}_{k:m}$ is defined as two cascaded filter matrices:

\begin{equation}
\begin{aligned}\label{cellular-model-2-eq1-receiver-ref-1}
&\mathbf{U}_{k:m}=\mathbf{G}_{k:m}^{\dag}{\Lambda}_{\mathcal{K}:m}\\
&\mathbf{G}_{k:m}=[{\mathbf{I}_{N_t}
\mathbf{I}_{N_t}}][{\mathbf{H}_{k-1}^{k:m}
\ \mathbf{H}_{k+1}^{k:m}}]^{-1}\\
&\text{So that}\ \ \mathbf{G}_{k:m}\in\mathbb{C}^{N_t\times
N_r},\mathbf{G}_{k:m}\mathbf{H}_j^{k:m}=\mathbf{I}_{N_t}\\
&\hspace{38mm}\forall j=k-1,k+1
\end{aligned}
\end{equation}

$\mathbf{G}_{k:m}$ is set to make all ICI from the two neighboring
cells aligned together. While ${\Lambda}_{k:m}$ is a common filter
set to be the same for all cells in the network.

To be feasible to determine $\mathbf{G}_{k:m}$ by the inverse
operation, the dimensions should satisfy $N_r\geq 2N_t$. Compared
with model 1, it is obvious that this condition is greatly loosened
to save a lot of antennas.

Step 2: Since the common receive filter
$\Lambda_{\mathcal{K}:m}\in\mathbb{C}^{N_t\times d}$ is set to be
the same for all cells of $\mathcal{K}$, the output stream of
(\ref{cellular-model-2-eq1}) is presented as following:

\begin{equation}
\begin{aligned}\label{cellular-model-2-eq1-receiver-ref-2}
&\Lambda_{\mathcal{K}:m}^{\dag}\mathbf{G}_{k:m}\mathbf{y}_{k:m}=\\
&\Lambda_{\mathcal{K}:m}^{\dag}\mathbf{G}_{k:m}\mathbf{H}_{k}^{k:m}\mathbf{V}_{k:m}\mathbf{x}_{k:m}
+\sum_{n\neq
m}^{n\in\mathcal{M}}\Lambda_{\mathcal{K}:m}^{\dag}\mathbf{G}_{k:m}\mathbf{H}_{k}^{k:m}\mathbf{V}_{k:n}\mathbf{x}_{k:n}\\
&\hspace{30mm}+\sum_{j=k-1,k+1}\sum_{n\in\mathcal{M}}\Lambda_{\mathcal{K}:m}^{\dag}\mathbf{V}_{j:n}\mathbf{x}_{j:n}
\end{aligned}
\end{equation}

To eliminate ICI and IUI terms in equation
(\ref{cellular-model-2-eq1-receiver-ref-2}) respectively, the
zero-forcing conditions should be set as:

\begin{equation}
\begin{aligned}\label{cellular-model-2-eq1-receiver-ref-2zf}
\Lambda_{\mathcal{K}:m}^{\dag}\mathbf{G}_{k:m}\mathbf{H}_{k}^{k:m}\mathbf{V}_{k:n}&=\mathbf{0}\ \ n\in\mathcal{M}\backslash\{m\}\\
\Lambda_{\mathcal{K}:m}^{\dag}\mathbf{V}_{j:n}&=\mathbf{0}\ \
j=k-1,k+1,n\in\mathcal{M}
\end{aligned}
\end{equation}

Compared with model 1, the condition is loosened.

Step 3: Then observe from $k$-th BS to choose the precoder $[\cdots\
\mathbf{V}_{k:m}\ \cdots]$, the zero-forcing conditions in equation
(\ref{cellular-model-2-eq1-receiver-ref-2zf}) are equivalently
presented with new indices in the following expression:

\begin{equation}
\begin{aligned}\label{cellular-model-2-eq1-receiver-ref-3}
\Lambda_{\mathcal{K}:n}^{\dag}\mathbf{G}_{k:n}\mathbf{H}_{k}^{k:n}\mathbf{V}_{k:m}&=\mathbf{0}\ \ n\in\mathcal{M}\backslash\{m\}\\
\Lambda_{\mathcal{K}:n}^{\dag}\mathbf{V}_{k:m}&=\mathbf{0}\ \
n\in\mathcal{M}
\end{aligned}
\end{equation}

Step 4: So for each $k$-th BS, the precoder could be determined by
only one inverse operation:

\begin{equation}
\begin{aligned}\label{cellular-model-2-eq1-receiver-ref-4}
\mathbf{B}_k&=\left[\begin{aligned}\left.
\begin{aligned}
&\ \ \ \ \ \ \ \ \ \ \vdots\\
&\Lambda_{\mathcal{K}:m}^{\dag}\mathbf{G}_{k:m}\mathbf{H}_{k}^{k:m}\\
&\ \ \ \ \ \ \ \ \ \ \vdots
\end{aligned}
\right\}{M}\\\left.
\begin{aligned}
&\vdots\\
&\Lambda_{\mathcal{K}:m}^{\dag}\\
&\vdots
\end{aligned}
\right\}{M}\end{aligned}
\right]\\
[\mathbf{V}_{k:1}\ \mathbf{V}_{k:2}\ \cdots\
\mathbf{V}_{k:M}]&=\text{The first}\ Md\ \text{columns of}\
\mathbf{B}_k^{-1}
\end{aligned}
\end{equation}

Check that $[\mathbf{V}_{k:1}\ \mathbf{V}_{k:2}\ \cdots\
\mathbf{V}_{k:M}]$ indeed satisfies the conditions in
(\ref{cellular-model-2-eq1-receiver-ref-3}).

\begin{Remark}
To be feasible to obtain $[\mathbf{V}_{k:1}\ \mathbf{V}_{k:2}\
\cdots\ \mathbf{V}_{k:M}]$ by an inverse operation, the dimensions
should satisfy $N_t\geq 2Md$. While in general cases, it is required
that $N_t\geq Md+2Md$ for the BS to handle interferences to two
adjacent cells. So this process naturally contains an
\textit{implicit} and \textit{inherent} alignment.
\end{Remark}

In summary, the four steps determine all receive filters, precoders
and second receive filters in sequence. Alignment is explicitly
implemented in step 1. Compared with model 1, less antennas are
required to implement alignment because of the loose connectivity.

\subsection{Approach from Receiver Side with Direct Interplay}

This approach is similar to the design of
\cite{design-IA-two-cell-IBC}. The alignment is jointly implemented
on both BS(transmitter) side and MS(receiver) side via intricate and
meticulous interplays and cooperations. The whole process takes two
steps as following:

Step 1: Regarding channels from the $k$-th BS to all MSs in $j$-th
cell, the equivalent receive filtered channel subspaces are set to
be equal to a random pre-determined reference subspace:

\begin{equation}
\begin{aligned}\label{cellular-model-2-eq1-joint-1}
\mathbf{U}_{j:1}^{\dag}\mathbf{H}_{k}^{j:1}=\mathbf{U}_{j:2}^{\dag}\mathbf{H}_{k}^{j:2}=\cdots=\mathbf{U}_{j:M}^{\dag}\mathbf{H}_{k}^{j:M}=(\mathbf{\Omega}_{k}^{j})^{\dag}\\
\hspace{50mm}j=k-1,k+1
\end{aligned}
\end{equation}

\begin{Remark}
Notice the step of (\ref{cellular-model-2-eq1-joint-1}) is an
\textit{explicit} alignment of ICI from the the $k$-th BS to all MSs
only in the two adjacent cells.
\end{Remark}

Observe from all MSs in $k$-th cell, (index changed) to satisfy the
condition (\ref{cellular-model-2-eq1-joint-1}), $\mathbf{U}_{k:m}$,
$\mathbf{\Omega}_{k-1}^{k}$ and $\mathbf{\Omega}_{k+1}^{k}$ could be
found by solving the following null space problem:

\begin{equation}\label{cellular-model-2-eq1-joint-2}
\begin{aligned}
&\left[{(\mathbf{\Omega}_{k-1}^{k})^{\dag}\
(\mathbf{\Omega}_{k+1}^{k})^{\dag}}\ \mathbf{U}_{k:1}^{\dag}\
\mathbf{U}_{k:2}^{\dag}\ \cdots\
\mathbf{U}_{k:M}^{\dag}\right]\bullet\\
&\ \ \left[
\begin{aligned}
&\begin{aligned}
&\mathbf{I}_{N_t} \cdots \mathbf{I}_{N_t}        \\
&\ \ \ \ \ \ \ \ \ \ \ \ \ \ \ \ \mathbf{I}_{N_t} \cdots \mathbf{I}_{N_t}       \\
\end{aligned}\\
&-\mathbf{H}_{k-1}^{k:1}\ \ \ \ \ \ -\mathbf{H}_{k+1}^{k:1}\\
&\ \ -\mathbf{H}_{k-1}^{k:2}\ \ \ \ \ \ -\mathbf{H}_{k+1}^{k:2}   \\
&\ \ \ \ \ \ddots\ \   \ \ \ \ \ \ \ \ \ \ \ddots\ \    \\
&\ \ \ \ \ -\mathbf{H}_{k-1}^{k:M} \ \ \ \ \ -\mathbf{H}_{k+1}^{k:M}
\end{aligned}
\right]=\mathbf{0}
\end{aligned}
\end{equation}

To be feasible to obtain $\mathbf{U}_{k:m}$,
$\mathbf{\Omega}_{k-1}^{k}$ and $\mathbf{\Omega}_{k+1}^{k}$ via a
null space operation, the dimensions should follow $2N_t+MN_r\geq
2MN_t+d$. Compared with model 1, the condition is greatly loosened,
and a lot of antennas could be saved.

Step 2: Thus for the $k$-th BS, the precoders are determined by the
following inverse operation to eliminate both IUI and ICI:

\begin{equation}
\begin{aligned}\label{cellular-model-2-eq1-joint-3}
[\ \cdots\ \mathbf{V}_{k:m}\ \cdots\ ]=\text{first }Md\text{ columns of }\\
\left[\begin{aligned}&\left.\begin{aligned}&\hspace{9mm}\vdots\\
&\mathbf{U}_{k:m}^{\dag}\mathbf{H}_{k}^{k:m}\\
&\hspace{9mm}\vdots\end{aligned}\right\}{m\in\mathcal{M}} \\
&
\begin{aligned}
&\hspace{6mm}{(\mathbf{\Omega}_{k}^{k-1})}^{\dag}\\
&\hspace{6mm}{(\mathbf{\Omega}_{k}^{k+1})}^{\dag}\end{aligned}\end{aligned}
\right]^{-1}
\end{aligned}
\end{equation}

To be feasible to obtain $[\ \cdots\ \mathbf{V}_{k:m}\ \cdots\ ]$
via the inverse operation, the dimensions should satisfy $N_t\geq
Md+2d$. Compared with model 1, the condition is loosened to save a
number of antennas.

Check the equation (\ref{cellular-model-2-eq1-joint-3}) and find
that $[\ \cdots\ \mathbf{V}_{k:m}\ \cdots\ ]$ indeed eliminate the
IUI in the cell and ICI to the two adjacent cells.

In summary, the two steps determine receive filters, aligned
subspace and precoders in sequence. Alignment is implemented on the
MS side and BS side. Compared with model 1, the requirement for
antennas are greatly reduced because of the loose connectivity.

\subsection{Approach from Transmitter Side with Direct Interplay}

This approach is proposed as a dual solution of the above approach
on receiver side. It implements the alignment on the BS side. The
whole process takes the following two steps:

Step 1: Regarding channels from the two adjacent $(k-1)$-th and
$(k+1)$-th BS to all MS in $k$-th cell, the two equivalent precoded
channel subspaces are set to be equal to a random predetermined
reference subspace:

\begin{equation}
\begin{aligned}\label{cellular-model-2-eq1-jointII-1}
&\mathbf{H}_{k-1}^{k:m}[\mathbf{V}_{k-1:1}\cdots\mathbf{V}_{k-1:M}]=\mathbf{H}_{k+1}^{k:m}[\mathbf{V}_{k+1:1}\cdots\mathbf{V}_{k+1:M}]=\mathbf{\Theta}^{k:m}
\end{aligned}
\end{equation}

\begin{Remark}
Notice the step of (\ref{cellular-model-2-eq1-jointII-1}) is an
\textit{explicit} alignment of ICI from the two adjacent BSs to
$m$-th MS in the $k$-th cell.
\end{Remark}

Observe from the side of all BSs, (index changed) to satisfy the
condition (\ref{cellular-model-2-eq1-jointII-1}),
$[\mathbf{V}_{k:1}\cdots\mathbf{V}_{k:M}]$ and
$\mathbf{\Theta}^{k:m}$ could be found by solving the following null
space problem:

\begin{equation}
\begin{aligned}\label{cellular-model-2-eq1-jointII-2}
\left[\begin{aligned}
&\left.\begin{aligned}&\begin{aligned}&\mathbf{I}_{N_r}\\&\mathbf{I}_{N_r}\end{aligned}
\hspace{18mm}{\left.\begin{aligned}&-\mathbf{H}_{K}^{1:m}\\\hspace{0mm}&-\mathbf{H}_{2}^{1:m}\end{aligned}\right\}}\\
&\ \ \ \ddots\hspace{25mm}\vdots\end{aligned}\right\}m\in\mathcal{M}\\
&\left.\begin{aligned}&\ \ \ \ \ \
\begin{aligned}&\mathbf{I}_{N_r}\\&\mathbf{I}_{N_r}\end{aligned}
\hspace{11mm}{\left.\begin{aligned}&-\mathbf{H}_{k-1}^{k:m}\\\hspace{0mm}&-\mathbf{H}_{k+1}^{k:m}\end{aligned}\right\}}\\
&\ \ \ \ \ \ \ \ \ \ddots\hspace{20mm}\vdots\end{aligned}\right\}m\in\mathcal{M}\\
&\left.\begin{aligned}&\ \ \ \ \ \ \ \ \ \ \ \
\begin{aligned}&\mathbf{I}_{N_r}\\&\mathbf{I}_{N_r}\end{aligned}
\hspace{4mm}{\left.\begin{aligned}&-\mathbf{H}_{K-1}^{K:m}\\\hspace{0mm}&-\mathbf{H}_{K+1}^{K:m}\end{aligned}\right\}}\\
&\hspace{19mm}\ddots\hspace{14mm}\vdots\end{aligned}\right\}m\in\mathcal{M}\end{aligned}\right]
\\\bullet\left[\begin{aligned}&\left.\begin{aligned}&\left.\begin{aligned}&\hspace{8mm}\vdots\\&\hspace{7mm}\mathbf{\Theta}^{k:m}\\&\hspace{8mm}\vdots\end{aligned}\right\}k\in\mathcal{K}\\&\hspace{10mm}\vdots\end{aligned}\right\}m\in\mathcal{M}\\&\left.\begin{aligned}&\hspace{10mm}\vdots\\&\left[\mathbf{V}_{k:1}\cdots\mathbf{V}_{k:M}\right]\\&\hspace{10mm}\vdots\end{aligned}\right\}k\in\mathcal{K}\end{aligned}\right]=\mathbf{0}
\end{aligned}
\end{equation}

To be feasible to obtain $[\mathbf{V}_{k:1}\cdots\mathbf{V}_{k:M}]$
and $\mathbf{\Theta}^{k:m}$ via a null space operation, the
dimensions should follow $KMN_r+KN_t\geq 2KMN_r+Md$. Compared with
model 1, the condition is greatly loosened.

Step 2: Thus for the $m$-th MS in $k$-th cell, its receive filter is
determined via a null space operation to eliminate both IUI and ICI:

\begin{equation}
\begin{aligned}\label{cellular-model-2-eq1-jointII-3}
\mathbf{U}_{k:m}^{\dag}\left[\
\mathbf{H}_{k}^{k:m}\big[\underbrace{\cdots\ \mathbf{V}_{k:n}\
\cdots}_{n\in\mathcal{M}\backslash\{m\}}\big] \ \
{\mathbf{\Theta}^{k:m}}\right]=\mathbf{0}
\end{aligned}
\end{equation}

To be feasible to obtain $\mathbf{U}_{k:m}$, the dimensions should
follow $N_r\geq (M-1)d+Md+d$. It is easy to check that the condition
(\ref{cellular-model-2-eq1-jointII-3}) directly eliminates both IUI
and ICI.

In summary, the two steps determine precoders, aligned subspaces,
and receive filters in sequence. Alignment is explicitly
implemented. Compared with model 1, the antennas for BS nodes are
greatly reduced because of the loose connectivity.

\subsection{Comparison of System Conditions}


Similar to the analyses for model 1, system conditions are listed
and compared for all the proposed five approaches of downlink
interference alignment in a Wyner-type cellular network with two
adjacent links for each cell. There are three aspects to
investigate: the usage of antennas which is related to available
DoF; the CSI required which may generate overhead in the system; the
computational complexity which impacts realistic performances.

\subsubsection{DoF and Antenna Range}

Similar to the analyses for model 1, the number of antennas plays a
key role to implement alignment, while the feasibility conditions
are not solved in general. The antennas could have different balance
between BS side and MS side to achieve alignment.

The aim of this work is exploit interference alignment with limited
provision of antennas at both BS side and MS side with a balance.
For all the five approaches, all the antenna configurations are
listed in Table \ref{cellular-model2-system-condition}, in which the
notations of 'A', 'B', 'C', 'D' and 'E' denote the five approaches
in sequence.

\begin{table}
\centering
\begin{tabular}{|c|c|c|}
  \hline
  Approach & Minimum BS Antenna. & Minimum MS Antenna  \\
  \hline
  A & $Md$ & $2Md+d$  \\
  \hline
  B & $2M^2d+Md$ & $Md$  \\
  \hline
  C & $2Md$ & $4Md$  \\
  \hline
  D & $(M+2)d$ & $\scriptsize\begin{aligned}2\frac{(M-1)}{M}(M+2)d\\+d/M\end{aligned}\normalsize$  \\
  \hline
  E & $\scriptsize\begin{aligned}2M^2d\\+(M/K)d\end{aligned}\normalsize$ & $2Md$  \\
  \hline
\end{tabular}\caption{Minimum Antenna Usage for BSs and MSs in All Approaches in Model 2}\label{cellular-model2-system-condition}
\end{table}

In summary, as shown in Table
\ref{cellular-model2-system-condition}, approach 'B' and 'E' need a
large number of antennas for BS nodes. Compared model 1 in Table
\ref{cellular-model1-system-condition}, the antenna requirement is
greatly reduced, especially for MS nodes.

\subsubsection{CSI Overhead}

Similar to the analyses for model 1, the proposed interference
alignment approaches require all the BS nodes and MS nodes to have
certain knowledge of channel state information (CSI).

The aim of this work is to compare the CSI requirements for the
network in different approaches. For the above five approaches, all
the CSI configurations are listed in Table
\ref{cellular-model2-system-condition-2}. The notations from 'A' to
'E' denote the five approaches in sequence. Other settings are also
the same as in model 1.

\begin{table}[htbp]\scriptsize
\centering
\begin{tabular}{|c|c|c|c|}
\hline
\multirow{2}{*}{Approach} & \multicolumn{3}{|c|}{required CSI at $k$-th BS} \\
\cline{2-4}
& Source & Content & Quantity\\
\hline
A & all intra-cell MSs & $\mathbf{U}_{k:m}^{\dag}\mathbf{H}_k^{k:m}$ & $M$ \\
\hline
B & adjacent inter-cell MSs & $\mathbf{\Psi}_{j:n}^{\dag}\mathbf{H}_k^{j:n}$ & $2M$ \\
\hline
C & all intra-cell MSs & $\mathbf{G}_{k:m}\mathbf{H}_{k}^{k:m}$ & $M$ \\
\hline
D & $\begin{aligned}\text{all intra-cell MSs} \\\text{adjacent inter-cell MSs}\end{aligned}$ & $\mathbf{U}_{k:m}^{\dag}\mathbf{H}_k^{k:m}$,$\mathbf{\Omega}_k^j$ & $M$,$2$ \\
\hline
E & adjacent inter-cell MSs & $\mathbf{H}_k^{j:m}$ & $\begin{aligned}&\hspace{5mm}2M\\&(\text{excl. backhaul})\end{aligned}$ \\
\hline \hline
\multirow{2}{*}{Approach} & \multicolumn{3}{|c|}{required CSI at $m$-th MS in $k$-th cell} \\
\cline{2-4}
& Source & Content & Quantity\\
\hline
A & adjacent inter-cell BSs & $\mathbf{H}_j^{k:m}\mathbf{\Phi}_j$ & $2$ \\
\hline
B & intra-cell BS & $\mathbf{H}_k^{k:m}\mathbf{V}_{k:n}$ & $(M-1)$ \\
\hline
C & adjacent inter-cell BSs & $\mathbf{H}_j^{k:m}$ & $2$ \\
\hline
D & adjacent inter-cell BSs & $\mathbf{H}_j^{k:m}$ & $\begin{aligned}&\hspace{9mm}2\\&(\text{excl. conferencing})\end{aligned}$ \\
\hline
E & intra-cell BS & $\mathbf{H}_k^{k:m}\mathbf{V}_{k:n}$,$\mathbf{\Theta}^{k:m}$ & $(M-1)$,$1$ \\
\hline
\end{tabular}\caption{CSI for BSs and MSs in All
Approaches in Model
2}\label{cellular-model2-system-condition-2}\normalsize
\end{table}

In summary, as shown in Table
\ref{cellular-model2-system-condition-2}, approach 'B' and 'E' need
a large amount of CSI. Compared with Table
\ref{cellular-model1-system-condition-2} of model 1, the CSI
requirement is lowered in general. While the requirements for
inter-cell CSI or intra-cell CSI are still the same for
corresponding nodes.

\subsubsection{Computational Complexity}

Similar to the analyses for model 1, all the nodes need to calculate
their coding matrices or filters with obtained CSI. The calculations
use different operations, and are commonly measured by the metric of
computational complexity.

The aim of this work is to compare the computational complexities
for the network using different approaches. For the above five
approaches, all the complexity configurations are listed in Table
\ref{cellular-model2-system-condition-3}. The notations from 'A' to
'E' denote the five approaches in sequence. Other settings are also
the same as model 1.

\begin{table}[htbp]
\centering
\begin{tabular}{|c|c|c|c|}
\hline
\multirow{2}{*}{Approach} & \multicolumn{3}{|c|}{expended Complexity at $k$-th BS } \\
\cline{2-4}
& Target & Operation & Scale \\
\hline
A & $\mathbf{\tilde V}_{k:m}$ & matrix inverse & $Md\times Md$ \\
\hline
B & $\mathbf{V}_{k:m}$ & null space & $N_t\times Md$ \\
\hline
C & $\mathbf{V}_{k:m}$ & matrix inverse & $2Md\times N_t$ \\
\hline
D & $\mathbf{V}_{k:m}$ & matrix inverse & $(M+2)d\times N_t$ \\
\hline
E & $\mathbf{V}_{k:m},\mathbf{\Theta}^{k:m}$ & null space & $(KMN_r+KN_t)\times Md$ \\
\hline \hline
\multirow{2}{*}{Approach} & \multicolumn{3}{|c|}{expended Complexity at $m$-th MS in $k$-th cell} \\
\cline{2-4}
& Target & Operation & Scale \\
\hline
A & $\mathbf{U}_{k:m}$ & null space & $N_r\times d$ \\
\hline
B & $\mathbf{\tilde U}_{k:m}$ & null space & $Md\times d$ \\
\hline
C & $\mathbf{G}_{k:m}$ & matrix inverse & $N_r\times 2N_t$ \\
\hline
D & $\mathbf{U}_{k:m}$,$\mathbf{\Omega}_j^k$ & null space & $(2N_t+MN_r)\times d$ \\
\hline
E & $\mathbf{U}_{k:m}$ & null space & $N_r\times d$ \\
\hline
\end{tabular}\caption{Complexity for BSs and MSs in All
Approaches in Model 2}\label{cellular-model2-system-condition-3}
\end{table}

In summary, as shown in Table
\ref{cellular-model2-system-condition-3}, approach 'D' and 'E' need
a large amount of computational complexities. Compared with model 1
in Table \ref{cellular-model1-system-condition-3}, however the
expended complexity is reduced greatly.


\section{Basic Design and Analysis of Model 3}


This section looks into model 3 as shown in Fig.
\ref{Cellular-IFBC-IA-3}, equation (\ref{cellular-model-3-eq1a}) and
(\ref{cellular-model-3-eq1b}). Similar to model 2, another
Wyner-type model is studied here. It has only one interference link
between the BS and MSs belonging to two neighboring cells, in a
cascaded fashion in the cyclic network. It could be also referred as
the Z-interference channel (ZIC) or cylic interference channel
\cite{capacity-K-user-cyclic-Gaussian-IFC},
\cite{sum-capacity-K-user-cascade-Gaussian-Z-IFC}. Previous
researches only focus on the capacity regions by using Han-Kobayashi
schemes. However in this work, the focus is on the DoF structure of
interference in this cellular network. Similar to model 1 and model
2, five approaches are proposed to implement the alignment as
following.

\textit{Clarification}: the notations are similar to the analyses
for model 1 and model 2. Some symbols are reused so long as they do
not cause any confusion. In the expression of the design process
referring to an arbitrary BS or MS in a set, the variable index may
change between successive steps accordingly in the discussion. They
may change in corresponding tables as well.

\subsection{Approach on Transmitter Side with Cascaded Coder}

This approach is similar to the design of \cite{downlink-IA} as used
in model 1 and model 2. Each BS uses a cascaded precoder to
zero-forcing ICI, and then deal with its own IUI. The whole process
takes four steps as following:

Step 1: Choose an auxiliary intermediate precoder
$\mathbf{\Phi}_{k}\in\mathbb{C}^{N_t\times (M\cdot d)}$ at $k$-th BS
(transmitter), and a second precoder $\mathbf{\tilde
V}_{k:m}\in\mathbb{C}^{(M\cdot d)\times d}$ preceding the first
chosen precoder. Then the actual precoder $\mathbf{V}_{k:m}$ is
composed of:

\begin{equation}
\begin{aligned}\label{cellular-model-3-eq1-transmit-V}
\mathbf{V}_{k:m}=\mathbf{\Phi}_{k}\mathbf{\tilde{V}}_{k:m} \ \ \
m\in\mathcal{M}, k\in\mathcal{K}
\end{aligned}
\end{equation}

(Notice the index refers to all equal nodes.)

Step 2: When the first precoder $\mathbf{\Phi}_{j}$ is randomly
picked, to enable zero-forcing the ICI from the only one adjacent
cell, all the $M^{\circ}$ cell-edge MSs in $k$-th cell should
determine $\mathbf{U}_{k:m^{\circ}}$ to satisfy the following
conditions (index changed):

\begin{equation}
\begin{aligned}\label{cellular-model-3-eq1-transmit-ZF-u}
\mathbf{U}_{k:m^{\circ}}^{\dag}\mathbf{{H}}_{k+1}^{k:m^{\circ}}\mathbf{\Phi}_{k+1}=\mathbf{0}
\ \ \ m^{\circ}\in\mathcal{M}^{\circ}
\end{aligned}
\end{equation}

To be feasible to find $\mathbf{U}_{k:m^{\circ}}$, the dimensions
should satisfy $N_r^{\circ}\geq Md+d$ for cell-edge users, where
$N_r^{\circ}$ is the number of receive antennas of any MS in the
edge set $\mathcal{M}^{\circ}$. While for cell-interior users, the
receive filters $\mathbf{U}_{k:m^{\star}}$ for $M^{\star}$
cell-interior MSs in $k$-th cell could be randomly generated. So
that it's only required the dimensions should satisfy
$N_r^{\star}\geq d$, where $N_r^{\star}$ is the number of receive
antennas of any MS in the interior set $\mathcal{M}^{\star}$.
Compared with analyses for model 1 and model 2, the requirements for
antennas are greatly loosened.

Step 3: Now observe each $\mathbf{\Phi}_{k}$ again, there're
$M^{\circ}$ zero-forcing conditions to satisfy (notice change of
index):

\begin{equation}
\begin{aligned}\label{cellular-model-3-eq1-transmit-ZF-phi}
\left[
\begin{aligned}
&\hspace{10mm}\vdots\\
&\mathbf{U}_{k-1:m^{\circ}}^{\dag}\mathbf{{H}}_k^{k-1:m^{\circ}}\\
&\hspace{10mm}\vdots \\
\end{aligned}
\right]\mathbf{\Phi}_{k}=\mathbf{0} \ \ \
m^{\circ}\in\mathcal{M}^{\circ}
\end{aligned}
\end{equation}

\begin{Remark}
check the condition (\ref{cellular-model-3-eq1-transmit-ZF-phi}).
Normally it is feasible to find $\mathbf{\Phi}_{k}$ only when
$N_t\geq Md+M^{\circ}d$. However, since
(\ref{cellular-model-3-eq1-transmit-ZF-u}) already guarantees this
condition, it is only required that $N_t\geq Md$ in this specific
design process. So this process naturally contains an
\textit{implicit} and \textit{inherent} alignment.
\end{Remark}

Step 4: Then each $k$-th BS knows its equivalent downlink channel
$\mathbf{U}_{k:m}\mathbf{{H}}_k^{k:m}\mathbf{\Phi}_{k}$ from each
intra-cell MS. It forms zero-forcing transmit streams eliminating
IUI, without knowing the actual interfering streams. The second
precoder matrix is obtained by an inverse operation:

\begin{equation}\label{cellular-model-3-eq1-transmit-Vtilde}
\begin{aligned}
&\left[\mathbf{\tilde{V}}_{k:1}\ \mathbf{\tilde{V}}_{k:2}\ \cdots\
\mathbf{\tilde{V}}_{k:M}\right]= \left[
\begin{aligned}
&\mathbf{U}_{k:1}^{\dag}\mathbf{{H}}_k^{k:1}\mathbf{\Phi}_{k}\\&\mathbf{U}_{k:2}^{\dag}\mathbf{{H}}_k^{k:2}\mathbf{\Phi}_{k}\\&\hspace{10mm}\vdots\\&\mathbf{U}_{k:M}^{\dag}\mathbf{{H}}_k^{k:M}\mathbf{\Phi}_{k}
\end{aligned}
\right]^{-1}
\end{aligned}
\end{equation}

In summary, the four steps determine intermediate precoders, receive
filters, and second precoders in sequence. Alignment is implicitly
contained in step 3. Compared with model 1 and model 2, it is
evident that less antennas are used to implement alignment, because
there is only one adjacent interfering path for each cell and only
cell-edge users are involved and affected.


\subsection{Approach on Receiver Side with Cascaded Coder: I}

This approach is proposed as a dual solution of the above approach
on transmitter side. Set the cascaded coders for the cell-edge users
on the receiver side. Then the whole process takes the following
four steps:

Step 1: Choose an auxiliary or intermediate receive filter matrix
$\mathbf{\Psi}_{k:m^{\circ}}\in\mathbb{C}^{N_r^{\circ}\times Md}$ at
the $m^{\circ}$-th cell-edge MS in $k$-th cell, and a second receive
filter $\mathbf{\tilde U}_{k:m^{\circ}}\in\mathbb{C}^{Md\times d}$
preceding the first chosen receive filter. Then the actual receive
filter $\mathbf{U}_{k:m^{\circ}}$ is composed of:

\begin{equation}
\begin{aligned}\label{cellular-model-3-eq1-receiver-U}
\mathbf{U}_{k:m^{\circ}}=\mathbf{\Psi}_{k:m^{\circ}}\mathbf{\tilde{U}}_{k:m^{\circ}}
\ \ \ m^{\circ}\in\mathcal{M^{\circ}}, k\in\mathcal{K}
\end{aligned}
\end{equation}

While for cell-interior users, any $m^{\star}$-th MS in the $k$-th
cell has a complete receive filter $\mathbf{U}_{k:m^{\star}},
m^{\star}\in\mathcal{M^{\star}}$.

Step 2: When every first receive filter
$\mathbf{\Psi}_{j:n^{\circ}}$ is randomly picked, to enable
zero-forcing ICI, the $k$-th BS should determine $\mathbf{V}_{k:m},
m\in\mathcal{M}$ which satisfies the condition:

\begin{equation}
\begin{aligned}\label{cellular-model-3-eq1-receiver-ZF-V}
\left[
\begin{aligned}
&\hspace{10mm}\vdots\\
&\mathbf{\Psi}_{k-1:n^{\circ}}^{\dag}\mathbf{{H}}_k^{k-1:n^{\circ}}\\
&\hspace{10mm}\vdots\\
\end{aligned}
\right][\mathbf{V}_{k:1}\ \mathbf{V}_{k:2}\ \cdots\
\mathbf{V}_{k:M}]=\mathbf{0} \ \ \ n\in\mathcal{M}^{\circ}
\end{aligned}
\end{equation}

To be feasible to determine $[\mathbf{V}_{k:1}\ \mathbf{V}_{k:2}\
\cdots\ \mathbf{V}_{k:M}]$, the dimensions should satisfy $N_t\geq
M^{\circ}Md+Md$. Compared with model 1 and model 2, the requirement
for antennas is greatly reduced.

Step 3: Now observe each
$\mathbf{\Psi}_{k:m^{\circ}},m^{\circ}\in\mathcal{M^{\circ}}$ again,
notice there're $M^{\circ}$ zero-forcing conditions to satisfy:

\begin{equation}
\begin{aligned}\label{cellular-model-3-eq1-receiver-ZF-psi}
\mathbf{\Psi}_{k:m^{\circ}}^{\dag}[\underbrace{\cdots\
\mathbf{{H}}_{k+1}^{k:m}\mathbf{V}_{k+1:n}\
\cdots}_{n\in\mathcal{M}}]=\mathbf{0} \ \ \
m^{\circ}\in\mathcal{M^{\circ}}
\end{aligned}
\end{equation}

\begin{Remark}
Check the condition of (\ref{cellular-model-3-eq1-receiver-ZF-psi}).
Notice in general it is only feasible to find
$\mathbf{\Psi}_{k:m^{\circ}}$ when the dimensions satisfy
$N_r^{\circ}\geq Md+Md$. However, since
(\ref{cellular-model-3-eq1-receiver-ZF-V}) already guarantees the
condition, it is only required that $N_r^{\circ}\geq Md$ in this
specific design process. So this process naturally contains an
\textit{implicit} and \textit{inherent} alignment.
\end{Remark}

Step 4: Then each cell-edge MS in the $k$-th cell knows its
equivalent channel
$\mathbf{\Psi}_{k:m^{\circ}}^{\dag}\mathbf{{H}}_k^{k:m^{\circ}}\mathbf{V}_{k:n},m^{\circ}\in\mathcal{M^{\circ}},n\in\mathcal{M}$.
Also each cell-interior MS in the $k$-th cell knows its equivalent
channel
$\mathbf{{H}}_k^{k:m^{\star}}\mathbf{V}_{k:n},m^{\star}\in\mathcal{M^{\star}},n\in\mathcal{M}$.
Also they do not know the actual interfering beams, they could form
zero-forcing receive filters to eliminate IUI. The second receive
filters $\mathbf{\tilde{U}}_{k:m^{\circ}}$ and
$\mathbf{{U}}_{k:m^{\star}}$ for cell-edge and cell-interior users
respectively are determined by an operation to find the null space:

\begin{equation}\label{cellular-model-3-eq1-receiver-Utilde}
\begin{aligned}
&\mathbf{\tilde{U}}_{k:m^{\circ}}^{\dag}\left[\ \underbrace{\cdots\
\mathbf{\Psi}_{k:m^{\circ}}^{\dag}\mathbf{{H}}_k^{k:m^{\circ}}\mathbf{V}_{k:n}\
\cdots}_{n\in\mathcal{M}\backslash\{m^{\circ}\}}\
\right]=\mathbf{0}\\
&\mathbf{{U}}_{k:m^{\star}}^{\dag}\left[\ \underbrace{\cdots\
\mathbf{{H}}_k^{k:m^{\star}}\mathbf{V}_{k:n}\
\cdots}_{n\in\mathcal{M}\backslash\{m^{\star}\}}\ \right]=\mathbf{0}
\end{aligned}
\end{equation}

To be feasible to determine $\mathbf{{U}}_{k:m^{\star}}$, the
dimensions should satisfy $N_r^{\star}\geq Md$.

In summary, the four steps determine intermediate receive filters,
precoders, second receive filters in sequence. Alignment is
implicitly contained in step 3. Compared with model 1 and model 2,
the usage of antennas is evidently reduced because the loose
connectivity only between adjacent BS and cell-edge MSs.

%

\subsection{Approach on Receiver Side with  Cascaded Coder: II}

This approach is similar to the design of
\cite{mixed-rank-compound-X-IA-MU}. Each cell-interior MS uses a
cascaded receive filter to align ICI from only one adjacent cell,
and deal with ICI and IUI in its own cell. The whole process takes
four steps as following:

Step 1: For each $m^{\circ}$-th ($m^{\circ}\in\mathcal{M^{\circ}}$)
cell-edge MS in $k$-th cell, the receive filter
$\mathbf{U}_{k:m^{\circ}}$ is defined as two cascaded filter
matrices:

\begin{equation}
\begin{aligned}\label{cellular-model-3-eq1-receiver-ref-1}
&\mathbf{U}_{k:m^{\circ}}=\mathbf{G}_{k:m^{\circ}}^{\dag}{\Lambda}_{\mathcal{K}:m^{\circ}}\\
&\mathbf{G}_{k:m^{\circ}}={\mathbf{H}_{k+1}^{k:m^{\circ}}}^{-1}\\
&\text{So that}\ \ \mathbf{G}_{k:m^{\circ}}\in\mathbb{C}^{N_t\times
N_r^{\circ}}
\end{aligned}
\end{equation}

$\mathbf{G}_{k:m^{\circ}}$ is set to make all ICI from the
neighboring cell aligned together. While
${\Lambda}_{\mathcal{K}:m^{\circ}}$ is a common filter set to be the
same for all cells in the network.

To be feasible to determine $\mathbf{G}_{k:m^{\circ}}$ by the
inverse operation, the dimensions should satisfy $N_r^{\circ}\geq
N_t$. While for each $m^{\star}$-th
($m^{\star}\in\mathcal{M^{\star}}$) cell-interior MS in $k$-th cell,
it is only required the dimensions should satisfy $N_r^{\star}\geq
d$. Compared with model 1 and model 2, it is obvious that these
conditions are greatly loosened to save a number of antennas.

Step 2: Since the common receive filter
$\Lambda_{\mathcal{K}:m^{\circ}}\in\mathbb{C}^{N_t\times d}$ is set
to be the same for all cells of $\mathcal{K}$, the output stream of
(\ref{cellular-model-3-eq1b}) is presented as following:

\begin{equation}
\begin{aligned}\label{cellular-model-3-eq1-receiver-ref-2}
&\Lambda_{\mathcal{K}:m^{\circ}}^{\dag}\mathbf{G}_{k:m^{\circ}}\mathbf{y}_{k:m^{\circ}}=\\
&\Lambda_{\mathcal{K}:m^{\circ}}^{\dag}\mathbf{G}_{k:m^{\circ}}\mathbf{H}_{k}^{k:m^{\circ}}\mathbf{V}_{k:m^{\circ}}\mathbf{x}_{k:m^{\circ}}
\\
&\hspace{30mm}+\sum_{n\neq
m^{\circ}}^{n\in\mathcal{M}}\Lambda_{\mathcal{K}:m^{\circ}}^{\dag}\mathbf{G}_{k:m^{\circ}}\mathbf{H}_{k}^{k:m^{\circ}}\mathbf{V}_{k:n}\mathbf{x}_{k:n}\\
&\hspace{30mm}+\sum_{n\in\mathcal{M}}\Lambda_{\mathcal{K}:m^{\circ}}^{\dag}\mathbf{V}_{k+1:n}\mathbf{x}_{k+1:n}
\end{aligned}
\end{equation}

To eliminate ICI and IUI terms in equation
(\ref{cellular-model-3-eq1-receiver-ref-2}) respectively, the
zero-forcing conditions should be set as:

\begin{equation}
\begin{aligned}\label{cellular-model-3-eq1-receiver-ref-2zf}
\Lambda_{\mathcal{K}:m^{\circ}}^{\dag}\mathbf{G}_{k:m^{\circ}}\mathbf{H}_{k}^{k:m^{\circ}}\mathbf{V}_{k:n}&=\mathbf{0}\ \ n\in\mathcal{M}\backslash\{m^{\circ}\}\\
\Lambda_{\mathcal{K}:m}^{\dag}\mathbf{V}_{k+1:n}&=\mathbf{0}\ \
n\in\mathcal{M}
\end{aligned}
\end{equation}

Compared with model 1 and model 2, the condition is loosened a lot.

While for each cell-interior $m^{\star}$-th
($m^{\star}\in\mathcal{M^{\star}}$) MS in $k$-th cell, the receiver
filter $\mathbf{U}_{k:m^{\star}}$ is in a complete form and randomly
chosen.

Step 3: Then observe from the $k$-th BS to choose the precoder
$[\cdots\ \mathbf{V}_{k:m}\ \cdots]$. The zero-forcing conditions in
equation (\ref{cellular-model-3-eq1-receiver-ref-2zf}) are
equivalently presented with new indices in the following
expressions:

\begin{equation}
\begin{aligned}\label{cellular-model-3-eq1-receiver-ref-3}
\mathbf{U}_{k:n}^{\dag}\mathbf{H}_k^{k:n}\mathbf{V}_{k:m}&=\mathbf{0}\
\
n\in\mathcal{M^{\star}}\backslash\{m\},m\in\mathcal{M}\\
\Lambda_{\mathcal{K}:n}^{\dag}\mathbf{G}_{k:n}\mathbf{H}_{k}^{k:n}\mathbf{V}_{k:m}&=\mathbf{0}\ \ n\in\mathcal{M^{\circ}},m\in\mathcal{M^{\star}}\\
\Lambda_{\mathcal{K}:n}^{\dag}\mathbf{G}_{k:n}\mathbf{H}_{k}^{k:n}\mathbf{V}_{k:m}&=\mathbf{0}\ \ n\in\mathcal{M^{\circ}}\backslash\{m\},m\in\mathcal{M^{\circ}}\\
\Lambda_{\mathcal{K}:n}^{\dag}\mathbf{V}_{k:m}&=\mathbf{0}\ \
n\in\mathcal{M}^{\circ},m\in\mathcal{M}
\end{aligned}
\end{equation}

The last condition in equation
(\ref{cellular-model-3-eq1-receiver-ref-3}) is due to the ICI to the
adjacent cell. Compared with model 1 and model 2, notice that the
conditions are different for cell-edge users and cell-interior
users.

Step 4: So for each $k$-th BS, the precoder should be determined by
only one inverse operation:

\begin{equation}
\begin{aligned}\label{cellular-model-3-eq1-receiver-ref-4}
&[\underbrace{\cdots\mathbf{V}_{k:m^{\star}}\cdots}_{m^{\star}\in\mathcal{M^{\star}}}
\underbrace{\cdots\mathbf{V}_{k:m^{\circ}}\cdots}_{m^{\circ}\in\mathcal{M^{\circ}}}]=\text{The
first}\ Md\ \text{columns of}\\
&\hspace{25mm}\left[\begin{aligned}\left.
\begin{aligned}
&\ \ \ \vdots\\
&\mathbf{U}_{k:m^{\star}}^{\dag}\mathbf{H}_k^{k:m^{\star}}\\
&\ \ \ \vdots
\end{aligned}
\right\}{m^{\star}\in\mathcal{M^{\star}}}\\\left.
\begin{aligned}
&\ \ \ \ \ \ \ \ \ \ \vdots\\
&\Lambda_{\mathcal{K}:m^{\circ}}^{\dag}\mathbf{G}_{k:m^{\circ}}\mathbf{H}_{k}^{k:m^{\circ}}\\
&\ \ \ \ \ \ \ \ \ \ \vdots
\end{aligned}
\right\}{m^{\circ}\in\mathcal{M^{\circ}}}\\\left.
\begin{aligned}
&\vdots\\
&\Lambda_{\mathcal{K}:n}^{\dag}\\
&\vdots
\end{aligned}
\right\}{n\in\mathcal{M}^{\circ}}\end{aligned}
\right]^{-1}\\
\end{aligned}
\end{equation}

Check that
$[\underbrace{\cdots\mathbf{V}_{k:m^{\star}}\cdots}_{m^{\star}\in\mathcal{M^{\star}}}
\underbrace{\cdots\mathbf{V}_{k:m^{\circ}}\cdots}_{m^{\circ}\in\mathcal{M^{\circ}}}]$
indeed satisfies the conditions in
(\ref{cellular-model-3-eq1-receiver-ref-3}).

\begin{Remark}
To be feasible to obtain
$[\underbrace{\cdots\mathbf{V}_{k:m^{\star}}\cdots}_{m^{\star}\in\mathcal{M^{\star}}}
\underbrace{\cdots\mathbf{V}_{k:m^{\circ}}\cdots}_{m^{\circ}\in\mathcal{M^{\circ}}}]$
by an inverse operation, the dimensions should satisfy $N_t\geq
Md+M^{\circ}d$. While in general cases, it is required that $N_t\geq
Md+Md$ for the BS to handle interferences to the adjacent cell. So
this process naturally contains an \textit{implicit} and
\textit{inherent} alignment.
\end{Remark}

In summary, the four steps determine receive filters, precoders and
second receive filters in sequence. Alignment is explicitly
presented in step 1. Compared with model 1 and model 2, less
antennas are required to implement alignment because of the loose
connectivity.

\subsection{Approach from Receiver Side with Direct Interplay}

This approach is similar to the design of
\cite{design-IA-two-cell-IBC}. The alignment is jointly implemented
on both BS(transmitter) side and MS(receiver) side via intricate and
meticulous interplays and cooperations. The whole process takes two
steps as following:

Step 1: Regarding channels from the $(k+1)$-th BS to all cell-edge
MSs in the $k$-th cell, the equivalent receive filtered channel
subspaces are set to be equal to a random predetermined reference
subspace:

\begin{equation}
\begin{aligned}\label{cellular-model-3-eq1-joint-1}
\cdots=\mathbf{U}_{k:m^{\circ}}^{\dag}\mathbf{H}_{k+1}^{k:m^{\circ}}=\cdots=(\mathbf{\Omega}_{k+1}^{k})^{\dag}\ \ \ \ \ \ m^{\circ}\in\mathcal{M^{\circ}}\\
\hspace{0mm}
\end{aligned}
\end{equation}

\begin{Remark}
Notice that the step of (\ref{cellular-model-3-eq1-joint-1}) is an
\textit{explicit} alignment of ICI from the $(k+1)$-th BS to all
cell-edge MSs in the adjacent cell.
\end{Remark}

Observe from all cell-edge MSs in the $k$-th cell, (index changed)
to satisfy the condition (\ref{cellular-model-3-eq1-joint-1}),
$\mathbf{U}_{k:m^{\circ}}$ and $\mathbf{\Omega}_{k+1}^{k}$ could be
found by solving the following null space problem:

\begin{equation}\label{cellular-model-3-eq1-joint-2}
\begin{aligned}
&\left[{(\mathbf{\Omega}_{k+1}^{k})^{\dag}}\ \ \cdots\
\mathbf{U}_{k:m^{\circ}}^{\dag}\ \cdots\
\right]\bullet\\
&\ \ \left[
\begin{aligned}
&\begin{aligned}
&\ \mathbf{I}_{N_t} \ \ \cdots \ \ \mathbf{I}_{N_t}        \end{aligned}\\
&\ \ddots\ \ \ \ \\
&\ \ \ \ -\mathbf{H}_{k+1}^{k:m^{\circ}} \\
&\ \ \ \ \ \ \ \ \ \ \ \ \ \ \ddots
\end{aligned} \right]=\mathbf{0}
\end{aligned}
\end{equation}

To be feasible to obtain $\mathbf{U}_{k:m^{\circ}}$ and
$\mathbf{\Omega}_{k+1}^{k}$ via a null space operation, the
dimensions should satisfy $N_t+M^{\circ}N_r^{\circ}\geq MN_t+d$.
While for cell-interior MSs, it is only required the dimensions
satisfy $N_r^{\star}\geq d$. Compared with model 1 and model 2, the
conditions are greatly loosened to save a number of antennas.

Step 2: Thus for the $k$-th BS, the precoders are determined by the
following inverse operation to eliminate both IUI and ICI:

\begin{equation}
\begin{aligned}\label{cellular-model-3-eq1-joint-3}
&[\underbrace{\cdots\mathbf{V}_{k:m^{\star}}\cdots}_{m^{\star}\in\mathcal{M^{\star}}}
\underbrace{\cdots\mathbf{V}_{k:m^{\circ}}\cdots}_{m^{\circ}\in\mathcal{M^{\circ}}}]=\text{The
first}\ Md\ \text{columns of}\\
&\hspace{25mm}\left[\begin{aligned}&\left.
\begin{aligned}
&\ \ \ \vdots\\
&\mathbf{U}_{k:m^{\star}}^{\dag}\mathbf{H}_k^{k:m^{\star}}\\
&\ \ \ \vdots
\end{aligned}
\right\}{m\in\mathcal{M^{\star}}}\\&\left.
\begin{aligned}
&\ \ \ \ \ \ \ \ \ \ \vdots\\
&\mathbf{U}_{k:m^{\circ}}^{\dag}\mathbf{H}_{k}^{k:m^{\circ}}\\
&\ \ \ \ \ \ \ \ \ \ \vdots
\end{aligned}
\right\}{m^{\circ}\in\mathcal{M^{\circ}}}\\
&\ \ \ \ \ \ \ \ {(\mathbf{\Omega}_{k}^{k-1})}^{\dag}\\
\end{aligned}
\right]^{-1}\\
\end{aligned}
\end{equation}

To be feasible to obtain
$[\underbrace{\cdots\mathbf{V}_{k:m^{\star}}\cdots}_{m^{\star}\in\mathcal{M^{\star}}}
\underbrace{\cdots\mathbf{V}_{k:m^{\circ}}\cdots}_{m^{\circ}\in\mathcal{M^{\circ}}}]$
via the inverse operation, the dimensions should satisfy $N_t\geq
Md+d$. Compared with model 1 and model 2, the condition is loosened.

Check the equation (\ref{cellular-model-3-eq1-joint-3}) and find
that
$[\underbrace{\cdots\mathbf{V}_{k:m^{\star}}\cdots}_{m^{\star}\in\mathcal{M^{\star}}}
\underbrace{\cdots\mathbf{V}_{k:m^{\circ}}\cdots}_{m^{\circ}\in\mathcal{M^{\circ}}}]$
indeed eliminate the IUI in the cell and ICI to the adjacent cell.

In summary, the two steps determine receive filters, aligned
subspace and precoders in sequence. Alignment is explicitly and
implicitly implemented on the MS side and BS side. Compared with
model 1 and model 2, the requirement for antennas is greatly reduced
because of the loose connectivity.


\subsection{Approach from Transmitter Side with Direct Interplay}

This approach is proposed as a dual solution of the above approach
on receiver side. It implements the alignment on the BS side. The
whole process takes the following two steps:

Step 1: Regarding channels from the $(k+1)$-th BS to all cell-edge
MSs in $k$-th cell, the equivalent precoded channel subspaces are
set to be equal to a random predetermined reference subspace:

\begin{equation}
\begin{aligned}\label{cellular-model-3-eq1-jointII-1}
&\mathbf{H}_{k+1}^{k:m^{\circ}}[\mathbf{V}_{k+1:1}\cdots\mathbf{V}_{k+1:M}]=\mathbf{\Theta}^{k:m^{\circ}}\
\ \ \ \ m^{\circ}\in\mathcal{M^{\circ}}
\end{aligned}
\end{equation}

\begin{Remark}
Notice the step of (\ref{cellular-model-3-eq1-jointII-1}) is
\textit{not} an alignment.
\end{Remark}

Observe from the side of all BSs, (index changed) to satisfy the
condition (\ref{cellular-model-3-eq1-jointII-1}),
$[\mathbf{V}_{k:1}\cdots\mathbf{V}_{k:M}]$ and
$\mathbf{\Theta}^{k:m^{\circ}}$ could be found by solving the
following null space problem:

\begin{equation}
\begin{aligned}\label{cellular-model-3-eq1-jointII-2}
\left[\begin{aligned}
&\left.\begin{aligned}&\begin{aligned}&\mathbf{I}_{N_r^{\circ}}\end{aligned}
\hspace{18mm}{\begin{aligned}&-\mathbf{H}_{2}^{1:m^{\circ}}\end{aligned}}\\
&\ \ \ \ddots\hspace{25mm}\ddots\end{aligned}\right\}m^{\circ}\in\mathcal{M^{\circ}}\\
&\left.\begin{aligned}&\ \ \ \ \ \
\begin{aligned}&\mathbf{I}_{N_r^{\circ}}\end{aligned}
\hspace{11mm}{\begin{aligned}\hspace{0mm}&-\mathbf{H}_{k+1}^{k:m^{\circ}}\end{aligned}}\\
&\ \ \ \ \ \ \ \ \ \ddots\hspace{20mm}\ddots\end{aligned}\right\}m^{\circ}\in\mathcal{M^{\circ}}\\
&\left.\begin{aligned}&\ \ \ \ \ \ \ \ \ \ \ \
\begin{aligned}&\mathbf{I}_{N_r^{\circ}}\end{aligned}
\hspace{4mm}{\begin{aligned}\hspace{0mm}&-\mathbf{H}_{1}^{K:m^{\circ}}\end{aligned}}\\
&\hspace{19mm}\ddots\hspace{14mm}\ddots\end{aligned}\right\}m^{\circ}\in\mathcal{M^{\circ}}\end{aligned}\right]
\\\bullet\left[\begin{aligned}&\left.\begin{aligned}&\left.\begin{aligned}&\hspace{8mm}\vdots\\&\hspace{7mm}\mathbf{\Theta}^{k:m^{\circ}}\\&\hspace{8mm}\vdots\end{aligned}\right\}k\in\mathcal{K}\\&\hspace{10mm}\vdots\end{aligned}\right\}m^{\circ}\in\mathcal{M^{\circ}}\\&\left.\begin{aligned}&\hspace{10mm}\vdots\\&\left[\mathbf{V}_{k:1}\cdots\mathbf{V}_{k:M}\right]\\&\hspace{10mm}\vdots\end{aligned}\right\}k\in\mathcal{K}\end{aligned}\right]=\mathbf{0}
\end{aligned}
\end{equation}

To be feasible to obtain $[\mathbf{V}_{k:1}\cdots\mathbf{V}_{k:M}]$
and $\mathbf{\Theta}^{k:m^{\circ}}$ via a null space operation, the
dimensions should follow $KM^{\circ}N_r^{\circ}+KN_t\geq
KM^{\circ}N_r^{\circ}+Md$. Obviously this actual condition $KN_t\geq
Md$ is a very loose contraint because the cell-edge MSs have enough
space of receive antennas. So for the BS, it is required the
dimensions satisfy the basic constraint $N_t\geq Md$.

Step 2: Thus for the $m$-th MS (cell-edge and cell-interior
respectively) in $k$-th cell, its receive filter is determined via a
null space operation to eliminate both IUI and ICI:

\begin{equation}
\begin{aligned}\label{cellular-model-3-eq1-jointII-3}
\mathbf{U}_{k:m^{\circ}}^{\dag}\left[\
\mathbf{H}_{k}^{k:m^{\circ}}\big[\underbrace{\cdots\
\mathbf{V}_{k:n}\
\cdots}_{n\in\mathcal{M}\backslash\{m^{\circ}\}}\big] \ \
{\mathbf{\Theta}^{k:m^{\circ}}}\right]=\mathbf{0}\ \
\ m^{\circ}\in\mathcal{M^{\circ}}\\
\mathbf{U}_{k:m^{\star}}^{\dag}\
\mathbf{H}_{k}^{k:m^{\star}}\big[\underbrace{\cdots\
\mathbf{V}_{k:n}\
\cdots}_{n\in\mathcal{M}\backslash\{m^{\star}\}}\big]=\mathbf{0}\ \
\ m^{\star}\in\mathcal{M^{\star}}
\end{aligned}
\end{equation}

To be feasible to obtain $\mathbf{U}_{k:m^{\circ}}$ for the
cell-edge MS, the dimension should follow $N_r^{\circ}\geq
(M-1)d+Md+d$. For the cell-interior MS to obtain
$\mathbf{U}_{k:m^{\star}}$, the dimension should follow
$N_r^{\star}\geq (M-1)d+d$. It is easy to check the conditions of
(\ref{cellular-model-3-eq1-jointII-3}) directly eliminates both IUI
and ICI.

In summary, the two steps determine precoders, aligned subspaces,
and receive filters in sequence. Alignment is not necessary.
Compared with model 1 and model 2, the requirement for antennas is
only reduced for cell-interior MSs.

\subsection{Comparison of System Conditions}


Similar to the analyses for model 1 and model 2, system conditions
are listed and compared for all the proposed five approaches of
downlink interference alignment in a Wyner-type cellular network
with only one adjacent link for each cell. Three aspects are
investigated including the usage of antennas, required CSI, and
computational complexity. Moreover, in this case, the design
deliberately considers distinguishing between cell-interior and
cell-edge users corresponding to the realistic settings of the
network \cite{multi-cell-downlink-capacity-coordinated-processing},
and actually it benefits the obtained DoF and saves part of the CSI
overhead.


\subsubsection{DoF and Antenna Range}

Similar to the analyses for model 1 and model 2, the number of
antennas determine the achievable DoF for all the nodes. The aim of
this work is to exploit alignment with limited antennas in the
specified Wyner-type network. For all the five approaches, all the
antenna configurations are listed in Table
\ref{cellular-model3-system-condition}, in which the notations of
'A', 'B', 'C', 'D' and 'E' denote the five approaches in sequence.

\begin{table}
\centering
\begin{tabular}{|c|c|c|}
  \hline
  Approach & Minimum BS Antenna. & $\begin{aligned}&\text{Minimum MS Antenna}\\&\ \ \ \ \ \ (N_r^{\star},N_r^{\circ})\end{aligned}$  \\
  \hline
  A & $Md$ & $(d,Md+d)$  \\
  \hline
  B & $(M^{\circ}+1)Md$ & $(Md,Md)$  \\
  \hline
  C & $Md+M^{\circ}d$ & $(d,Md+M^{\circ}d)$  \\
  \hline
  D & $(M+1)d$ & $(d,\begin{aligned}({M^2}/{M^{\circ}})d\end{aligned})$  \\
  \hline
  E & $Md$ & $(Md,2Md)$  \\
  \hline
\end{tabular}\caption{Minimum Antenna Usage for BSs and MSs in All Approaches in Model 3}\label{cellular-model3-system-condition}
\end{table}

In summary, as shown in Table
\ref{cellular-model3-system-condition}, approach 'B' needs a large
number of antennas for BS nodes. Compared with model 2 in Table
\ref{cellular-model2-system-condition}, the antenna requirement is
greatly reduced. Cell-interior MSs generally need less antennas than
cell-edge MSs.

\subsubsection{CSI Overhead}

Similar to the analyses in model 1 and model 2, the proposed
alignment approaches require all the BS nodes and MS nodes to have
different knowledge of channel state information (CSI). The aim of
this work is to compare the CSI requirements of different
approaches. For the five approaches, all the CSI configurations are
listed in Table \ref{cellular-model3-system-condition-2}. The
notations from 'A' to 'E' denote the five approaches in sequence.
Other settings are the same as model 1 and model 2.

\begin{table}[htbp]\scriptsize
\centering
\begin{tabular}{|c|c|c|c|}
\hline
\multirow{2}{*}{Approach} & \multicolumn{3}{|c|}{$\begin{aligned}&\text{required CSI at $k$-th BS from Interior and Edge MSs}\\&\hspace{20mm}(m^{\star},m^{\circ})\end{aligned}$} \\
\cline{2-4}
& Source & Content & Quantity\\
\hline
A & all intra-cell MSs & $\mathbf{U}_{k:m}^{\dag}\mathbf{H}_k^{k:m}$ & $M$ \\
\hline
B & adjacent inter-cell MSs & $(/,\mathbf{\Psi}_{k-1:n}^{\dag}\mathbf{H}_k^{k-1:n})$ & $M^{\circ}$ \\
\hline
C & all intra-cell MSs & $\begin{aligned}&(\mathbf{U}_{k:m^{\star}}^{\dag}\mathbf{H}_{k}^{k:m^{\star}},\\&\mathbf{G}_{k:m^{\circ}}\mathbf{H}_{k}^{k:m^{\circ}})\end{aligned}$ & $(M^{\star},M^{\circ})$ \\
\hline
D & $\begin{aligned}\text{all intra-cell MSs}\\\text{adjacent inter-cell MSs}\end{aligned}$ & $\begin{aligned}&(\mathbf{U}_{k:m^{\star}}^{\dag}\mathbf{H}_k^{k:m^{\star}},\\&\mathbf{U}_{k:m^{\circ}}^{\dag}\mathbf{H}_k^{k:m^{\circ}}),\mathbf{\Omega}_{k}^{k-1}\end{aligned}$ & $\begin{aligned}&(M^{\star},\\&M^{\circ}),1\end{aligned}$ \\
\hline
E & adjacent inter-cell MSs & $(/,\mathbf{H}_k^{k-1:m^{\circ}})$ & $\begin{aligned}M^{\circ}(\text{excl.} \\\text{backhaul})\end{aligned}$ \\
\hline \hline
\multirow{2}{*}{Approach} & \multicolumn{3}{|c|}{$\begin{aligned}&\text{required CSI at $m$-th MS in $k$-th cell}\\&\hspace{15mm}(m^{\star},m^{\circ})\end{aligned}$} \\
\cline{2-4}
& Source & Content & Quantity\\
\hline
A & adjacent inter-cell BS & $(/,\mathbf{H}_{K+1}^{k:m^{\circ}}\mathbf{\Phi}_{K+1})$ & $1$ \\
\hline
B & intra-cell BS & $\begin{aligned}&(\mathbf{H}_k^{k:m^{\star}}\mathbf{V}_{k:n},\\&\mathbf{\Psi}_{k:m^{\circ}}^{\dag}\mathbf{H}_k^{k:m^{\circ}}\mathbf{V}_{k:n})\end{aligned}$ & $\begin{aligned}(M-1,\\M-1)\end{aligned}$ \\
\hline
C & adjacent inter-cell BS & $(/,\mathbf{H}_{k+1}^{k:m^{\circ}})$ & $1$ \\
\hline
D & adjacent inter-cell BS & $\mathbf{H}_{k+1}^{k:m^{\circ}}$ & $\begin{aligned}&1(\text{excl.} \\&\text{conferencing})\end{aligned}$ \\
\hline
E & intra-cell BS & $\begin{aligned}&(\mathbf{H}_k^{k:m^{\star}}\mathbf{V}_{k:n},\\&\mathbf{H}_k^{k:m^{\circ}}\mathbf{V}_{k:n}),\mathbf{\Theta}^{k:m^{\circ}}\end{aligned}$ & $\begin{aligned}&(M-1,\\&M-1),1\end{aligned}$ \\
\hline
\end{tabular}\caption{CSI for BSs and MSs in All
Approaches in Model
3}\label{cellular-model3-system-condition-2}\normalsize
\end{table}

In summary, as shown in Table
\ref{cellular-model3-system-condition-2}, approach 'B' and 'E' need
a large amount of CSI. Compared with model 2 in Table
\ref{cellular-model2-system-condition-2}, the CSI requirement is
lowered in general. Each BS require less CSI from cell-interior MSs
than from cell-edge MSs. While cell-interior MSs require less CSI
than cell-edge MSs do.

\subsubsection{Computational Complexity}

Similar to the analyses for model 1 and model 2, all the nodes need
to calculate coding matrices or filters with obtained channel
knowledge. The calculations apply different operations and are
universally measured by the metric of computational complexity. The
aim of this work is to compare the computational complexities for
the proposed approaches. For all the five approaches, all the
complexity configurations are listed in Table
\ref{cellular-model3-system-condition-3}. The notations from 'A' to
'E' denote the five approaches in sequence. Other settings are the
same as model 1 and model 2.

\begin{table}[htbp]
\centering
\begin{tabular}{|c|c|c|c|}
\hline
\multirow{2}{*}{Approach} & \multicolumn{3}{|c|}{expended Complexity at $k$-th BS } \\
\cline{2-4}
& Target & Operation & Scale \\
\hline
A & $\mathbf{\tilde V}_{k:m}$ & matrix inverse & $Md\times Md$ \\
\hline
B & $\mathbf{V}_{k:m}$ & null space & $N_t\times Md$ \\
\hline
C & $\mathbf{V}_{k:m}$ & matrix inverse & $(M+M^{\circ})d\times N_t$ \\
\hline
D & $\mathbf{V}_{k:m}$ & matrix inverse & $(M+1)d\times N_t$ \\
\hline
E & $\mathbf{V}_{k:m},\mathbf{\Theta}^{k:m^{\circ}}$ & null space & $\begin{aligned}&(KM^{\circ}N_r^{\circ}+KN_t)\\&\hspace{10mm}\times Md\end{aligned}$ \\
\hline \hline
\multirow{2}{*}{Approach} & \multicolumn{3}{|c|}{$\begin{aligned}&\text{expended Complexity at $m$-th MS in $k$-th cell}\\&\hspace{15mm}(m^{\star},m^{\circ})\end{aligned}$} \\
\cline{2-4}
& Target & Operation & Scale \\
\hline
A & $(/,\mathbf{U}_{k:m^{\circ}})$ & null space & $N_r^{\circ}\times d$ \\
\hline
B & $(\mathbf{U}_{k:m^{\star}},\mathbf{\tilde U}_{k:m^{\circ}})$ & null space & $(N_r^{\star}\times d,N_r^{\circ}\times d)$ \\
\hline
C & $(/,\mathbf{G}_{k:m^{\circ}})$ & matrix inverse & $N_r^0\times N_t$ \\
\hline
D & $(/,(\mathbf{U}_{k:m^{\circ}},\mathbf{\Omega}_j^k))$ & null space & $(N_t+M^{\circ}N_r^{\circ})\times d$ \\
\hline
E & $(\mathbf{U}_{k:m^{\star}},\mathbf{U}_{k:m^{\circ}})$ & null space & $(N_r^{\star}\times d,N_r^{\circ}\times d)$ \\
\hline
\end{tabular}\caption{Complexity for BSs and MSs in All
Approaches in Model 3}\label{cellular-model3-system-condition-3}
\end{table}

In summary, as shown in Table
\ref{cellular-model3-system-condition-3}, approach 'D' and 'E' need
a large amount of computational complexities. Compared with model 2
in Table \ref{cellular-model2-system-condition-3}, the expended
complexities are reduced. Cell-interior MSs require less
complexities than cell-edge MSs.


\section{Advanced Design and Analysis of Pragmatic Models}

%

The previous sections 5.3, 5.4 and 5.5 discuss comprehensive and
basic approaches for the three models. However, they are not as
perfect solutions as real cellular networks desire. These basic
solutions need to be further specialized and enhanced.

This section crystalizes the designs and analyses in previous
investigations and looks into pragmatic models which are fit to
current cellular systems. Advanced schemes are proposed to obtain
better performance for these typical models in the discussion.

\subsection{Motivations, Models, and Strategies}


In previous sections, for model 1, model 2, and model 3, elaborate
designs and analyses are proposed, which are non-trivial extensions
from two-cell scenarios and schemes as in
\cite{downlink-IA,mixed-rank-compound-X-IA-MU,design-IA-two-cell-IBC}
to general multi-cell scenarios and schemes. However, new issues
come out in different angles. First, all the approaches applied for
each model mainly implement alignment on either the BS side or on
the MS side. If the alignment happens on both sides within certain
antenna range, there is potential advantage for the achievable DoF
while the implementation is not explicit. Second, all the approaches
only consider alignment between IUI and ICI as the conventional
design, while if there are more than two cells, the alignment
between ICI and ICI from different interfering cells are potentially
exploitable. Third, all the approaches require cooperations of BSs
or MSs in all cells with a large amount of unrealistic global CSI.
Actually it could be improved in realistic settings by introducing a
sequential process to construct alignment in a more local way.

Proper models are needed in order to be fit to the general framework
of cellular networks. Model 1, model 2 and model 3 are
comprehensively studied in previous sections. While in the following
part, only model 2 and model 3 are considered. The reasons are as
following. Model 1 is a full connected $K$-cell network. When $K$ is
large or the network is large, the connectivity is not realistic due
to path loss. Furthermore this model is not tractable to analyze or
design because it needs many antennas and CSI exchange. Model 2 is a
cyclic Wyner network with two adjacent links for each cell. It
represents a usual scenario of cellular networks, which has multiple
cells and appropriate interfering links between adjacent cells.
Model 3 is a cyclic Wyner network with one adjacent link for each
cell and only cell-edge users are interfered. This model is simple
but typical. In the following discussion for model 2 and model 3,
all cell indices from 1 to $K$ are in the set $\mathcal{K}$ in a
circular manner, i.e. modulo operations are applied to all cell
indices in $\mathcal{K}$.


Appropriate strategies are necessary for advanced schemes. From
previous investigations, system conditions of usage of antennas, CSI
and complexity are shown in Table
\ref{cellular-model2-system-condition},
\ref{cellular-model2-system-condition-2},
\ref{cellular-model2-system-condition-3}. There three factors need a
balance in the advanced design. In the proposed five approaches, the
first approach is a proper basis. It applies cascaded precoders on
the BS side. It is proved to be compatible to practical cellular
systems to eliminate ICI. While conventionally the cascaded
precoders are randomly chosen as in \cite{downlink-IA}, it could be
further explored in advanced schemes. Besides, the roles of BSs and
MSs need a balance. Usually BSs have more capability and resources,
for example, CSI exchange is more practical via BS backhaul
mechanism than via MS conferencing mechanism. However in the
broadcast channels to MSs the scheme becomes cumbersome if too many
MSs are activated. In summary, the system needs to find a tradeoff
with all the above concerns.


%
%
%
%
%

\subsection{Advanced Approach for Model 2}

Model 2 is discussed here. The approach is similar to the design of
\cite{downlink-IA} with cascaded precoders on the BS side. However,
the alignment of ICI could be implemented through different novel
options. The advanced process are illustrated as following:

Step 1: Choose an auxiliary intermediate precoder matrix
$\mathbf{\Phi}_{k}\in\mathbb{C}^{N_t\times (M\cdot d)}$ at $k$-th BS
(transmitter), and a second precoder $\mathbf{\tilde
V}_{k:m}\in\mathbb{C}^{(M\cdot d)\times d}$ preceding the first
chosen precoder. Then the actual precoder $\mathbf{V}_{k:m}$ is
composed of:

\begin{equation}
\begin{aligned}\label{cellular-model-2-ADV-eq1-transmit-V}
\mathbf{V}_{k:m}=\mathbf{\Phi}_{k}\mathbf{\tilde{V}}_{k:m} \ \ \
m\in\mathcal{M}, k\in\mathcal{K}
\end{aligned}
\end{equation}

(Notice the index refers to all equal nodes)

Step 2: When the first precoder $\mathbf{\Phi}_{j}$ is randomly
picked, to enable zero-forcing the ICI from the two adjacent cells,
all $M$ users in $k$-th cell should determine $\mathbf{U}_{k:m}$ to
satisfy the following conditions (index changed)

\begin{equation}
\begin{aligned}\label{cellular-model-2-ADV-eq1-transmit-ZF-u}
\mathbf{U}_{k:m}^{\dag}[\
\mathbf{{H}}_{k-1}^{k:m}\mathbf{\Phi}_{k-1}\
\mathbf{{H}}_{k+1}^{k:m}\mathbf{\Phi}_{k+1}\ ]=\mathbf{0} \ \ \
m\in\mathcal{M}
\end{aligned}
\end{equation}

In previous basic schemes, to be feasible to find
$\mathbf{U}_{k:m}$, the dimensions should satisfy $N_r\geq 2Md+d$.
This requirement is loosened compared with model 1 because the
connectivity of the network is lowered. However, even based on the
same model of model 2, the antennas could be also further saved with
a novel strategy, i.e. as long as the following new condition is
satisfied:

\begin{equation}\label{cellular-model-2-ADV-step2a-1}
\mathrm{span}[\mathbf{{H}}_{k-1}^{k:m}\mathbf{\Phi}_{k-1}]=\mathrm{span}[\mathbf{{H}}_{k+1}^{k:m}\mathbf{\Phi}_{k+1}]
\ \ \ m\in\mathcal{M}
\end{equation}

When the new condition holds, it is only required that the
dimensions satisfy $N_r\geq Md+d$, so that antennas could be saved.
Observe equation (\ref{cellular-model-2-ADV-step2a-1}) and important
features are concluded in the following remark:

\begin{Remark}
The purpose and essence of condition
(\ref{cellular-model-2-ADV-step2a-1}) is to implement an extra
alignment in advance on the BS side in addition to the following
alignment on the MS side. This alignment is between ICI and ICI from
different cells. In principle, this extra condition is natural and
reasonable, because $\mathbf{\Phi}_k$ is randomly selected in the
conventional scheme as in \cite{downlink-IA} while in fact the
potential could be exploited by selecting appropriate values.
\end{Remark}

Observe the structure of the condition
(\ref{cellular-model-2-ADV-step2a-1}) in the whole cyclic network,
it is easy to decompose it into two groups of equations:

\begin{equation}\label{cellular-model-2-ADV-step2a-2}
\begin{aligned}
&\mathrm{span}[\mathbf{{H}}_{i}^{i+1:m}\mathbf{\Phi}_{i}]=\mathrm{span}[\mathbf{{H}}_{i+2}^{i+1:m}\mathbf{\Phi}_{i+2}]
\ \ \ \\&\hspace{10mm}i\in\mathcal{K}^{ev}\triangleq\left\{\begin{aligned}&\{2,4,6,\ldots,K\}\ \ K\text{ is even}\\&\{2,4,6,\ldots,K-1\}\ \ \ \ K\text{ is odd}\end{aligned}\right.\\
&\mathrm{span}[\mathbf{{H}}_{j}^{j+1:m}\mathbf{\Phi}_{j}]=\mathrm{span}[\mathbf{{H}}_{j+2}^{j+1:m}\mathbf{\Phi}_{j+2}]
\ \ \
\\&\hspace{10mm}j\in\mathcal{K}^{od}\triangleq\left\{\begin{aligned}&\{1,3,5,\ldots,K-1\}\ \ K\text{ is even}\\&\{1,3,5,\ldots,K\}\ \ K\text{ is odd}\end{aligned}\right.
\end{aligned}
\end{equation}

It means all the conditions in the cells of even numbers are
connected and so are the the cells of odd numbers. Since the indices
in equation (\ref{cellular-model-2-ADV-step2a-2}) are actually
operated in a cyclic mode, e.g. $K\equiv 0$, to explicitly specify
the boundary connections between cells at the end of the network,
the following equations are highlighted:

\begin{equation}\label{cellular-model-2-ADV-step2a-2-bound}
\begin{aligned}
&\hspace{0mm}\text{For }i\in\mathcal{K}^{ev},\\
&\ \ \ \left\{\begin{aligned}&\mathrm{span}[\mathbf{{H}}_{K}^{1:m}\mathbf{\Phi}_{K}]=\mathrm{span}[\mathbf{{H}}_{2}^{1:m}\mathbf{\Phi}_{2}]\ \ \ \ \ \ \ \ K\text{ is even}\\&\mathrm{span}[\mathbf{{H}}_{K-1}^{K:m}\mathbf{\Phi}_{K-1}]=\mathrm{span}[\mathbf{{H}}_{1}^{K:m}\mathbf{\Phi}_{1}]\ \ \ \ K\text{ is odd}\end{aligned}\right.\\
&\hspace{0mm}\text{For }j\in\mathcal{K}^{od},\\
&\ \ \
\left\{\begin{aligned}&\mathrm{span}[\mathbf{{H}}_{K-1}^{K:m}\mathbf{\Phi}_{K-1}]=\mathrm{span}[\mathbf{{H}}_{1}^{K:m}\mathbf{\Phi}_{1}]\
\ \ K\text{ is
even}\\&\mathrm{span}[\mathbf{{H}}_{K}^{1:m}\mathbf{\Phi}_{K}]=\mathrm{span}[\mathbf{{H}}_{2}^{1:m}\mathbf{\Phi}_{2}]\
\ \ \ \ \ \ \ \ K\text{ is odd}\end{aligned}\right.
\end{aligned}
\end{equation}

In order to satisfy conditions (\ref{cellular-model-2-ADV-step2a-2})
and (\ref{cellular-model-2-ADV-step2a-2-bound}), in the following
discussion, five options of designs are proposed and the system
conditions are analyzed respectively. For step 2, the five options
are marked as: 2a, 2b, 2c, 2d, 2e. Correspondingly, for the
subsequent step 3, the five options are marked as: 3a, 3b, 3c, 3d,
3e. For convenience, the following steps are presented in pairs,
e.g. 2a combined with 3a.

\subsubsection{Step 2a,3a}
{Step 2a}: A naive and direct implementation of condition
(\ref{cellular-model-2-ADV-step2a-1}) is as following:

\begin{equation}\label{cellular-model-2-ADV-step2a-1-naive}
\mathbf{{H}}_{k-1}^{k:m}\mathbf{\Phi}_{k-1}=\mathbf{{H}}_{k+1}^{k:m}\mathbf{\Phi}_{k+1}
\ \ \ m\in\mathcal{M}
\end{equation}

To simplify the discussion, just assume $K$ is even here. So that
correspondingly, the conditions of
(\ref{cellular-model-2-ADV-step2a-2}) and
(\ref{cellular-model-2-ADV-step2a-2-bound}) are implemented as the
following equations of $\mathbf{\Phi}_k$:

\begin{equation}\label{cellular-model-2-ADV-step2a-3}
\begin{aligned}
\left[\begin{aligned}&\
\vdots\hspace{9mm}\vdots\\&\mathbf{{H}}_{2}^{3:m}\
-\mathbf{{H}}_{4}^{3:m}\\&\
\vdots\hspace{9mm}\vdots\\&\hspace{11mm}\vdots\hspace{9mm}\vdots\\&\hspace{10mm}\mathbf{{H}}_{4}^{5:m}\
-\mathbf{{H}}_{6}^{5:m}\\&\hspace{11mm}\vdots\hspace{9mm}\vdots\\&\hspace{23mm}\ddots\\&\
\vdots\hspace{27mm}\vdots\\&-\mathbf{{H}}_{2}^{1:m}\hspace{15mm}\mathbf{{H}}_{K}^{1:m}\
\\&\ \vdots\hspace{27mm}\vdots\end{aligned}\right]
\left[\begin{aligned}&\mathbf{\Phi}_{2}\\
&\mathbf{\Phi}_{4}\\&\ \
\vdots\\&\mathbf{\Phi}_K\end{aligned}\right]=\mathbf{0}\\
\left[\begin{aligned}&\
\vdots\hspace{9mm}\vdots\\&\mathbf{{H}}_{1}^{2:m}\
-\mathbf{{H}}_{3}^{2:m}\\&\
\vdots\hspace{9mm}\vdots\\&\hspace{11mm}\vdots\hspace{9mm}\vdots\\&\hspace{10mm}\mathbf{{H}}_{3}^{4:m}\
-\mathbf{{H}}_{5}^{4:m}\\&\hspace{11mm}\vdots\hspace{9mm}\vdots\\&\hspace{23mm}\ddots\\&\
\vdots\hspace{27mm}\vdots\\&-\mathbf{{H}}_{1}^{K:m}\hspace{15mm}\mathbf{{H}}_{K-1}^{K:m}\
\\&\ \vdots\hspace{27mm}\vdots\end{aligned}\right]
\left[\begin{aligned}&\mathbf{\Phi}_{1}\\
&\mathbf{\Phi}_{3}\\&\ \
\vdots\\&\mathbf{\Phi}_{K-1}\end{aligned}\right]=\mathbf{0}\\
\end{aligned}
\end{equation}

To be feasible to obtain $\mathbf{\Phi}_k$, the dimensions should
satisfy $N_tK/2\geq MN_rK/2+Md$, i.e. $N_t\geq MN_r+2Md/K$. When
$K\gg M$, it becomes $N_t\geq MN_r$.

Step 3a: Now observe each $\mathbf{\Phi}_{k}$ again, notice there
are 2$M$ zero-forcing conditions to satisfy just as the same as in
previous section in (\ref{cellular-model-2-eq1-transmit-ZF-phi}).
Normally it is feasible to find $\mathbf{\Phi}_{k}$ only when
$N_t\geq2Md+Md$. However, if considering
(\ref{cellular-model-2-ADV-eq1-transmit-ZF-u}) already guarantees
the condition, it is only required that $N_t\geq Md$ in this
process. If further considering
(\ref{cellular-model-2-ADV-step2a-3}), as just mentioned, it is
required $N_t\geq MN_r$. Notice the mentioned condition $N_r\geq
Md+d$. So in summary, when $M\geq 2$, there is no \textit{implicit}
alignment on the MS side; only when $M=1$, there is
\textit{implicit} alignment on the MS side.


\subsubsection{Step 2b,3b}
$\ $

Step 2b: Notice the above step of
(\ref{cellular-model-2-ADV-step2a-3}) has a severe drawback. It
needs global CSI and mass calculations which are not reasonable for
practical systems. To overcome this drawback, the following process
is proposed to implement alignment in a successive manner. According
to the condition (\ref{cellular-model-2-ADV-step2a-2}), every
$\mathbf{\Phi}_k$ is obtained one by one.

First, two arbitrary cells are set as the initial cells, say $i=2$,
$j=1$ for the cells of even numbers and odd numbers respectively.
Choose random intermediate precoders $\mathbf{\Phi}_i$ and
$\mathbf{\Phi}_j$ for both of them. Second, determine receive
filters $\mathbf{U}_{i+1:m}$ and $\mathbf{U}_{j+1:m}$ to satisfy the
conditions (\ref{cellular-model-2-ADV-eq1-transmit-ZF-u}), i.e.
$\mathbf{U}_{i+1:m}^{\dag}\mathbf{H}_{i}^{i+1:m}\mathbf{\Phi}_i=\mathbf{0}$,
$\mathbf{U}_{j+1:m}^{\dag}\mathbf{H}_{j}^{j+1:m}\mathbf{\Phi}_j=\mathbf{0}$.
It's only required that $N_r\geq Md+d$ because the alignment
condition (\ref{cellular-model-2-ADV-step2a-1-naive}) guarantees
zero-forcing by setting $\mathbf{\Phi}_{i+2}$ and
$\mathbf{\Phi}_{j+2}$ in the following cells. Third, according to
(\ref{cellular-model-2-ADV-step2a-1-naive}), $\mathbf{\Phi}_{i+2}$
and $\mathbf{\Phi}_{j+2}$ could be obtained by (pseudo) inverse
operations:

\begin{equation}\label{cellular-model-2-ADV-step2a-4}
\begin{aligned}
\mathbf{\Phi}_{i+2}=\left[\left.\begin{aligned}&\ \ \ \
\vdots\\&\mathbf{{H}}_{i+2}^{i+1:m}\\&\ \ \ \
\vdots\end{aligned}\right\}m\in\mathcal{M}\right]^{-1}\left[\left.\begin{aligned}&\
\ \ \ \vdots\\&\mathbf{{H}}_{i}^{i+1:m}\\&\ \ \ \
\vdots\end{aligned}\right\}m\in\mathcal{M}\right]\mathbf{\Phi}_{i}\\
\mathbf{\Phi}_{j+2}=\left[\left.\begin{aligned}&\ \ \ \
\vdots\\&\mathbf{{H}}_{j+2}^{j+1:m}\\&\ \ \ \
\vdots\end{aligned}\right\}m\in\mathcal{M}\right]^{-1}\left[\left.\begin{aligned}&\
\ \ \ \vdots\\&\mathbf{{H}}_{j}^{j+1:m}\\&\ \ \ \
\vdots\end{aligned}\right\}m\in\mathcal{M}\right]\mathbf{\Phi}_{j}
\end{aligned}
\end{equation}

To be feasible to obtain $\mathbf{\Phi}_{i+2}$ and
$\mathbf{\Phi}_{j+2}$ by inverse operations, the dimensions should
satisfy $N_t\geq MN_r$. Finally, after completing the whole
procedure for all cells, all the other intermediate precoders
$\mathbf{\Phi}_{i+4}$, $\mathbf{\Phi}_{j+4}$, $\cdots$,
$\mathbf{\Phi}_{K}$, $\mathbf{\Phi}_{1}$, $\cdots$,
$\mathbf{\Phi}_{i-2}$, $\mathbf{\Phi}_{j-2}$ are obtained. However,
notice that the boundary condition
(\ref{cellular-model-2-ADV-step2a-2-bound}) is not satisfied. This
issue could be ignored considering two reasons. The network may be
not circular. The loss of DoF is relatively small to be neglected
when $K$ is large.

Step 3b: Now observe each $\mathbf{\Phi}_{k}$ again, notice there
are 2$M$ zero-forcing conditions to satisfy just as the same as in
previous section in (\ref{cellular-model-2-eq1-transmit-ZF-phi}).
Normally it is feasible to find $\mathbf{\Phi}_{k}$ only when
$N_t\geq2Md+Md$. However, if further considering
(\ref{cellular-model-2-ADV-step2a-4}), as just mentioned, it is
required $N_t\geq MN_r$. The requirement is the same as the above
step 3a.

\subsubsection{Step 2c,3c}
$\ $

Step 2c: Notice the step of (\ref{cellular-model-2-ADV-step2a-4})
has a drawback. It needs a large number of antennas, i.e. $N_t$ is
large. To overcome this drawback, the following process is proposed.
Compared with Step 2b, $\mathbf{\Phi}_k$ is also obtained in a
successive manner, but it utilizes another operation and involves
the MS side as well.

First, two arbitrary cells are set as initial cells, say $i=2$,
$j=1$ for the cells of even numbers and odd numbers respectively.
Choose random intermediate precoders $\mathbf{\Phi}_i$ and
$\mathbf{\Phi}_j$. Second, determine receive filters
$\mathbf{U}_{i+1:m}$ and $\mathbf{U}_{j+1:m}$ to satisfy half of the
condition (\ref{cellular-model-2-ADV-eq1-transmit-ZF-u}), i.e.
$\mathbf{U}_{i+1:m}^{\dag}\mathbf{H}_{i}^{i+1:m}\mathbf{\Phi}_i=\mathbf{0}$,
$\mathbf{U}_{j+1:m}^{\dag}\mathbf{H}_{j}^{j+1:m}\mathbf{\Phi}_j=\mathbf{0}$.
It's only required that $N_r\geq Md+d$ as mentioned. Third,
according to the other half of the condition
(\ref{cellular-model-2-ADV-eq1-transmit-ZF-u}),
$\mathbf{\Phi}_{i+2}$ and $\mathbf{\Phi}_{j+2}$ could be obtained by
alternative null space operations:

\begin{equation}\label{cellular-model-2-ADV-step2a-5}
\begin{aligned}
&\mathbf{\Phi}_{i+2}=\mathrm{null}(\left[\left.\begin{aligned}&\hspace{10mm}\vdots\\&\mathbf{U}_{i+1:m}^{\dag}\mathbf{{H}}_{i+2}^{i+1:m}\\&\hspace{10mm}\vdots\end{aligned}\right\}m\in\mathcal{M}\right])\\
&\mathbf{\Phi}_{j+2}=\mathrm{null}(\left[\left.\begin{aligned}&\hspace{10mm}\vdots\\&\mathbf{U}_{j+1:m}^{\dag}\mathbf{{H}}_{j+2}^{j+1:m}\\&\hspace{10mm}\vdots\end{aligned}\right\}m\in\mathcal{M}\right])
\end{aligned}
\end{equation}

\begin{Remark}To be feasible to obtain $\mathbf{\Phi}_{i+2}$ and $\mathbf{\Phi}_{j+2}$
 by null space operations, the dimensions should satisfy $N_t\geq Md+Md$.
  So that BS antennas only require $N_t\geq 2Md$, which is
enormously reduced compared with Step 2a and Step 2b. Thanks to the
successive implicit alignment involving precoders, receive filters
in a cooperative way. This successive manner obtains BS precoder
$\mathbf{\Phi}_{k}$ and MS receive filter $\mathbf{U}_{k:m}$
alternatively like 'bounces' between two sides, saving both huge
antenna usage and global CSI exchange resulting in a chain
connecting all the coders.
\end{Remark}

Finally, after completing the whole procedure for all cells, all the
other intermediate precoders $\mathbf{\Phi}_{i+4}$,
$\mathbf{\Phi}_{j+4}$, $\cdots$, $\mathbf{\Phi}_{K}$,
$\mathbf{\Phi}_{1}$, $\cdots$, $\mathbf{\Phi}_{i-2}$,
$\mathbf{\Phi}_{j-2}$ are obtained. As mentioned in Step 2b, the
boundary condition (\ref{cellular-model-2-ADV-step2a-2-bound}) could
be ignored.

Step 3c: Now observe each $\mathbf{\Phi}_{k}$ again, notice there
are 2$M$ zero-forcing conditions to satisfy just as in previous
section in (\ref{cellular-model-2-eq1-transmit-ZF-phi}). Normally,
it is required $N_t\geq 2Md+Md$. However, further consider the
condition (\ref{cellular-model-2-ADV-eq1-transmit-ZF-u}) to obtain
$\mathbf{U}_{k:m}$ and the operation
(\ref{cellular-model-2-ADV-step2a-5}) to obtain $\mathbf{\Phi}_{k}$,
the alignment is successfully implemented on both BS side and MS
side by zero-forcing:

\begin{equation}\label{cellular-model-2-ADV-step2c-align}
\begin{aligned}
&(\mathbf{U}_{k-1:m}^{\dag}\mathbf{H}_{k}^{k-1:m})\mathbf{\Phi}_k=(\mathbf{U}_{k+1:m}^{\dag}\mathbf{H}_{k}^{k+1:m})\mathbf{\Phi}_k\ \ \ m\in\mathcal{M}\\
&\mathbf{U}_{k:m}^{\dag}(\mathbf{H}_{k-1}^{k:m}\mathbf{\Phi}_{k-1})=\mathbf{U}_{k:m}^{\dag}(\mathbf{H}_{k+1}^{k:m}\mathbf{\Phi}_{k+1})\
\ \ m\in\mathcal{M}
\end{aligned}
\end{equation}

In summary, for BS antennas, it is only required $N_t\geq 2Md$
instead of previous requirement $N_t\geq 3Md$; for MS antennas, it
is only required $N_r\geq Md+d$ instead of previous requirement
$N_r\geq 2Md+d$. The antenna usage is greatly saved to obtain more
DoF.

\subsubsection{Step 2d,3d}
$\ $

Step 2d: Since Step 2b needs a large number of antennas and Step 2c
needs successive CSI exchange between BS side and MS side as well as
a lot of computations, the following process is proposed to further
reduce the usage of antennas and to focus on the BS side. It applies
a robust measure to implement the alignment in a flexible and
compatible way. The new process deals with the condition of
(\ref{cellular-model-2-ADV-step2a-2}) by utilizing a metric called
\textit{chordal distance}.

For each BS, the precoder $\mathbf{\Phi}_k$ is set to be only
selected from a finite predefined set named as $\mathfrak{B}_k$.
Then the whole procedure is similar to Step 2b. First, two initial
cells $2$ and $1$ are set for even numbers and odd numbers
respectively. Choose $\mathbf{\Phi}_i$ and $\mathbf{\Phi}_j$ for
them. Second, determine receive filters $\mathbf{U}_{i+1:m}$ and
$\mathbf{U}_{j+1:m}$. Third, obtain $\mathbf{\Phi}_{i+2}$ and
$\mathbf{\Phi}_{j+2}$. Instead of the inverse operations in
(\ref{cellular-model-2-ADV-step2a-4}), the new operations are
approximations by selecting the most aligned ICI with the metric of
chordal distance:

\begin{equation}\label{cellular-model-2-ADV-step2d-align}
\begin{aligned}
&\mathbf{\Phi}_{i+2}=\arg\min_{\mathbf{\Phi}_{i+2}\in\mathfrak{B}_{i+2}}\sum_{m=1}^MD_c^2(\mathbf{H}_i^{i+1:m}\mathbf{\Phi}_i,\mathbf{H}_{i+2}^{i+1:m}\mathbf{\Phi}_{i+2})\\
&\mathbf{\Phi}_{j+2}=\arg\min_{\mathbf{\Phi}_{j+2}\in\mathfrak{B}_{j+2}}\sum_{m=1}^MD_c^2(\mathbf{H}_j^{j+1:m}\mathbf{\Phi}_j,\mathbf{H}_{j+2}^{j+1:m}\mathbf{\Phi}_{j+2})
\end{aligned}
\end{equation}

In equation (\ref{cellular-model-2-ADV-step2d-align}),
$D_c(\cdot,\cdot)$ denotes the {chordal distance} used as the metric
of alignment as in
\cite{IA-oppo-selection-3-user-IFC,linear-precoder-K-user}.
Mathematically, the chordal distance between two $N_r\times Md$
matrices $\mathbf{P}$ and $\mathbf{Q}$ is defined as:

%



\begin{equation}
\begin{aligned}\label{cellular-chordal-distance}
&D^2_c(\mathbf{P},\mathbf{Q})=D^2_c(\mathbb{O}(\mathbf{P}),\mathbb{O}(\mathbf{Q}))\\
&=\frac{1}{2}\|\mathbb{O}(\mathbf{P})\mathbb{O}(\mathbf{P})^{\dag}-\mathbb{O}(\mathbf{Q})\mathbb{O}(\mathbf{Q})^{\dag}\|_F\\
&=Md-\mathrm{Trace}(\mathbb{O}(\mathbf{P})^{\dag}\mathbb{O}(\mathbf{Q})\mathbb{O}(\mathbf{Q})^{\dag}\mathbb{O}(\mathbf{P}))
\end{aligned}
\end{equation}

where $\mathbb{O}(\mathbf{P})$ and $\mathbb{O}(\mathbf{Q})$ are
orthonormal bases of $\mathbf{P}$ and $\mathbf{Q}$ respectively.


To be feasible to obtain $\mathbf{\Phi}_{i+2}$ and
$\mathbf{\Phi}_{j+2}$ by selections of chordal distances, it is only
required the dimensions should satisfy $N_t\geq Md$ rather than
$N_t\geq MN_r$ in Step 2b or $N_t\geq 2Md$ in Step 2c. In principle,
the condition (\ref{cellular-model-2-ADV-step2a-1-naive})
$\mathbf{{H}}_{k-1}^{k:m}\mathbf{\Phi}_{k-1}=\mathbf{{H}}_{k+1}^{k:m}\mathbf{\Phi}_{k+1}$
is too tight for Step 2b, and the condition
(\ref{cellular-model-2-ADV-step2d-align}) actually relaxes it to
$\mathrm{span}[\mathbf{{H}}_{k-1}^{k:m}\mathbf{\Phi}_{k-1}]=\mathrm{span}[\mathbf{{H}}_{k+1}^{k:m}\mathbf{\Phi}_{k+1}]$.
However, when $N_t<2Md$, the alignment could not be perfect and
there is still residue interference for sure which degrades the
performance severely. Anyway, it provides an accessible way to
approach alignment under the situation with few antennas.

Step 3d: Now observe each $\mathbf{\Phi}_{k}$ again, notice there
are 2$M$ zero-forcing conditions to satisfy just as the same as in
previous section in (\ref{cellular-model-2-eq1-transmit-ZF-phi}).
The conditions are partially satisfied by
(\ref{cellular-model-2-ADV-step2d-align}).
 Other requirements are the same as Step 3b and 3c.

In summary, Step 2d applies the precoder selection method. It has a
few advantages. Antenna requirement is significantly lowered; CSI is
reduced by only exchanging precoder indices; computational
complexity to satisfy condition
(\ref{cellular-model-2-ADV-step2a-2}) is saved a lot; the system
complicacy is simplified. It also has a few disadvantages. The
residue inter-cell interference (ICI) definitely causes rate and DoF
loss, although the loss is not severe if considering large path loss
of ICI. While conventional opportunistic selection method only
selects one best MS in each cell with other MSs inactive
\cite{IA-oppo-selection-3-user-IFC,transmit-BF-inter-cell-IA-selection-CoMP}.
When there are enough number of MSs, the performance of the selected
MS is is always good because of the selection diversity. In this
work the process enable all the MSs in all cells to be active. The
selection is proceeded on the precoder matrices at BSs to approach
alignment. Actually it takes a tradeoff between the usage of
antennas and the system performance. Compared with pure analytical
designs, the system conditions are simplified. Although it is
unlikely to be optimal, the effort of this process is to ensure acceptable performance. 

\subsubsection{Step 2e,3e}
$\ $

Step 2e: Since Step 2c applies a different successive approach
between BS side and MS side and still uses a lot of antennas, the
following process is proposed to improve them by applying a robust
measure to implement the alignment of condition
(\ref{cellular-model-2-ADV-eq1-transmit-ZF-u}) in a flexible and
compatible way. The new process utilizes a metric called
\textit{interference leakage}.


The whole procedure is similar to Step 2c. First, two initial cells
$2$ and $1$ are set for even numbers and odd numbers respectively.
Choose $\mathbf{\Phi}_i$ and $\mathbf{\Phi}_j$ for them. Second,
determine receive filters $\mathbf{U}_{i+1:m}$ and
$\mathbf{U}_{j+1:m}$. Third, obtain $\mathbf{\Phi}_{i+2}$ and
$\mathbf{\Phi}_{j+2}$. Instead of the null space operations in
(\ref{cellular-model-2-ADV-step2a-5}), the new operations are
approximations by selecting the minimized ICI with the metric of
 interference leakage:

\begin{equation}\label{cellular-model-2-ADV-step2e-align}
\begin{aligned}
&\mathbf{\Phi}_{i+2}=\arg\min_{\mathbf{\Phi}_{i+2}^{\dag}\mathbf{\Phi}_{i+2}=\mathbf{I}_{Md}}\sum_{m=1}^M{L_{IF}}(\mathbf{U}_{i+1:m},\mathbf{H}_{i+2}^{i+1:m}\mathbf{\Phi}_{i+2})\\
&\mathbf{\Phi}_{j+2}=\arg\min_{\mathbf{\Phi}_{j+2}^{\dag}\mathbf{\Phi}_{j+2}=\mathbf{I}_{Md}}\sum_{m=1}^M{L_{IF}}(\mathbf{U}_{j+1:m},\mathbf{H}_{j+2}^{j+1:m}\mathbf{\Phi}_{j+2})
\end{aligned}
\end{equation}

In equation (\ref{cellular-model-2-ADV-step2e-align}),
${L_{IF}}(\cdot,\cdot)$ denotes {interference leakage} as the metric
of alignment as in \cite{IA-oppo-selection-3-user-IFC,approaching-capacity-IA}.
Mathematically, interference leakage on the receive matrix
$\mathbf{P}$ from the interference source matrix $\mathbf{Q}$ is
defined as:

\begin{equation}
\begin{aligned}\label{cellular-interference-leakage}
&L_{IF}(\mathbf{P},\mathbf{Q})=\mathrm{Trace}((\mathbf{P}^{\dag}\mathbf{Q})(\mathbf{P}^{\dag}\mathbf{Q})^{\dag})
\end{aligned}
\end{equation}

While the solution to the minimization problem
(\ref{cellular-model-2-ADV-step2e-align}) is given by:

\begin{equation}\label{cellular-model-2-ADV-step2e-align-solution}
\begin{aligned}
&\mathbf{\Phi}_{i+2}=\nu_{Md}\Big\{\sum_{m=1}^M(\mathbf{H}_{i+2}^{i+1:m})^{\dag}\mathbf{U}_{i+1:m}\mathbf{U}_{i+1:m}^{\dag}\mathbf{H}_{i+2}^{i+1:m}\Big\}\\
&\mathbf{\Phi}_{j+2}=\nu_{Md}\Big\{\sum_{m=1}^M(\mathbf{H}_{j+2}^{j+1:m})^{\dag}\mathbf{U}_{j+1:m}\mathbf{U}_{j+1:m}^{\dag}\mathbf{H}_{j+2}^{j+1:m}\Big\}
\end{aligned}
\end{equation}

where $\nu_{Md}\{\cdot\}$ is composed of $Md$ eigenvectors
corresponding to the smallest $Md$ eigenvalues of the matrix.



To be feasible to obtain $\mathbf{\Phi}_{i+2}$ and
$\mathbf{\Phi}_{j+2}$ by leakage minimization, it is only required
the dimensions should satisfy $N_t\geq Md$ rather than $N_t\geq 2Md$
in Step 2c. Because the condition
(\ref{cellular-model-2-ADV-step2a-5}) actually guarantees the
interference leakage is absolutely zero, which is a tight
constraint. However, the robust measure allows an imperfect
alignment and residue interference leakage as well as corresponding
DoF loss in the practice.

Step 3e: Now observe each $\mathbf{\Phi}_{k}$ again, the
zero-forcing conditions in
(\ref{cellular-model-2-eq1-transmit-ZF-phi}) are partially
satisfied. Other requirements are the same as Step 3c.


In summary, this robust measure of interference leakage minimization
is adaptable to a wider range of antenna configurations. The
performance of rate and DoF are controllable and amendable regarding
the usage of antennas. When there are enough antennas, perfect
alignment could be achieved. When there are less antennas, partial
alignment with residue interference is obtained.

$\ $

Step 4: Following Step 2 and Step 3 (with options: a,b,c,d,e
respectively), then each $k$-th BS knows its equivalent downlink
channel $\mathbf{U}_{k:m}\mathbf{{H}}_k^{k:m}\mathbf{\Phi}_{k}$ from
each intra-cell MS. It forms zero-forcing transmit beams which could
eliminate IUI between the MSs, while it does not need to know the
actual interfering beams. The second precoder matrix is obtained by
the following inverse operation:

\begin{equation}\label{cellular-model-2-ADV-eq1-transmit-Vtilde}
\begin{aligned}
&\left[\mathbf{\tilde{V}}_{k:1}\ \mathbf{\tilde{V}}_{k:2}\ \cdots\
\mathbf{\tilde{V}}_{k:M}\right]= \left[
\begin{aligned}
&\mathbf{U}_{k:1}^{\dag}\mathbf{{H}}_k^{k:1}\mathbf{\Phi}_{k}\\&\mathbf{U}_{k:2}^{\dag}\mathbf{{H}}_k^{k:2}\mathbf{\Phi}_{k}\\&\hspace{10mm}\vdots\\&\mathbf{U}_{k:M}^{\dag}\mathbf{{H}}_k^{k:M}\mathbf{\Phi}_{k}
\end{aligned}
\right]^{-1}
\end{aligned}
\end{equation}

\subsection{Advanced Approach for Model 3}

Model 3 is discussed here. In this Wyner-type network, there is only
one adjacent link for each cell. So that a new approach is
specifically proposed to fit this model. This advanced process takes
four steps as following:

Step 1: Choose a random receive filter $\mathbf{U}_{k:m^{\star}}$
for each $m^{\star}$-th MS among all $M^{\star}$ cell-interior MSs
in the $k$-th cell. Also choose a random receive filter
$\mathbf{U}_{k:m^{\circ}}$ for each $m^{\circ}$-th MS among all
$M^{\circ}$ cell-edge MSs in the $k$-th cell. To be feasible to
obtain $\mathbf{U}_{k:m^{\star}}$ and $\mathbf{U}_{k:m^{\circ}}$ for
available transmissions, it is only required the dimensions should
satisfy $N_r^{\star}\geq d$ and $N_r^{\circ}\geq d$.


Step 2: Because of the special structure of model 3, it is not
necessary to apply intermediate precoder $\mathbf{\Phi}_{k}$ at
$k$-th BS as in \cite{downlink-IA}. So that the precoders need to
satisfy all the zero-forcing conditions directly. Each
$\mathbf{V}_{k:m}$ for the $m$-th user in the $k$-the cell has
$(M^{\circ}+M-1)$ zero-forcing conditions as following:

\begin{equation}
\begin{aligned}\label{cellular-model-3-ADV-eq1-transmit-ZF-phi}
\left[
\begin{aligned}
&\left.
\begin{aligned}
&\hspace{10mm}\vdots\\
&\mathbf{U}_{k-1:m^{\circ}}^{\dag}\mathbf{{H}}_k^{k-1:m^{\circ}}\\
&\hspace{10mm}\vdots \\
\end{aligned}
\right\} m^{\circ}\in\mathcal{M}^{\circ}\\
&\left.
\begin{aligned}
&\hspace{10mm}\vdots\\
&\mathbf{U}_{k:n}^{\dag}\mathbf{{H}}_k^{k:n}\\
&\hspace{10mm}\vdots \\
\end{aligned}
\right\} n\in\mathcal{M}\backslash\{m\}
\end{aligned}
\right]\mathbf{V}_{k:m}=\mathbf{0} \ \ \
\end{aligned}
\end{equation}

The condition (\ref{cellular-model-3-ADV-eq1-transmit-ZF-phi})
represents that the ICI from the $k$-th BS to cell-edge MSs in the
$(k-1)$-th cell is zero and the IUI between the $m$-th MS and other
MSs in the $k$-th cell is zero.

Step 3: Then each $k$-th BS knows its equivalent downlink channel
$\mathbf{U}_{k-1:m^{\circ}}\mathbf{{H}}_k^{k-1:m^{\circ}}$ from MSs
in the $(k-1)$-th cell and $\mathbf{U}_{k:m}\mathbf{{H}}_k^{k:m}$
from its own intra-cell MSs. It forms zero-forcing transmit beams to
eliminate both ICI and IUI together as following:

\begin{equation}\label{cellular-model-3-ADV-eq1-transmit-Vtilde}
\begin{aligned}
&[\mathbf{V}_{k:1}\ \mathbf{V}_{k:2}\ \cdots\ \mathbf{V}_{k:M}]= \text{The last }Md\text{ columns of} \\
&\left[
\begin{aligned}
&\left.
\begin{aligned}
&\hspace{10mm}\vdots\\
&\mathbf{U}_{k-1:m^{\circ}}^{\dag}\mathbf{{H}}_k^{k-1:m^{\circ}}\\
&\hspace{10mm}\vdots \\
\end{aligned}
\right\} m^{\circ}\in\mathcal{M}^{\circ}\\
&\left.
\begin{aligned}
&\hspace{10mm}\vdots\\
&\mathbf{U}_{k:m}^{\dag}\mathbf{{H}}_k^{k:m}\\
&\hspace{10mm}\vdots \\
\end{aligned}
\right\} m\in\mathcal{M}
\end{aligned}
\right]^{-1}
\end{aligned}
\end{equation}

Check condition (\ref{cellular-model-3-ADV-eq1-transmit-ZF-phi}) and
condition (\ref{cellular-model-3-ADV-eq1-transmit-Vtilde}). Both of
them require the dimensions should satisfy $N_t\geq M^{\circ}d+Md$.

Step 4: Observe the receive filters $\mathbf{U}_{k:m^{\circ}}$ and
$\mathbf{U}_{k:m^{\star}}$ again, notice there are $(2M-1)$ and
$(M-1)$ zero-forcing conditions to satisfy respectively.

\begin{equation}
\begin{aligned}\label{cellular-model-3-ADV-eq1-transmit-ZF-u}
&\mathbf{U}_{k:m^{\circ}}^{\dag}[\underbrace{\cdots\mathbf{{H}}_{k}^{k:m^{\circ}}\mathbf{V}_{k:n}\cdots}_{n\in\mathcal{M}\backslash\{m^{\circ}\}}\
\underbrace{\cdots\mathbf{{H}}_{k+1}^{k:m^{\circ}}\mathbf{V}_{k+1:m}\cdots}_{m\in\mathcal{M}}]=\mathbf{0}\\
&\mathbf{U}_{k:m^{\star}}^{\dag}[\underbrace{\cdots\mathbf{{H}}_{k}^{k:m^{\star}}\mathbf{V}_{k:n}\cdots}_{n\in\mathcal{M}\backslash\{m^{\star}\}}]=\mathbf{0}
\end{aligned}
\end{equation}

\begin{Remark}
Check the condition (\ref{cellular-model-3-ADV-eq1-transmit-ZF-u}).
Normally, it is feasible to find $\mathbf{U}_{k:m^{\circ}}$ and
$\mathbf{U}_{k:m^{\star}}$ only when the dimensions satisfy
$N_r^{\circ}\geq (M-1)d+Md+d$ and $N_r^{\star}\geq (M-1)d+d$
respectively. However, since
(\ref{cellular-model-3-ADV-eq1-transmit-ZF-phi}) and
(\ref{cellular-model-3-ADV-eq1-transmit-Vtilde}) already guarantee
this condition, it is only required that $N_r^{\circ}\geq d$ and
$N_r^{\star}\geq d$ in this specific design process. So this process
naturally contains an \textit{implicit} and \textit{inherent}
alignment.
\end{Remark}

In summary, the four steps determine receive filters and precoders
in one time because of the connectivity is loose in this network.
Alignment is implemented on the receiver side.

\subsection{Comparison of System Conditions}

Similar to the basic designs and analyses for model 1, model 2, and
model 3, system conditions are listed and compared here for the
advanced approaches for model 2 and model 3. Three aspects are
investigated including the usage of antennas, CSI exchange, and
computational complexity.


\subsubsection{DoF and Antenna Range}

For the advanced approach of model 2, there are five options of Step
2 and Step 3. All the antenna configurations for these options are
listed in Table \ref{cellular-model2-ADV-system-condition}
respectively. The notations 'a','b','c','d','e' denote them in
sequence. Compared with the antenna configurations in Table
\ref{cellular-model2-system-condition} for the basic design of model
2, the antenna usage is greatly reduced.

\begin{table}
\centering
\begin{tabular}{|c|c|c|}
  \hline
  Option & Minimum BS Antenna. & Minimum MS Antenna  \\
  \hline
  a & $MN_r+2Md/K$ & $Md+d$  \\
  \hline
  b & $MN_r$ & $Md+d$  \\
  \hline
  c & $2Md$ & $Md+d$  \\
  \hline
  d & $Md$ & $Md+d$\\
  \hline
  e & $Md$ & $Md+d$  \\
  \hline
\end{tabular}\caption{Minimum Antenna Usage for BSs and MSs in the Advanced Approach for Model 2}\label{cellular-model2-ADV-system-condition}
\end{table}

For the advanced approach of model 3, the antenna configuration is
listed in Table \ref{cellular-model3-ADV-system-condition}, which is
denoted as 'F' for consistency. Compared with the antenna
configuration in Table \ref{cellular-model3-system-condition} for
the basic design of model 3, the antenna usage is greatly reduced.

\begin{table}
\centering
\begin{tabular}{|c|c|c|}
  \hline
  Approach & Minimum BS Antenna. & $\begin{aligned}&\text{Minimum MS Antenna}\\&\ \ \ \ \ \ (N_r^{\star},N_r^{\circ})\end{aligned}$  \\
  \hline
  F & $M^{\circ}d+Md$ & $(d,d)$  \\
  \hline
\end{tabular}\caption{Minimum Antenna Usage for BSs and MSs in the Advanced Approach for Model 3}\label{cellular-model3-ADV-system-condition}
\end{table}

In summary, the number of antennas plays a key role to implement
alignment and the antennas could have different balance between BS
side and MS side to achieve alignment. The aim of this work is
exploit interference alignment with limited provision of antennas.
Among the five options of model 2, options 'c', 'd', and 'e' use
less antennas for the BS side. Notice options 'd' and 'e' apply
robust measures so that there are residue interferences due to
minimizing the number of antennas. For model 3, BS side uses extra
antennas to deal with interference while MS side uses the minimum
number of antennas.

\subsubsection{CSI Overhead}


For the advanced approach of model 2, there are five options of Step
2 and Step 3. All the CSI configurations for these options are
listed in Table \ref{cellular-model2-ADV-system-condition-2}. The
notations 'a','b','c','d','e' denote them in sequence. Compared with
the CSI overhead in Table \ref{cellular-model2-system-condition-2}
for the basic design of model 2, the CSI requirement is lowered
much.

\begin{table}[htbp]\scriptsize
\centering
\begin{tabular}{|c|c|c|c|}
\hline
\multirow{2}{*}{Option} & \multicolumn{3}{|c|}{$\begin{aligned}&\hspace{6mm}\text{required extra CSI at $k$-th BS}\\&(\text{besides: all intra-cell MSs, }\mathbf{U}_{k:m}^{\dag}\mathbf{H}_k^{k:m},\ M)\end{aligned}$} \\
\cline{2-4}
& Source & Content & Quantity\\
\hline
a & $\begin{aligned}&\text{adjacent inter-cell MSs}\\&\ \ \text{(excl. backhaul)}\end{aligned}$ & $\mathbf{H}_k^{k-1:m},\mathbf{H}_k^{k+1:m}$ & $M,M$ \\
\hline
b & $\begin{aligned}&\ \ \ \ \ \text{adjacent inter-cell MSs} \\ &\text{preceding-adjacent inter-cell MSs}\end{aligned}$ & $\begin{aligned}&\mathbf{H}_{k-2}^{k-1:m}\mathbf{\Phi}_{k-2}\\&\ \ \ \ \mathbf{H}_k^{k-1:m}\end{aligned}$ & $M,M$ \\
\hline
c & adjacent inter-cell MSs & $\mathbf{U}_{k-1:m}^{\dag}\mathbf{H}_{k}^{k-1:m}$ & $M$ \\
\hline
d & $\begin{aligned}&\ \ \ \ \ \text{adjacent inter-cell MSs} \\ &\text{preceding-adjacent inter-cell MSs}\end{aligned}$ & $\begin{aligned}&\mathbf{H}_{k-2}^{k-1:m}\mathbf{\Phi}_{k-2}\\&\ \ \ \ \mathbf{H}_k^{k-1:m}\end{aligned}$ & $M,M$ \\
\hline
e & adjacent inter-cell MSs & $\mathbf{U}_{k-1:m}^{\dag}\mathbf{H}_{k}^{k-1:m}$ & $M$ \\
\hline \hline
\multirow{2}{*}{Option} & \multicolumn{3}{|c|}{required CSI at $m$-th MS in $k$-th cell} \\
\cline{2-4}
& Source & Content & Quantity\\
\hline
a & adjacent inter-cell BS & $\mathbf{H}_{k-1}^{k:m}\mathbf{\Phi}_{k-1}$ & $1$ \\
\hline
b & adjacent inter-cell BS & $\mathbf{H}_{k-1}^{k:m}\mathbf{\Phi}_{k-1}$ & $1$ \\
\hline
c & adjacent inter-cell BS & $\mathbf{H}_{k-1}^{k:m}\mathbf{\Phi}_{k-1}$ & $1$ \\
\hline
d & adjacent inter-cell BS & $\mathbf{H}_{k-1}^{k:m}\mathbf{\Phi}_{k-1}$ & $1$ \\
\hline
e & adjacent inter-cell BS & $\mathbf{H}_{k-1}^{k:m}\mathbf{\Phi}_{k-1}$ & $1$ \\
\hline
\end{tabular}\caption{CSI for BSs and MSs in the Advanced Approach for Model 2}\label{cellular-model2-ADV-system-condition-2}\normalsize
\end{table}

For the advanced approach of model 3, the CSI configuration is
listed in Table \ref{cellular-model3-ADV-system-condition-2}, which
is denoted as 'F' for consistency. Compared with the CSI overhead in
Table \ref{cellular-model3-system-condition-2} for the basic design
of model 3, the CSI requirement is lowered much.

\begin{table}[htbp]\scriptsize
\centering
\begin{tabular}{|c|c|c|c|}
\hline
\multirow{2}{*}{Approach} & \multicolumn{3}{|c|}{$\begin{aligned}&\text{required CSI at $k$-th BS from Interior and Edge MSs}\\&\hspace{20mm}(m^{\star},m^{\circ})\end{aligned}$} \\
\cline{2-4}
& Source & Content & Quantity\\
\hline
F & $\begin{aligned}&\ \ \ \text{all intra-cell MSs}\\&\text{adjacent inter-cell MSs}\end{aligned}$ & $\begin{aligned}&(\mathbf{U}_{k:m^{\star}}^{\dag}\mathbf{H}_k^{k:m^{\star}},\mathbf{U}_{k:m^{\circ}}^{\dag}\mathbf{H}_k^{k:m^{\circ}})\\&\ \ \ \ \ \ (/,\mathbf{U}_{k-1:m^{\circ}}^{\dag}\mathbf{H}_k^{k-1:m^{\circ}})\end{aligned}$ & $\begin{aligned}&(M^{\star},M^{\circ})\\&\ \ (/,M^{\circ})\end{aligned}$ \\
\hline \hline
\multirow{2}{*}{Approach} & \multicolumn{3}{|c|}{$\begin{aligned}&\text{required CSI at $m$-th MS in $k$-th cell}\\&\hspace{15mm}(m^{\star},m^{\circ})\end{aligned}$} \\
\cline{2-4}
& Source & Content & Quantity\\
\hline
F & / & $(/,/)$ & $/$ \\
\hline
\end{tabular}\caption{CSI for BSs and MSs in the Advanced Approach for Model
3}\label{cellular-model3-ADV-system-condition-2}\normalsize
\end{table}

In summary, the proposed interference alignment approaches require
all the BS nodes and MS nodes to have certain knowledge of channel
state information (CSI). The aim of this work is look for the lowest
CSI requirement for the network in different approaches. Among the
five options of model 2, options 'c' and 'e' need less CSI for the
BS side. For model 3, only BS side needs CSI.

\subsubsection{Computational Complexity}



For the advanced approach of model 2, there are five options of Step
2 and Step 3. All the complexity configurations are listed in Table
\ref{cellular-model2-ADV-system-condition-3}. The notations
'a','b','c','d','e' denote them in sequence. Compared with the
computational complexity in Table
\ref{cellular-model2-system-condition-3} for the basic design of
model 2, the expended complexity is reduced a lot.

\begin{table}[htbp]
\centering
\begin{tabular}{|c|c|c|c|}
\hline
\multirow{2}{*}{Option} & \multicolumn{3}{|c|}{$\begin{aligned}&\text{Expended complexity at }k\text{-th BS}\\&(\text{besides: }\mathbf{\tilde V}_{k:m}, \text{matrix inverse},Md\times Md)\end{aligned}$} \\
\cline{2-4}
& Target & Operation & Scale \\
\hline
a & $\mathbf{\Phi}_{k}$ & null space & $\begin{aligned}N_tK/2\times Md\\(\text{for }K/2\text{ BSs})\end{aligned}$ \\
\hline
b & $\mathbf{\Phi}_{k}$ & $\begin{aligned}&\ \ \ \text{matrix inverse}\end{aligned}$ & $N_t\times MN_r$ \\
\hline
c & $\mathbf{\Phi}_{k}$ & null space & $N_t\times Md$ \\
\hline
d & $\mathbf{\Phi}_{k}$ & chordal distance & $N_r\times Md\times |\mathfrak{B}_k|\times M$ \\
\hline
e & $\mathbf{\Phi}_{k}$ & trace & $d\times Md\times |\mathfrak{B}_k|\times M$ \\
\hline \hline
\multirow{2}{*}{Option} & \multicolumn{3}{|c|}{Expended complexity at $m$-th MS in $k$-th cell} \\
\cline{2-4}
& Target & Operation & Scale \\
\hline
a & $\mathbf{U}_{k:m}$ & null space & $N_r\times d$ \\
\hline
b & $\mathbf{U}_{k:m}$ & null space & $N_r\times d$ \\
\hline
c & $\mathbf{U}_{k:m}$ & null space & $N_r\times d$ \\
\hline
d & $\mathbf{U}_{k:m}$ & null space & $N_r\times d$ \\
\hline
e & $\mathbf{U}_{k:m}$ & null space & $N_r\times d$ \\
\hline
\end{tabular}\caption{Complexity for BSs and MSs in the Advanced Approach for Model 2}\label{cellular-model2-ADV-system-condition-3}
\end{table}

For the advanced approach of model 3, all the complexity
configurations are listed in Table
\ref{cellular-model3-ADV-system-condition-3}, which is denoted as
'F' for consistency. Compared with the complexity in Table
\ref{cellular-model3-system-condition-3} for the basic design of
model 3, the expended complexity is reduced a lot.

\begin{table}[htbp]
\centering
\begin{tabular}{|c|c|c|c|}
\hline
\multirow{2}{*}{Approach} & \multicolumn{3}{|c|}{Expended complexity at $k$-th BS } \\
\cline{2-4}
& Target & Operation & Scale \\
\hline
F & $\mathbf{V}_{k:m}$ & matrix inverse & $(M^{\circ}d+Md)\times N_t$ \\
\hline \hline
\multirow{2}{*}{Approach} & \multicolumn{3}{|c|}{$\begin{aligned}&\text{Expended complexity at $m$-th MS in $k$-th cell}\\&\hspace{15mm}(m^{\star},m^{\circ})\end{aligned}$} \\
\cline{2-4}
& Target & Operation & Scale \\
\hline
F & $(/,/)$ & / & / \\
\hline
\end{tabular}\caption{Complexity for BSs and MSs in the Advanced
Approach for Model 3}\label{cellular-model3-ADV-system-condition-3}
\end{table}

In summary, all the nodes need to calculate their coding matrices or
filters with obtained CSI. The aim of this work is to look for the
minimum computational complexities for the network in different
approaches. Among the five options for model 2, option 'c' needs a
small computational complexity. For model 3, only BS side needs
computational complexity.

\subsection{Numerical Results}

%

The previous parts of this work give detailed designs and analyses
in various models with different approaches. System conditions and
performance are listed and compared. The result indicates that some
of them are quite fit to be applicable to current cellular network
frameworks. In the following part, numerical results are illustrated
to evaluate the rates and DoF performance impacted by key factors.
Model 2 is intensively studied and model 3 is also investigated.
Advanced approaches are compared with some basic approaches. An main
focus is on the usage of antennas and achievable DoF.

All the figures show the total achievable rates as the SNR grows so
that DoF is well illustrated. To present it in a intuitive way, the
SNR applies a $\log_2$ operation so that DoF is exactly the slope of
the corresponding curve. For all the discussion, the network is set
to have $K=6$ cells totally. Each cell has $M=3$ MSs and each MS has
$d=2$ streams.

\subsubsection{Model 2 in Different Cases}

\begin{figure}[htpb]
  \begin{center}
    \includegraphics[width=4.5in]{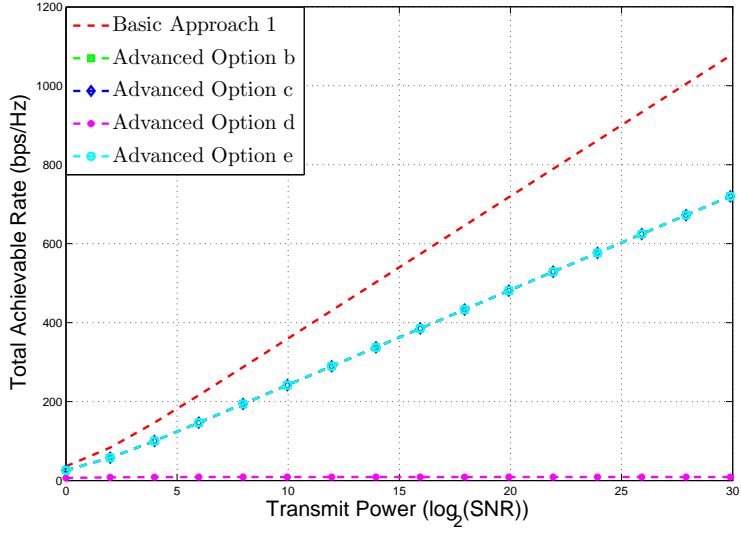}
  \end{center}
  \caption{Rates of Model 2 in Different Cases of Approaches}\label{Cellular-IFBC-IA-Numerical-Results}
\end{figure}

For model 2, numerical results of five cases are compared as shown
in Fig. \ref{Cellular-IFBC-IA-Numerical-Results}. In the first case,
the basic approach 1 is applied as a reference. So that the antennas
are set as $N_t=Md=6$ for each BS and $N_r=2Md+d=14$ antennas for
each MS. In the second case, option 'b' of the advanced approach is
applied. The antennas are set as $N_t=MN_r=24$ for each BS and
$N_r=Md+d=8$ for each MS. In the third case, option 'c' of the
advanced approach is applied. The antennas are set as $N_t=2Md=12$
for each BS and $N_r=Md+d=8$ for each MS. In the fourth case, option
'd' of the advanced approach is applied. The antennas are set as
$N_t=Md=6$ for each BS and $N_r=Md+d=8$ for each MS. In the fifth
case, option 'e' of the advanced approach is applied. The antennas
are set as $N_t=Md=6$ antennas for each BS and $N_r=Md+d=8$ antennas
for each MS.

From Fig. \ref{Cellular-IFBC-IA-Numerical-Results}, observe that the
basic approach and the options 'b', 'c', and 'e' of the advanced
approach obtain effective DoF for the network while the option 'd'
of the advanced approach has zero DoF. The slope of the basic
approach is higher than the advanced approach, because the basic
approach applies a perfect alignment for the whole network
symmetrically while the advanced approach ignores the DoF loss of
the boundary cells in the network as mentioned although the antenna
usage is greatly reduced. The option 'd' of the advanced approach
causes residue interference due to finite precoder selections so
that it is not an absolute alignment in term of DoF.

\subsubsection{Model 2 with Cascaded Precoders regarding Antenna Number and Codebook Size}

\begin{figure}[htpb]
  \begin{center}
    \includegraphics[width=4.5in]{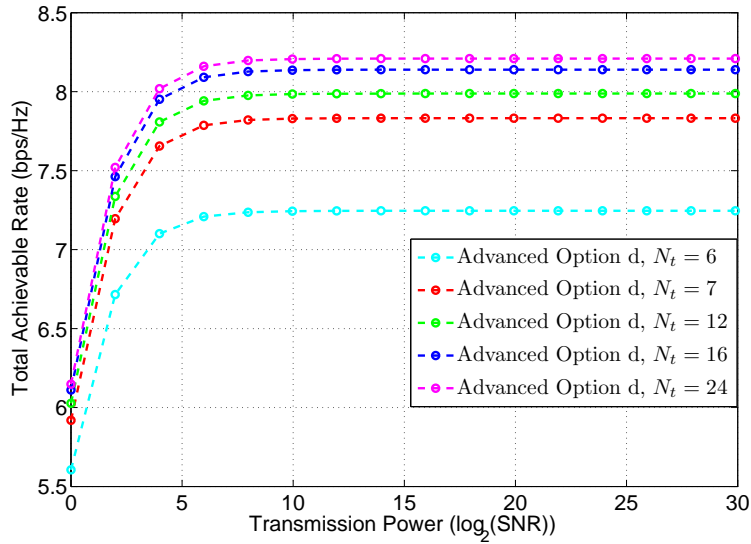}
  \end{center}
  \caption{Rates of Model 2 with Cascaded Precoders regarding Antenna Number}\label{Cellular-IFBC-IA-Numerical-Results-2Nt}
\end{figure}

\begin{figure}[htpb]
  \begin{center}
    \includegraphics[width=4.5in]{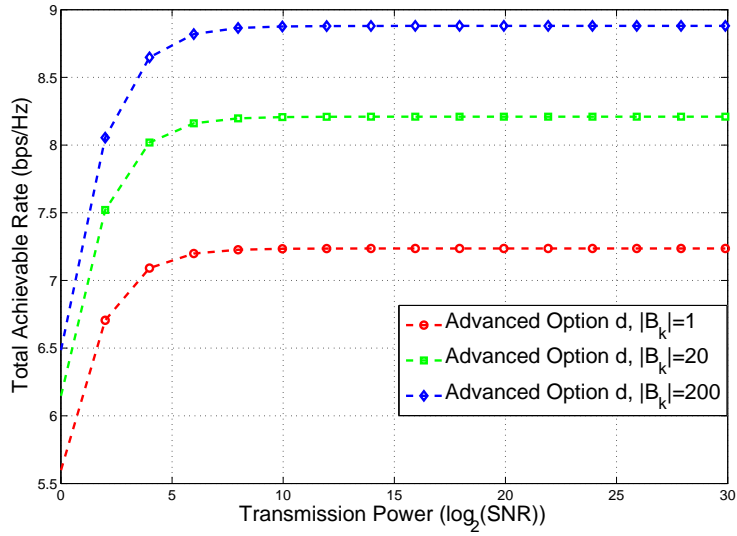}
  \end{center}
  \caption{Rates of Model 2 with Cascaded Precoders regarding Codebook Size}\label{Cellular-IFBC-IA-Numerical-Results-2Bk}
\end{figure}

Although the DoF is zero for the option 'd' of the advanced approach
for model 2 as mentioned above, it is necessary to look into its
rate performance since this options is proposed as a robust measure
to obtain performance under the condition of scarce antennas.
Intuitively, two factors impact the final performance. One is the
number of antennas, the other is the size of codebooks. Numerical
results are compared in Fig.
\ref{Cellular-IFBC-IA-Numerical-Results-2Nt} and Fig.
\ref{Cellular-IFBC-IA-Numerical-Results-2Bk} respectively.

In Fig. \ref{Cellular-IFBC-IA-Numerical-Results-2Nt}, option 'd' of
the advanced approach is applied and different numbers of BS
antennas $N_t$ are compared. The antennas are set as $N_r=Md+d=8$
for each MS and $N_t=6,7,12,16,24$ for each BS. Notice the key
values $N_t=Md=6$, $N_t=2Md=12$ and $N_t=MN_r=24$. Then Fig.
\ref{Cellular-IFBC-IA-Numerical-Results-2Nt} shows five curves
corresponding to different values of $N_t$. From Fig.
\ref{Cellular-IFBC-IA-Numerical-Results-2Nt}, observe that when
there are more extra BS antennas, the residue interference is lower.
So that the rates increase because the alignment is better.

In Fig. \ref{Cellular-IFBC-IA-Numerical-Results-2Bk}, option 'd' of
the advanced approach is applied and different sizes of the
selection set of $|\mathfrak{B}_k|$ are compared. The antennas are
set as $N_r=8$ for each MS and $N_t=12$ for each BS. The sizes of
$|\mathfrak{B}_k|$ are set as 1, 20, and 200. Then Fig.
\ref{Cellular-IFBC-IA-Numerical-Results-2Bk} shows three curves
corresponding to different sizes of $|\mathfrak{B}_k|$. From
\ref{Cellular-IFBC-IA-Numerical-Results-2Bk}, observe that when the
size of the selection set grows, the residue interference decreases.
So that the rates increase because the alignment is better.

\subsubsection{Model 2 with Interference Leakage regarding Abundant and Scarce BS antennas}

\begin{figure}[htpb]
  \begin{center}
    \includegraphics[width=4.5in]{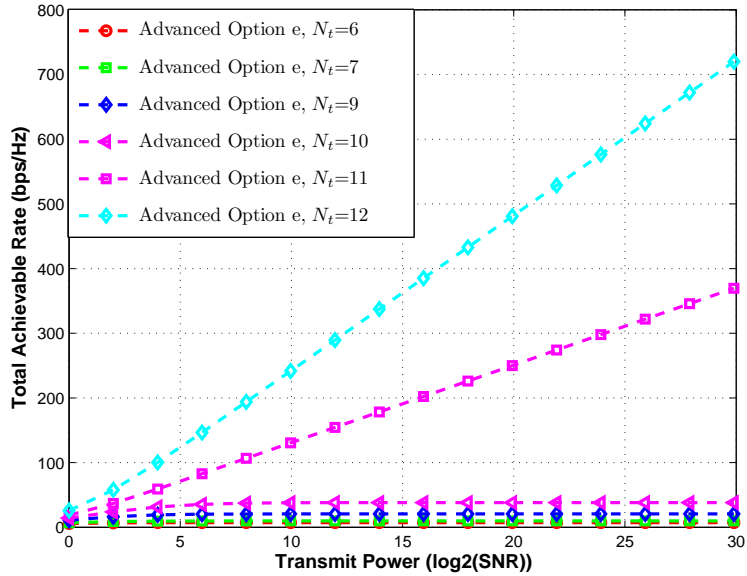}
  \end{center}
  \caption{Rates of Model 2 with Interference Leakage with Abundant and Scarce BS antennas}\label{Cellular-IFBC-IA-Numerical-Results-3p1}
\end{figure}

Since option 'd' and option 'e' are two robust measures of alignment
for model 2, and the above results show that option 'd' has only
zero DoF with precoder selection, the following results investigate
option 'e' could obtain appropriate DoF and rates depending on the
number of BS antennas. In Fig.
\ref{Cellular-IFBC-IA-Numerical-Results-3p1}, different numbers of
antennas $N_t$ are compared from $N_t=Md=6$ to $N_t=Md+Md=12$. It is
set as $N_t=6, 7, 9, 10, 11 \text{\ and\ } 12$. Then Fig.
\ref{Cellular-IFBC-IA-Numerical-Results-3p1} show six curves
corresponding to different values of $N_t$. From Fig.
\ref{Cellular-IFBC-IA-Numerical-Results-3p1}, observe that when
$N_t=12$ the network obtains full designed DoF and when $N_t=11$,
the network still obtains half designed DoF. When $N_t<11$, the
network has zero DoF due to residue interference.

\begin{figure}[htpb]
  \begin{center}
    \includegraphics[width=4.5in]{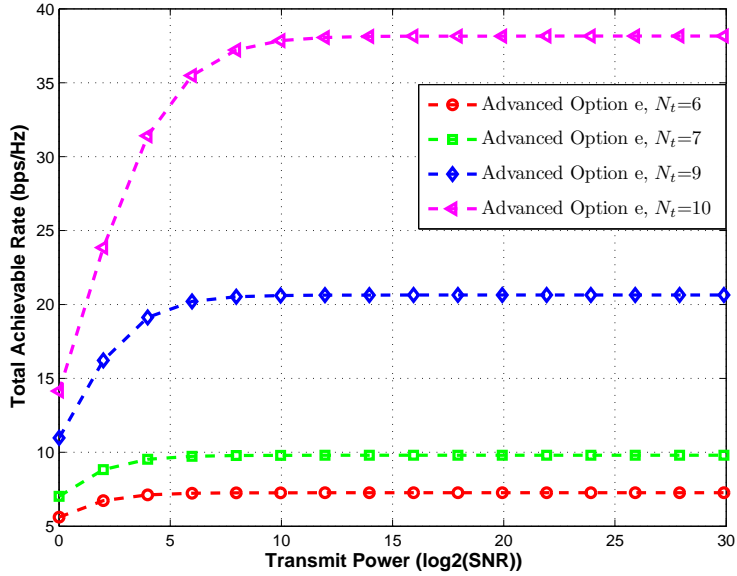}
  \end{center}
  \caption{Rates of Model 2 with Interference Leakage with Scarce BS antennas}\label{Cellular-IFBC-IA-Numerical-Results-3p2}
\end{figure}

Fig. \ref{Cellular-IFBC-IA-Numerical-Results-3p2} shows the four
curves of $N_t=6,7,9,10$ from Fig.
\ref{Cellular-IFBC-IA-Numerical-Results-3p1} separately in a zoomed
view. From Fig. \ref{Cellular-IFBC-IA-Numerical-Results-3p2},
observe that when the BS antennas grows, the residue interference
decreases. So that the rates increase because the alignment is
getting better.

\subsubsection{Model 3 with Basic and Advanced Approaches}

\begin{figure}[htpb]
  \begin{center}
    \includegraphics[width=4.5in]{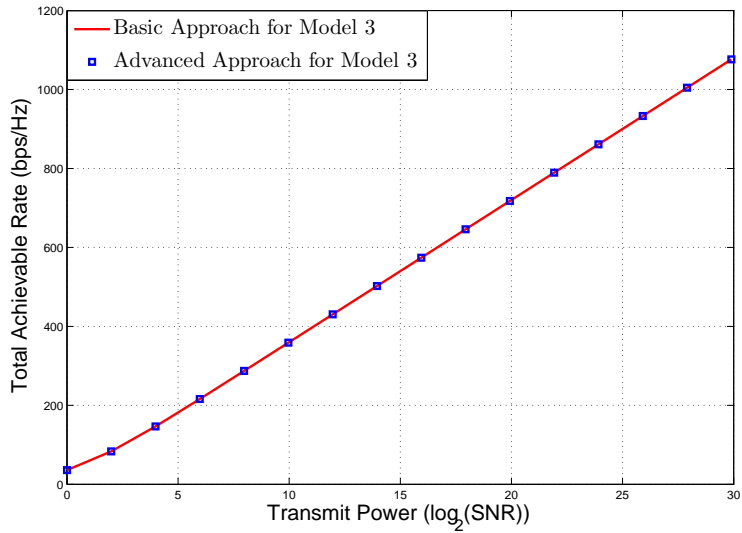}
  \end{center}
  \caption{Rates of Model 3 with Basic and Advanced Approaches}\label{Cellular-IFBC-IA-Numerical-Results-model3}
\end{figure}

For model 3, numerical results of the basic approach and advanced
approach are compared as in Fig.
\ref{Cellular-IFBC-IA-Numerical-Results-model3}. The network also
has $K=6$ cells. However, each cell has $M^{\star}=3$ cell-interior
MSs and $M^{\circ}=2$ cell-edge MSs. Each user has $d=2$ streams. In
the first case, the basic approach 1 is applied as a reference, i.e.
with cascaded precoders on BS side. The antennas are set as
$N_t=Md=10$ for each BS, $N_r=d=2$ for each cell-interior MS and
$N_r=Md+d=12$ for each cell-edge MS. In the second case, the
advanced approach is applied. The antennas are set as
$N_t=M^{\circ}d+Md=14$ for each BS, $N_r=d=2$ for each cell-interior
MS and $N_r=d=2$ for each cell-edge MS. From Fig.
\ref{Cellular-IFBC-IA-Numerical-Results-model3}, observe that both
the basic approach and the advanced approach obtain full DoF.
Nevertheless, the advanced approach has a lower usage of antennas in
the whole.

\section{Conclusion}
This work generalizes the design and analysis of interference
alignment in downlink channels of the multi-cell multi-user network.
In such a complicated network, it is necessary to deal with both
inter-user interference (IUI) and inter-cell interference (ICI).
Conventional alignment schemes for peer-to-peer interference
channels (IC) are not available to handle cellular networks. Also
conventional alignment schemes for two-cell networks are not robust
enough to settle multi-cell networks. Therefore, novel approaches
are proposed and compared for typical multi-cell networks. For
practical concerns, the Wyner-type networks are particularly
investigated. According to the broadcast nature of the cellular
networks, the alignment could be implemented from both the BS side
and the MS side to achieve a reasonable balance. The alignment
process is also in a dynamic manner to form all the coders. System
conditions of antennas, CSI, and complexity are compared
deliberately. Advanced approaches exploit the network structure to
further save the usage of antennas and also explore robust measures
to be adaptable to practical systems.

\bibliographystyle{IEEEtran}
\bibliography{Thesis}

\end{document}